\documentclass{ws-ijmpe}
\usepackage[super,compress]{cite}
\usepackage[dvips,breaklinks]{hyperref}
\hypersetup{colorlinks,urlcolor=black,citecolor=black,linkcolor=black,filecolor=black}
\usepackage{breakurl}
\usepackage{url}
\usepackage{cite}
\newcommand{\eq}{\begin{eqnarray}}
\newcommand{\en}{\end{eqnarray}}
\def\bfs{\mbox{{\bf $\sigma$}}}

\newcommand{\ba}[1]{\begin{eqnarray} \label{(#1)}}
\newcommand{\ea}{\end{eqnarray}}

\newcommand{\newc}{\newcommand}

\newc{\lra}{\leftrightarrow}
\newc{\beq}{\begin{equation}}
\newc{\eeq}{\end{equation}}
\newc{\barr}{\begin{eqnarray}}
\newc{\earr}{\end{eqnarray}}

\usepackage[normalem]{ulem}
\def\be{\begin{equation}}
\def\ee{\end{equation}}
\def\dg{\dagger}
\def\bee{\begin{eqnarray}}
\def\eee{\end{eqnarray}}

\newcommand{\lf}{\left(}
\newcommand{\rh}{\right)}

 \newcommand{\gh}[1]{\gamma^{#1}}
 \newcommand{\gd}[1]{\gamma_{#1}}

\usepackage{comment}
\newcommand{\ket}[1]{\left|#1\right\rangle}
\newcommand{\bra}[1]{\left\langle #1\right|}

\begin{document}   
%

\topmargin -0.50in

\title{Neutrinoless double beta decay and neutrino mass}

\author{J.D. VERGADOS}

\address{ARC Centre of Excellence in Particle Physics (CoEPP),
Department of Physics,
University of Adelaide,
Adelaide SA 5005,
Australia\footnote{Permanent address: Theoretical Physics Division, University of Ioannina, 
GR 451 10 Ioannina, Greece.},\\Center for Axion and Precision Physics Research, Institute
for Basic Science (IBS), Daejeon 34141, Republic of Korea  
and \\
Board of Trustees, Technical Educational Institute, Kozani, Greece.}

\author{H. EJIRI}
\address{RCNP, Osaka University, Osaka, 567-0047, Japan and\\
  Nuclear Science, Czech Technical University, Brehova, Prague, Czech Republic.}
  \author{F \v Simkovic}
\address{Laboratory of Theoretical Physics, JINR,
141980 Dubna, Moscow region, Russia  and\\
 Department of Nuclear Physics and Biophysics, 
Comenius University, Mlynsk\'a dolina F1, SK--842 15
Bratislava, Slovakia and\\Czech Technical University in Prague, 128-00 Prague, Czech Republic.}

\date{\today}

\maketitle
$$\mbox{Abstract}$$
\begin{abstract}
  The  observation of neutrinoless double beta decay will have  important consequences. First it will signal that lepton number is not conserved and the neutrinos are Majorana particles. Second, it represents our best hope for determining the absolute neutrino mass scale at the level of a  few tens of meV. 
 To achieve the last goal, however, certain hurdles have to be overcome involving particle, nuclear and experimental physics.\\
Particle physics is important since it provides the mechanisms for neutrinoless double beta decay. In this review we emphasize the light neutrino mass mechanism.\\
Nuclear	physics is important for extracting the useful information from the data. One must accurately  evaluate the relevant nuclear matrix elements, a formidable task.
To this end, we review the recently developed sophisticated nuclear structure approaches, employing different methods  and techniques of calculation.
We also examine the question of quenching of the axial vector coupling constant, which may have important consequences on the size of the nuclear matrix elements.\\
From an experimental point of view it is challenging, since the life times are extremely long and one has to 
fight against formidable backgrounds. One needs large isotopically enriched sources and detectors 
with good energy resolution and very low background.\\
\end{abstract}

PCCS numbers:13.15.+g, 14.60.Pq, 14.60.St, 21.60.Cs, 21.60.Jz, 23.40.-s, 14.60.Pq , 23.40.Bw, 24.80.+y, 29.90.+r


%
%
%
\section[A brief history of double beta decay]{A brief history of double beta decay}
%

%
%
Double beta decay (DBD), namely the non exotic two-neutrino double-beta decay 
($2\nu\beta\beta$-decay)
\begin{equation}
(A, Z) \to (A, Z + 2) + e^- + e^- + {\overline\nu}_e + {\overline\nu}_e,
\label{eq:a2}   
\end{equation}
was first considered in publication \cite{GOEPMAY}
of Maria Goeppert-Mayer in 1935, following the suggestion of  Eugene Wigner about one year after 
the Fermi weak interaction theory appeared. In this  work
 an expression for the $2\nu\beta\beta$-decay rate was 
derived and a half-life of  $10^{17}$ years was estimated, assuming a Q-value 
of about 10 MeV. 

Two years later (1937) Ettore  Majorana formulated a new theory of 
neutrinos, whereby the neutrino $\nu$ and the antineutrino $\overline\nu$ are indistinguishable, 
and suggested antineutrino induced $\beta^-$-decay for experimental
verification of this hypothesis \cite{emajorana}. 
In 1939, Wolfgang Furry considered for the first time
neutrinoless double beta decay
($0\nu\beta\beta$-decay),
\begin{eqnarray}
(A,Z) \rightarrow (A,Z+2) + e^- + e^-,
\label{eq:a}   
\end{eqnarray}
The available energy $\Delta$ is equal to the $Q$-value of the reaction, i.e. the mass difference of the ground states of the two atoms involved.\\
In 1952 Henry Primakoff  \cite{Pr52}
calculated the electron-electron angular correlations and electron energy spectra 
for both the $2\nu\beta\beta$-decay and the $0\nu\beta\beta$-decay, producing a
useful tool for distinguishing between the two processes.
At that time, however,  nothing was known about the chirality suppression of the 
$0\nu\beta\beta$-decay. It was believed that, due to a considerable phase-space
advantage, the $0\nu\beta\beta$-decay mode dominates the double beta decay rate. 
Starting in 1950 this phenomenon was exploited in early geochemical, radiochemical 
and counter experiments.
It was found that the measured lower limit on the 
$\beta\beta$-decay half-life far exceeds the values expected for this process,
$T^{0\nu}_{1/2} \sim 10^{12}-10^{15}$ years. In 1955 the Raymond Davis experiment \cite{davisno} searching for the antineutrinos from reactor via nuclear reaction
${\overline{\nu}}_e + {^{37}Cl} \rightarrow {^{37}Ar} + e^-$, produced a zero result.
The above experiments were interpreted as proof that the neutrino 
was not a Majorana particle, but a Dirac particle. This prompted the
introduction of the lepton number to distinguish the neutrino from its
antiparticle. The assumption of lepton number conservation allows the 
$2\nu\beta\beta$-decay but forbids the $0\nu\beta\beta$-decay, in which lepton
number is changed by two units. 

 The first geochemical observation of the $\beta\beta$-decay,
with an estimated half-life $T_{1/2} (^{130}Te) = 1.4 \times 10^{21}$
years, was announced by Ingram and Reynolds in 1950 \cite{ing}. 
Extensive studies have been made by Gentner and Kirsten \cite{kir67,kir67b} and others \cite{tak66},\cite{sir72} on such rare-gass isotopes as $^{82}$Kr, 
$^{128}$Xe, and $^{130}$Xe, which are $\beta\beta$-decay products of $^{82}$Se $^{128}$Te, and $^{130}$Te, 
respectively,  obtaining half lives around 10$^{21}$y for $^{130}$Te. 


Within the Standard Model (SM) it became apparent that the assumption of lepton number
conservation led to the neutrino being strictly massless.
With the development of Grand Unified Theories (GUT's) of the electroweak and strong 
interactions, it was realized that lepton number conservation 
was the result of a global symmetry not of a gauge symmetry and had to be broken at some level. 
In such models one could distinguish between the neutrinos produced in weak interactions (flavor neutrinos) and the eigenstates of the world Hamiltonian. The latter eigenstates can naturally be Majorana neutrinos, while  Dirac type eigenstates could arise as a special case. The flavor neutrinos could still  be of Dirac type, if the Majorana phases of the eigenstates are all the same, in agreement with Davis experiment \cite{davisno}.
Thus, through the pioneering work of Kotani and his group
		\cite {DTNOT},  the interest in 
 $0\nu\beta\beta$-decay experiments was revived and  brought it again to the attention of the nuclear physics 
	community.

The  $0\nu\beta\beta$-decay, which involves the emission of 
two electrons and no neutrinos, has been found to be more than a tool in studying lepton number violating processes. Schechter and Valle
proved that, if the $0\nu\beta\beta$-decay 
takes place, regardless of the mechanism causing it, the neutrinos are Majorana particles 
with non-zero mass \cite{SVa82,klko96}. It was also recognized that 
the GUT's and R-parity violating
SUSY models offer a plethora of the $0\nu\beta\beta$-decay 
mechanisms triggered by exchange of neutrinos, neutralinos, 
gluinos, leptoquarks, etc. \cite{FKSS97,FKS98,WKS99}.

The experimental effort concentrated on high $Q_{\beta\beta}$ isotopes,
in particular on $^{48}$Ca, $^{76}$Ge, $^{82}$Se, $^{96}$Zr, $^{100}$Mo,
$^{116}$Cd, $^{130}$Te, $^{136}$Xe and $^{150}$Nd \cite{zdes02,eji05,AEE08,RMP08}.  
In 1987 the first actual
laboratory observation of the two neutrino double beta decay ($2\nu\beta\beta$-decay) was done 
for $^{82}$Se by M. Moe and collaborators \cite{ell87}, who used a time
projection chamber. Within the next few years, experiments employing 
counters were able to detect $2\nu\beta\beta$-decay of many nuclei. 
In addition, the experiments searching for the signal of the $0\nu\beta\beta$-decay
pushed by many orders of magnitude the experimental 
lower limits for the $0\nu\beta\beta$-decay half-life of different nuclei.


A great leap forward was achieved, when, early measurements of neutrinos produced in the sun, in the 
atmosphere, and by accelerators, suggested that neutrinos may 
oscillate from one "flavor" (electron, muon, and tau) to 
another,  expected if the neutrinos are massive and non degenerate in mass. Non-zero neutrino mass can be accommodated by fairly straightforward 
extensions of the SM of particle physics. Thus now, starting in 1998, we have convincing evidence about the existence of non zero neutrino masses in  SuperKamiokande \cite{SUPERKAMIOKANDE}, SNO, \cite{SOLAROSC} KamLAND \cite{KAMLAND} and other experiments. Such experiments, however, cannot determine the absolute scale of neutrino mass. 
%
So the determination of the scale of neutrino mass has been directed to other methods, such as  cosmological observations, $\beta$-decay experiments and, especially if the scale happens to be in the meV range, to
 $0\nu\beta\beta$-decay, see, e.g., the recent review \cite{ROP12}.

So far the $2\nu\beta\beta$-decay has been recorded for eleven nuclei 
($^{48}$Ca, $^{76}$Ge, $^{82}$Se, $^{96}$Zr, $^{100}$Mo, $^{116}$Cd, 
$^{128}$Te, $^{130}$Te, $^{150}$Nd, $^{136}$Xe, $^{238}$U) \cite{zdes02,eji05,AEE08}.  
In addition, the $2\nu\beta\beta$-decay 
 to the first $0^+$ excited state of the daughter 
nucleus has been observed  in the case of the targets  $^{100}$Mo and $^{150}$Nd. Furthermore the two-neutrino double electron capture
process in $^{130}Ba$ has been recorded. 

Neutrinoless double beta decay has not yet been confirmed. The strongest limit recently obtained is $T_{1/2}>1.1 \times 10^{26}$y by Gando {\it et al} \cite{Xelimit} (see section \ref{sect:exp} for details). 
\begin{figure}[!t]
\centerline{\psfig{file=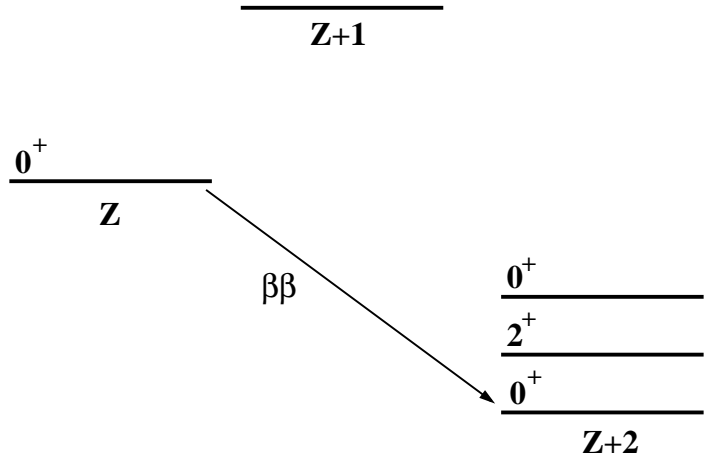,width=0.8\textwidth}}
\caption{Schematic diagrams of $\beta \beta $ decays in nuclear femto (10$^{-15}$m) laboratories,
where single $\beta$-decay is forbidden, while neutrinoless double-beta decay is allowed. 
\label{fig:6.1}}
\end{figure}


If the neutrinos are Majorana particles other related processes in which the charge of the nucleus
is decreased by two units may also occur, if they happen to be allowed by
energy and angular momentum conservation laws, e.g. double positron emission, electron positron conversion,
resonant neutrinoless double electron capture ($0\nu$ECEC) \cite{WINTER,Ver83,BeRuJar83,DK93}.
This is an interesting topic with a lot of theoretical
work\cite{SujWy04,lukas,SIMKO11,SimKriv09,verg11} and it appears
promising in view of progress in accurately determining the $Q$-values \cite{AUDI03} needed
to establish the condition of resonance employing Penning traps  \cite{BLAUM06,penning1,BNW10}.
Recently, the accuracy of $Q$-values at around 100 eV was achieved
\cite{DEC11,redshaw07,redshaw09,SCIE09,RAKH09,kolhinen,mount10,elis2,EliNov,elis4,elis5,droese11},  
which has already allowed to exclude some of isotopes from the list of the most promising 
candidates (e.g., $^{112}$Sn and $^{164}$Er) for searching the $0\nu$ECEC. In spite of the fact
that  an increased experimental activity in the field of the resonant $0\nu$ECEC
\cite{BELLI11,barab1,barab2,belli09,rukh11,FPS11,BELLI11b}  in the case of $^{106}$Cd \cite{rukh11}
and $^{112}$Sn \cite{barab2}. Resonant $0\nu$ECEC has some  important advantages with respect
to experimental signatures and background conditions, but we do not know of any experiment under
way in this direction and we are not going to review this field further.
%

Anyway  $0\nu\beta\beta$-decay (Eq. (\ref {eq:a})),
seventy five years after it was first conceived, seems to be the most
likely to yield the  information \cite{Ver86,HS84,DTK85,Tom91,SC98,FS98,Ver02,RODEJ11} we seek.


%
%
%
\section[Motivation for pursuing  neutrinolless double beta decay]{Motivation for pursuing  neutrinolless double beta decay}
%

		From a nuclear physics \cite {SC98,FS98,SDSJ97,RCN95,CNPR96,SSDV92,CPZ90}
		point of view, calculating the relevant nuclear matrix elements is indeed
		a challenge. First almost all nuclei, which can undergo double beta decay,
		are far from closed shells and some of them are even deformed. One thus faces
		a formidable task. Second the nuclear matrix elements are  small compared to
		 a canonical value, like the one associated with the matrix element to the (energy non
		allowed) double Gamow-Teller resonance or a small fraction of some appropriate sum
		rule. Thus, effects which are normally negligible, become important here.
		Third, in many models the dominant mechanism for $0\nu\beta\beta$-decay
		may not involve intermediate light neutrinos, but very heavy particles.
		Thus one must be able to cope with the short distance behavior of the
		relevant operators (see section \ref{secnme} for details).


		It is also important from a particle physics point of view. The recent discovery of neutrino oscillations
                \cite{SUPERKAMIOKANDE,AHARMIN,CHOOZ,ARAKI,dayabay,reno12}  have given the first evidence of the fact that the neutrinos are massive, which necessitates to go   
		beyond the Standard Model (SM) of particle physics
		 More specifically these experiments showed that the neutrinos are admixed and determined all three mixing angles (for a global analysis see, e.g.,  \cite{SCHWETZ,Capozzi14}). Furthermore they determined one  square mass difference $\Delta m_{21}^2$ and the absolute value of the other. i.e.  $\Delta m_{32}^2$ or $\Delta m_{31}^2$. Future neutrino oscillations in matter are expected to fix the unknown sign. Neutrino oscillations, however, cannot  determine:
		\begin{itemize}
		\item Whether the neutrinos are Majorana or Dirac particles. \\
		 It is obviously important to proceed further and decide on this important issue.
		 Neutrinoless double beta decay can achieve this, even though, as we have mentioned,  there might
		be other lepton violating processes  contributing to  $0\nu\beta\beta$-decay. It is known
 that whatever lepton number violating process  gives rise to $0\nu\beta\beta$-decay,
		it can be used to generate a Majorana mass for the neutrino \cite{SVa82}. 
		\item The scale of the neutrino masses.\\
		 This task can be accomplished by astrophysical observations or  via other experiments involving  low energy weak decays, like triton decay or electron capture, or the $0\nu\beta\beta$-decay.  It seems that for a  neutrino mass  scale in meV, ($10^{-3}$ eV), region, the best process to achieve this is the $0\nu\beta\beta$-decay. The extraction of neutrino masses from such observations will be discussed in detail and compared with each other later (see section \ref{sec:extrnumass}) \cite{ROP12}. This mechanism, however, is not the only one, which can induce  $0\nu\beta\beta$-decay. 
		If, however, $0\nu\beta\beta$ is ever found to occur, it will be possible to disentangle the most important neutrino mass contribution, involving the neutrino mass scale, from the other mechanisms, provided that data on a number of targets become available. 
		 \end{itemize}
		 The neutrino hierarchy, i.e. whether the neutrinos are almost degenerate, the normal hierarchy ($\Delta m_{32}^2>0)$ or the inverted hierarchy ($\Delta m_{31}^2<0)$, can also be inferred from double beta decay.	
		 For details on such issues see  recent reviews \cite{ROP12,G-GM08,PaesRod15,HHOPS15}.
		 As we have mentioned to extract useful information from the $0\nu \beta\beta$ decay one must know the nuclear matrix elements\footnote{It is not possible to deduce the expected neutrino mass from $0\nu \beta\beta$  employing Baysian statistics \cite{baysian15}  based on
a Markov chain Monte Carlo \cite{MCMC02}, since one cannot avoid the issue related to other possible mechanisms contributing to the process.}. Efforts to this end can be summarized as follows:\\
1. Shell model calculations. These have a long history 
		\cite{CPZ90,SSDV92,Ver76,HSS82,SV83,ZBR90,ZB93} in double beta decay
		calculations. In recent years it has lead to large matrix calculations  in
		traditional as well as Monte Carlo types of formulations
		\cite{SDSJ97,RCN95,CNPR96,Retal96,NSM96,KDL97,edf}. For a more complete set of
		references as well as a discussion of the appropriate
                effective interactions see Ref. \cite{SC98}).\\
2. QRPA calculations. There have been a number of such calculations covering almost all nuclear
		targets \cite{VZ86,CAT87,MBK88,EVJP91,RFSK91,GV92,SC93,CS94,SSVP97,SPF98,cheoun,MUT97,SRFV13}.
                 These involve a number of collaborations, but the most
                 extensive and complete calculations in one way or another
                 include the Tuebingen group.
		 We also have seen some refinements of QRPA, like proton neutron pairing,
		inclusion of 
		renormalization effects due to Pauli principle corrections
                \cite{TS95,SSF96} and isospin restoration\cite{TBC13}. Other less conventional approaches, like
                operator expansion techniques have also been employed
                \cite{FS98}.\\
  3.  Other nuclear models. Recently, calculations based on the Projected 
                Hartree-Fock-Bogoliubov (PHFB) method \cite{phfb}, the Interacting Boson Model (IBM)  \cite{IBM09,BaKotIac15}
                and the Energy Density Functional (EDF) method  \cite{edf} and  relativistic EDF (REDF) \cite{REDF15}
                entered the field of such calculations. \\The above schemes, in 
                conjunction with the other
		improvements mentioned above, offer some optimism in our efforts for obtaining
		nuclear matrix
		elements  accurate enough to allow us to extract reliable values of the lepton
		violating parameters from the data.\\
		The experimental results will be examined and discussed in section \ref{sect:exp}.

%
%
%
\section[The  neutrino mass and mixing]{The  neutrino mass and mixing}
\label{sec:numass}
%
                
		Within the SM of elementary particles, with the particle content of 
                the gauge bosons $A_{\mu},Z_{\mu}$ and $W_{\mu}^{\pm,0}$, the Higgs scalar isodoublet 
                $\Phi=(\phi^0,\phi^-)$ (and its conjugate $\Phi^{*}$) and the fermion fields arranged in:
		\begin{itemize}
		\item  Isodoublets: $(u_{\alpha L},d_{\alpha L})$ and $(\nu_{\alpha L},e_{\alpha L})$ 
                for quarks and leptons respectively and
		\item Isosinglets: $u_{\alpha R},d_{\alpha R}$ and $e_{\alpha R}$ 
		\end{itemize}
		where $\alpha$ is a family index taking three values.  In the context of the standard model (SM) the neutrinos are massless. They can not obtain mass after the symmetry breaking, like the quarks and the charged leptons do, since the right handed neutrino is absent.
\subsection{Neutrino mass}
\label{Neutrinomass}		
		The minimal extension of the SM that would yield mass for the neutrino is to introduce an isosinglet right handed neutrino. Then one can have a Dirac mass term  arising via coupling of the leptons and Higgs \cite{ROP12,PaesRod15,HHOPS15}. The existence of Dirac neutrinos is fine within the above minimal extension, but this is not interesting from our point of view, since the particle cannot be the same with its antiparticle and, thus, it cannot lead to neutrinoless double beta decay. 
		  Furthermore in grand unified theories one is faced with the problem that these neutrinos are going to be very heavy with a mass similar to that of up quarks, which is clearly unacceptable. So such a model is inadequate\footnote{There may exist light Dirac neutrinos in theories formulated in extra dimensions, see e.g. the recent review by Smirnov \cite{SMIRNOV04}. If these neutrinos do not couple to the usual leptons they are of little interest to us. If they do and it so happens that  the standard neutrinos are Majorana, they also become Majorana, except in the case of very fine tuning.}.  The next extension of the SM is to introduce a Majorana type mass involving the isosinglet neutrinos and an additional isosinglet Higgs field, which can acquire a large vacuum expectation value, an idea essentially put forward by Weinberg \cite{WEINBERG79} long time ago. Thus the neutrino mass matrix becomes \cite{ROP12}:
	\beq
	{\mathcal M}=
		\left(\bar{\nu}_L,\bar{\nu}_L^c\right)\left(\begin{array}{cc}0&\quad m^D
		\\
		(m^D)^T& m_R
		\end{array}\right ) \left(\begin{array}{c}\nu_R\\ \nu_R^c  \end{array} \right)
		\label{Eq:see-saw}
	\eeq
Thus, provided that the Majorana mass matrix has only very large eigenvalues, one obtains an effective Majorana $3\times 3$ matrix:
\beq
{\mathcal M}_{\nu}=-\bar {\nu}_L(m^D)^T M_R^{-1}m_D\nu^c_R,
\label{Eq:MajMat}
\eeq
which can provide small neutrino masses provided that the eigenvalues of the matrix $M_R$ are sufficiently large. $M_R$ can be arbitrarily large and is identified  the total lepton number violating (LNV) scale indicated by $m_{LNV}$ .  This  new scale, commonly associated with the  vacuum expectation of the isosinglet, does not affect the low energy scale arising from the vacuum expectation value of the standard Higgs particles. \\This is the celebrated see-saw mechanism. More precisely  there exist three see-saw types ( see, e.g., Abada {\it et al} \cite{SeeSaw07} for a summary and more recently \cite{L-PMP15,HuLiSm15}). Furthermore in some of these versions \cite{HuLiSm15} new contributions
to  neutrinoless double beta decay are claimed. A more systematic
 decomposition of the neutrinoless double beta decay
operator has also recently appeared  \cite{BonHirOtaWint}.

 It is possible for the heavy Majorana neutrinos not to be equally heavy. By appropriately arranging the corresponding Dirac coupling, it is possible to get one Majorana neutrino in the keV scale, which couples very weakly to the three light neutrinos. This is the so called sterile neutrino.\\
Anyway with the see-saw mechanism the neutrino flavors get admixed, the resulting eigenstates are Majorana particles and lepton number violating interactions, like $0\nu\beta \beta$ decay, become possible.

Other extensions of the SM, which do not require the presence of right-handed neutrinos, are possible in which a light $3\times 3$ Majorana mas matrix $m_{\nu}$ can be generated  via an isotriplet scalar acquiring a vacuum expectation value or via its couplings to two isodoublet scalar fields   at tree  or at the one  loop level. It is also possible to achieve this via generic diagrams involving Weinberg's idea \cite{WEINBERG79, Pascoli08}. By introducing exotic  isosinglet or isotriplet fermions  or in the context of  R-parity violating supersymmetry \cite{GKSA06}.
 We will not consider these possibilities, since they have been examined elsewhere \cite{ROP12}. We should mention, however, that there exist models which generate Majorana masses at the 2-loop level,  first proposed long time ago \cite{Twoloop80} as an economic way of getting neutrino Majorana mass. In one such approach  \cite{Twoloop15}, reinventing  the extended Majorana matrix \cite{Twoloop91}, now coined inverse see-saw, it is claimed that, using the available neutrino oscillation data, the full Majorana matrix can be determined leading to a prediction of the Majorana phases.

\subsection[Neutrino mixing]{Neutrino mixing}
\label{sec:numixing}
We have seen above that in general the neutrino mass matrix, Eq. (\ref{Eq:see-saw}), is a complex symmetric matrix. It can, however, be diagonalized by separate left and write unitary transformations \cite{Ver86} \cite{ROP12}:
 \barr
	S_L\leftrightarrow
\left(\nu^0_L,\nu^{0c}_L\right )&=&\left (\begin{array}{lr}
	S^{(11)}& S^{(12)}\\
                    S^{(21)}& S^{(11)}                  
\end{array}                      
\right )
\left (\begin{array}{c}
	                                  \nu'_L\\
	                                  N'_L               
\end{array}                      
\right ),\nonumber\\
	S_R\leftrightarrow
\left(\nu^{0c}_R,\nu^0_{R} \right )&=&\left (\begin{array}{lr}
	\left(S^{(11)} \right )^{*}& \left(S^{(12)} \right )^{*}\\
                    \left(S^{(21)} \right )^{*}& \left(S^{(22)} \right )^{*}                 
\end{array}                      
\right )
\left (\begin{array}{c}
	                                  \nu'_R\\
	                                  N'_R                
\end{array}                      
\right )
\label{Eq:mixing}
\earr
where we have added the superscript 0 to stress that they are the states entering the weak interactions. $S^{(ij)}$ are $3 \times 3$ matrices with $S^{(11)}$ and $S^{(22)}$ approximately unitary, while $S^{(12)}$ and $S^{(21)}$ are very small. 
$(\nu'_L,N'_L )$ and $(\nu'_R,N'_R)$ are the eigenvectors from the left and the right  respectively. 
Thus the neutrino mass in the new basis takes the form:
\beq
{\mathcal M}_{\nu}=\sum_{j=1}^3 \left ( m_j \bar{\nu}'_{jL} \nu'_{jR}+M_j \bar{N}'_{jL} N'_{jR} \right )+H.C.
\eeq
This matrix can be brought into standard form by writing:
$$
m_j=|m_i|e^{-i \alpha_j},\quad M_j=|M_i|e^{-i \Phi_j}
$$
and defining:
$$
\nu_j=\nu'_{jL}+e^{-i \alpha_j}\nu'_{jR}\quad N_j=\nu'_{jL}+e^{-i \Phi_j}N'_{jR}
$$
Then
\beq
{\mathcal M}_{\nu}=\sum_{j=1}^3 \left (| m_j| \bar{\nu}_{j} \nu_{j}+|M_j| \bar{N}_{j} N_{j} \right )
\eeq
Note, however, that:
\barr
\nu^c&=&\nu'_{jR}+e^{i \alpha_j}\nu'_{jL}=e^{i \alpha_j}\left (\nu'_{jL}+e^{-i \alpha_j}\nu'_{jR} \right )=e^{i \alpha_j}\nu_j\nonumber\\
N^c&=&N'_{jR}+e^{i \Phi_j}N'_{jr}=e^{i \Phi_j}\left (\nu'_{jR}+e^{-i \Phi_j}N'_{jR} \right )=e^{i \Phi_j}N_j
\label{Eq:Majphase}
\earr
i.e. they are Majorana neutrinos with the given Majorana phases. Furthermore
$$
\nu_{iL}=\nu'_{iL},\quad \nu_{iR}=e^{-i \alpha_j}\nu'_{iR},\quad N_{iL}=N'_{iL},\quad N_{iR}=e^{-i \Phi_j}\nu'_{iR}
$$
The first  of Eqs. (\ref{Eq:mixing}) remains unchanged, while the second can now be written as 
\beq
S_R\leftrightarrow
\left(\nu^{0c}_R,\nu^0_R \right )=\left (\begin{array}{lr}
	\left(S^{(11)} \right )^{*}& \left(S^{(12)} \right )^{*}\\
                    \left(S^{(21)}\right )^{*}& \left(S^{(22)} \right )^{*}                  
\end{array}                      
\right )
\left (\begin{array}{c}
	                                  e^{i \alpha}\nu_R\\
	                                  e^{i \Phi}N_R                
\end{array}                      
\right )
\label{Eq:mixing2}
\eeq
where $e^{i \alpha}$ and  $e^{i \Phi}$ are diagonal matrices containing the above Majorana phases.

 The full parametrization of matrix $\mathcal{U}$ includes 15 rotational
 angles and 10 Dirac and 5 Majorana CP violating phases. 

 The neutrinos interact with the charged leptons via the charged current (see below). So the effective coupling of the neutrinos to the charged leptons involves the mixing of the electrons $S^e$. Thus  the standard mixing matrix appearing in the absence of right-handed neutrinos is:
 \beq
 U_{PMNS}=U=U^{(11)}=\left (S^{(e)} \right )^{+} S^{(11)}
 \eeq
  The other entries are defined analogously:
  \beq
 U^{(ij)}=\left (S^{(e)} \right )^{+} S^{(ij)},\quad (ij)=(12),(21),(22)
 \eeq
 In particular the usual electronic neutrino is written as:
 \beq
 \nu_{eL}=\sum_j\left (U^{(11)}_{ej}\nu_{jL} +U^{(12)}_{ej}N_{jL} \right ),\quad \nu^c_{eL}=\sum_j\left (U^{(21)}_{ej}\nu_{jL} +U^{(22)}_{ej}N_{jL} \right )
 \label{Eq:leftnu}
 \eeq
  \barr
 \nu^c_{eR}&=&\sum_j\left (U^{(11)}_{ej})^*e^{\alpha_j}\nu_{jR} +(U^{(12)}_{ej})^*e^{\Phi_j}N_{jR} \right ),\nonumber\\ \nu_{eR}&=&\sum_j\left (U^{(21)}_{ej})^*e^{\alpha_j}\nu_{jR} +(U^{(22)}_{ej})^*e^{\Phi_j}N_{jR} \right )
 \label{Eq:rightnu}
 \earr
 In other words the left handed neutrino may have a small heavy  component, while the  right handed neutrino may have a small light component. Note also that the neutrinos appearing in weak interactions can be Majorana particles in the special case that all Majorana phases are the same.

Unfortunately the above notation is not unique. For the reader's convenience we mention that sometimes the notation \beq
U^{(11)}\rightarrow U, \,U^{(12)}\rightarrow S,\,U^{(21)}\rightarrow T,\,U^{(22)}\rightarrow V
\label{maticazmies}
\eeq
is employed \cite{Xing}.
It is also possible to decompose the $6\times 6$ mixing matrix  as follows \cite{Xing}
\bee
 {\mathcal{U}} &=& \left(
\begin{array}{ll}
\mathbf{1} & \mathbf{0}\\
 \mathbf{0} & U_0\\
 \end{array}
\right)
\left(
\begin{array}{ll}
A & R\\
S & B\\
 \end{array}
\right)
\left(
\begin{array}{ll}
V_0 & \mathbf{0}\\
 \mathbf{0} & \mathbf{1}\\
 \end{array}
\right),
\label{maticaXing}
 \eee
 where $\mathbf{0}$ and $\mathbf{1}$ are the $3\times 3$ zero and identity matrices, respectively. 
 The parametrization of matrices
 A, B, R and S and corresponding orthogonality relations are given in \cite{Xing}. In the limit case
 A = $\mathbf{1}$, B =$\mathbf{1}$, R = $\mathbf{0}$ and S = $\mathbf{0}$ there would be a separate mixing of
 heavy and light neutrinos, which would participate only in left and right-handed currents, respectively. In that
 case only the neutrino mass mechanism of the $0\nu\beta\beta$-decay would be allowed and exchange
 neutrino momentum dependent mechanisms associated with the $W_L$-$W_R$ exchange and $W_L$-$W_R$ mixing would be forbidden.
 If masses of heavy neutrinos are above the TeV scale, the mixing angles responsible for mixing of light and
 heavy neutrinos are small. By neglecting the mixing between different generations of light and heavy neutrinos
 A, B, R and S matrices can be approximated as follows:
\begin{eqnarray}
  A \approx \mathbf{1}, B \approx \mathbf{1}, R \approx \frac{m_D}{m_{LNV}}\mathbf{1}, S \approx -\frac{m_D}{m_{LNV}}\mathbf{1}.
\label{whynot}
\end{eqnarray}
Here, $m_D$ represents energy scale of charged leptons and $m_{LNV}$ is the total lepton number violating scale,
which corresponds to masses of heavy neutrinos.
For sake of simplicity the same mixing angle is assumed for each generation of mixing of light and heavy neutrinos.
We see that $U_0$ can be identified to a good approximation with the PMNS matrix and $V_0$ is its analogue for
heavy neutrino sector. Both of them are almost unitary matrices of order 1, but unrelated to each other. Since $V_0$ is unknown, sometimes it is assumed that the structure of
$V_0$ is the same one as $U_0$. 

The situation is very much simplified, if the mixing between the light and heavy neutrinos is small \cite{ROP12} by diagonalizing the matrix given by Eq. (\ref{Eq:MajMat}). Then the mixing is described by the Pontecorvo-Maki-Nakagawa-Sakata neutrino mixing matrix $U_{PMNS}$, which is parametrized 
by
 \beq
 U_{PMNS} = \left(
\begin{array}{lll}
 c_{12} c_{13} & c_{13} s_{12} & e^{-i \delta } s_{13} \\
 -c_{23} s_{12}-e^{i \delta } c_{12} s_{13} s_{23} & c_{12}
   c_{23}-e^{i \delta } s_{12} s_{13} s_{23} & c_{13} s_{23} \\
 s_{12} s_{23}-e^{i \delta } c_{12} c_{23} s_{13} & -e^{i \delta }
   c_{23} s_{12} s_{13}-c_{12} s_{23} & c_{13} c_{23}
\end{array}
\right),
\label{pmns}
 \eeq
 where
 \beq
 c_{ij}\equiv \cos{(\theta_{ij})}, \quad s_{ij}\equiv \sin{(\theta_{ij})}.
 \eeq
$\theta_{12}$, $\theta_{13}$ and $\theta_{23}$
and three mixing angles and $\delta$ is the 
CP-violating phase. Sometimes the Majorana phases are absorbed into the mixing matrix. Then the above  matrix is  multiplied from the right by the diagonal matrix, e.g.  $\mbox{diag}(e^{i\alpha_1},e^{i\alpha_2},1)$.


The theoretical goal is to drive the above matrix on the basis of suitable extensions of the Standard Model (SM) as mentioned above \cite{ROP12}, which are not going to be discussed here. In recent years, especially after the first neutrino oscillation experiments, a different approach based on symmetries has been adopted. In this approach one extends the symmetry $G_s$ of the SM to a larger symmetry $G\supset G_s\times G_f$, where $G_f$ is called flavor or horizontal symmetry. Since there exist only three generations a natural candidate  $G_f=SU_f(3)$. This leads to the phenomenologically successful discreet symmetry $A_4$  \cite{Ma01}, which is isomorphic to  the set of the even permutations of 4 objects, which has 12 generators.
An avalanche of papers involving this symmetry as well as its subsequent extensions and  breaking, when $\theta_{13}$ was found to be non zero,  followed, see  e.g. the recent article \cite{BAM-VK15} and the reviews \cite{DKN14,AltFer10,OSVV16} with relevance  to $0\nu \beta \beta$-decay. Applications of this approach to neutrino masses relevant to $0\nu \beta \beta$ have recently begun to develop. e.g.  \cite{DisSym15,FonHir15,VienNgocKhoi15}\\

One therefore would like to see the $SU_f$ as a gauge symmetry, spontaneously broken without surviving Golstone bosons. The first step in this direction has been made \cite{Vergados16}. In fact this has been shown to be possible by considering  quadratic and quartic scalar potentials, which are $SU_(3)$ and $SO(3($ invariant, 
constructed by exploiting the full symmetry chain $SU(3) \supset SO(4)\subset A4$. There are sufficient $A_4$ singlets, $\underline{1}$ ($ A_4$ invariant ) as well as  of the type $\underline{1}'$ and  $\underline{1}''$, which can cause the spontaneous symmetry breaking down to $A_4$. The minimum set of an  $SU(3)$ $\underline{10}$ (decouplet)  and $\underline{10}^*$  together with the adjoined $\underline{8}$ (octet) is sufficient for this purpose. Attempts to embed the $G_s\times SU_f(3)$ to a higher Grand Unified Symmetry (GUT) are also currently under way.
%
%
%
\section[The absolute scale of the neutrino mass]{Attempts at measuring the  absolute scale of the neutrino mass}
\label{sec:extrnumass}
%

The neutrino oscillation data, accumulated over many years, converge towards a minimal
three-neutrino framework, where known flavor states $(\nu_e,~\nu_\mu,~\nu_\tau)$ are
expressed as a quantum superpositions of three massive states $\nu_i$ (i=1,2,3) with masses $m_i$.
With the discovery of neutrino oscillations quite a lot of information
regarding the neutrino sector has become available
(e.g., for  recent reviews see \cite{G-GM08,MohEtal07}).
More specifically we know: 
	\begin{itemize}
	\item The mixing angles $\theta_{12}$ and $\theta_{23}$ and $\theta_{13}$.
	\item We know the two independent mass-squared differences,
          which can be chosen as follows:$\delta m^2 = m^2_2 - m^2_1$ and $\Delta m^2 = m_3^2 - (m_1^2+m_2^2)/2$.
        \item A limited information is available also about the Dirac CP-violating phase $\delta$\cite{lisiglob}.
        \end{itemize}
Neutrino oscillation experiments cannot tell us about the absolute scale of
neutrino masses. The measured two neutrino mass squared differences suggest
two scenarios for neutrino mass pattern:
i) {\it Normal Spectrum-NS}: $m_{1} < m_{2} < m_{3}$:
\begin{equation}
m_1=m_0,~~~m_2=\sqrt{m_0^2 + \delta m^2},~~~~m_3=\sqrt{m_0^2 + \Delta m^2 +  \frac{\delta m^2}{2}};
\end{equation}  
ii) {\it Inverted Spectrum-IS}, $m_{3} < m_{1} < m_{2}$:
\begin{equation}
m_1=\sqrt{m_0^2 + \Delta m^2 -  \frac{\delta m^2}{2}},~~~
m_2=\sqrt{m_0^2 + \Delta m^2 +  \frac{\delta m^2}{2}},~~~m_3=m_0.
\end{equation}
Here, $m_{0}=m_{1}(m_{3})$ is the lightest neutrino mass. Given the type of neutrino mass

The current values of neutrino oscillations parameters obtained by a global fit of
results coming from experiments using neutrinos from solar, atmospheric, accelerator
and reactor sources are presented in Ref.\cite{lisiglob}. This combined analysis allows
to constrain the previously unknown CP phase $\delta$. Concerning the type of
spectrum (sign($\Delta m^2$)), there is no indication in favor of normal or inverted 
mass ordering. However, assuming NS there is a hint about the other unknown,
namely a preference is found in favor of the first $\theta_{23}$ octant 
($\theta_{23} < \pi/4$ at $\sim$ 95\% C.L.). We note that a similar results were
obtained also by a global fit performed by Gonzalez-Garcia et al. \cite{globfit16},
who considered a different definition of two mass squared differences.

The absolute scale $m_0$ of neutrino mass can in principle be determined by the following observations:
 \begin{itemize}
 \item Neutrinoless double beta decay.\\
  As we shall see later (section \ref{sec:numec}) the 
  effective light neutrino mass $m_{\beta\beta}$ (sometimes denoted as Majorana neutrino mass)
  extracted in such experiments  is given as follows \cite {ROP12,Ver86,Pascoli05,PaesRod15,HHOPS15}:
	\beq
	 m_{\beta\beta}   =  |\sum^{3}_k~ (U^{(11)}_{ek})^2 ~ m_k| 
           = |c^2_{12}c^2_{13} e^{2i\alpha_1} m_1 +  c^2_{13} s^2_{12}e^{2i\alpha_2} m_2 + s^2_{13} m_3|.            
	\label{eq:1.5a}   
	\eeq
\item The  neutrino mass extracted from ordinary beta decay, e.g. from triton decay \cite{Katrin,otten,Mare}.\\
	\beq
	m_\beta = \sqrt{ \sum^{3}_k~ |U^{(11)}_{ek}|^2  m^2_k}
                = \sqrt{c^2_{12}c^2_{13} m^2_1 +  c^2_{13} s^2_{12} m^2_2 + s^2_{13} m^2_3}.            
	\label{eq:1.5b}   
	\eeq
assuming, of course, that the three neutrino states cannot be resolved. 
\item From astrophysical and cosmological observations (see, e.g., the recent summary\cite{Abarajan11}).\\
	\beq
\Sigma ~ =  \sum^{3}_k   m_k\leq m_{\mbox{\tiny astro}} 
	\label{eq:1.5c}   
	\eeq
	The current limit on $\Sigma$  depends on the type of observation\cite{Abarajan11}.  In our previous report \cite{ROP12} we used  the CMB primordial specrum which gives  1.3 eV,  CMB+distance 0.58 eV, galaxy distribution and and lensing 
        of galaxies  0.6 eV. On the other hand  the largest photometric red shift survey yields
   0.28 eV \cite{ThAbdaLah10}.  Since then various analysis have been performed. 
	It is worth stressing the following 2 $\sigma$ C. L. upper limits: 0.17 eV obtained
in Ref. \cite{SDSS05} by combining CMB data of the Wilkinson
Microwave Anisotropy Probe, galaxy clustering and
the Lyman-alpha forest of the Sloan Digital Sky Survey
(SDSS); 0.18 eV of Ref. \cite{RSPD14} using Planck and Wiggle Z
galaxy clustering data; 0.14 eV obtained in Ref. \cite{CSVB14} by
combining Lyman-alpha SDSS data with Planck; 0.153 eV
obtained in Ref. \cite{Planck15} by using Planck temperature and polarization
measurements including a prior on the Hubble parameter, Supernovae and Baryonic Acoustic Oscillations
(BAOs). It has recently been shown in Ref.  \cite{DMVV15} that the variation of the astrophysical data affects the range of the expected neutrino mass of the $0\nu \beta \beta$-decay.
 \end{itemize}
The above mass combinations entering triton decay  and cosmological constraints are not going to be discussed here, since they can be found in an earlier review \cite{ROP12}. We will discuss here only those relevant for $0\nu\,\beta\beta$ decay.

The value Majorana neutrino mass $m_{\beta\beta}$ can be predicted
in the limit of the normal and inverted hierarchies:
\begin{enumerate}
\item Normal Hierarchy (NH): $m_1\ll m_2 \ll m_3$:\\
In this case for the neutrino masses we have
$$m_1 \ll \sqrt{\delta m^2}.~~~m_2 \simeq \sqrt{\delta m^2},
~~~m_3 \simeq  \sqrt{\Delta m^2}. $$
Neglecting a small contribution of $m_1$ to $m_{\beta\beta}$ we find
\begin{equation}
|s^2_{12}  c^2_{13} \sqrt{\delta m^2}
- s^2_{13} \sqrt{\Delta m^2}|
\le m_{\beta\beta} \le
s^2_{12}  c^2_{13} \sqrt{\delta m^2} + s^2_{13} \sqrt{\Delta m^2}.
\end{equation}
Using the best-fit values of the mass squared differences and the mixing
angles we find
\begin{equation}
  1.4~ \text{meV} \le m_{\beta\beta} \le 3.6 ~\text{meV}.
  \label{mbbNH}
\end{equation}
\item  Inverted Hierarchy (IH): $m_3 \ll m_1 < m_2$:\\
  In the IH scenario
  $$m_3 \ll \sqrt{\Delta m^2}, ~~~~m_1\simeq m_2 \simeq \sqrt{\Delta m^2}.$$
We find
\begin{equation}
|1 -2 s^2_{12}| c^2_{13} \sqrt{\Delta m^2}
\le m_{\beta\beta} \le c^2_{13} \sqrt{\Delta m^2}.
\end{equation}
Using the best-fit values of the parameters  we find the following
range for $m_{\beta\beta}$ in the case of the IH:
\begin{equation}
  20~ \text{meV}~ \le m_{\beta\beta} \le ~49~ \text{meV}.
  \label{mbbIH}
\end{equation}  
\end{enumerate}

\begin{figure}[!t]
 \vspace*{1.2cm}
\centerline{\psfig{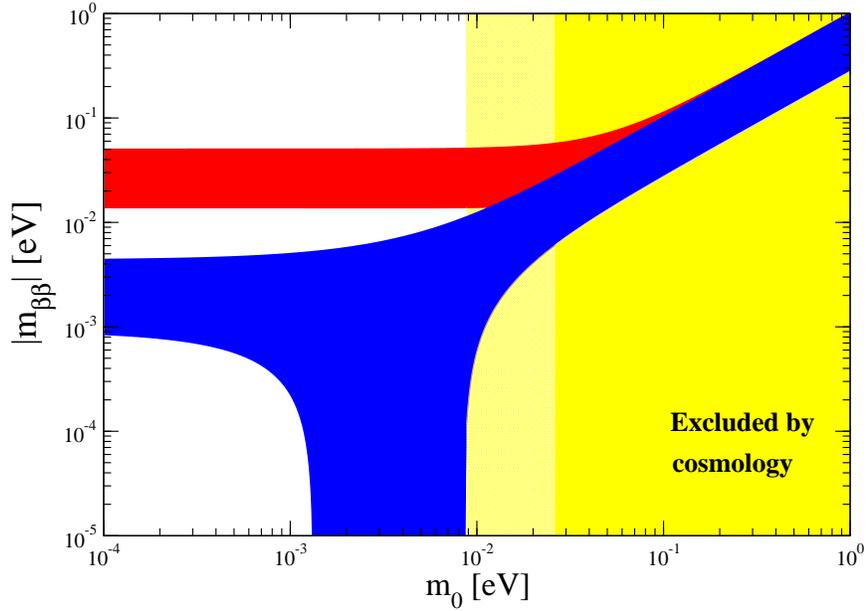}}
\caption{(Color online) 
  Updated predictions on $m_{\beta\beta}$ from neutrino oscillations versus the lightest neutrino mass
  $m_0$ in the two cases of normal (the blue region) and inverted (the red region) spectra. The 3$\sigma$ values of
  neutrino oscillation parameters are considered\protect\cite{lisiglob}.  The excluded region by cosmological
  data ($\Sigma < 110$ meV \protect\cite{Vissani16}) is ($m_0 >$ 26 meV (NS), 87 meV (IS)) presented in yellow. 
\label{mbbvac}
}
\end{figure}

In Fig. \ref{mbbvac} the updated prediction on the Majorana neutrino mass is plotted 
as function of the lightest neutrino mass $m_0$. The 3$\sigma$ values of neutrino
oscillations parameters $\theta_{12}$, $\theta_{13}$, $\delta m^2$ and $\Delta m^2$
are taken into account\protect\cite{lisiglob}. The two Majorana phases $\alpha_{1,2}$ 
are assumed to be arbitrary. The constraint from the cosmological data ($\Sigma < 110$ meV\cite{Vissani16})
on the lightest neutrino mass ($m_0 >$ 26 meV (NS), 87 meV (IS))
is displayed.

\subsection{The effect of nuclear enviroment on Majorana neutrino mass}

 It has recently been  proposed that the neutrino mixing and masses
in a nucleus can differ significantly from those in vacuum, if there exist  exotic particles, 
preferably scalars, which do interact with neutrinos. The related nuclear matter effect on
the $0\nu\beta\beta$-decay rate can be calculated in the mean field approach \cite{fmsprl}.

The effective  four-fermion neutrino-quark lepton number violating Lagrangian 
with the operators of the lowest dimension can be written as
\begin{eqnarray}\label{EffLag}
{\cal L}_{\rm eff} &=&  \frac{1}{\Lambda^{2}_{LNV}}   
\sum\limits_{i, j, q}\left(g^{q}_{ij} \overline{\nu_{Li}^{C}}  \nu_{Lj} ~  \bar{q}  q 
+ \mbox{H.c.}\right),
\end{eqnarray}
where the fields $\nu_{Li}$ are the active neutrino left-handed flavor states, $g^{q}_{ij}$ 
are their dimensionless  couplings to the scalar quark currents with $i,j=e,\mu, \tau$.

For sake of simplicity we consider case of scalar coupling.
In this case the effective Majorana mass  \cite{fmsprl}
 is
\begin{eqnarray}
m_{\beta \beta } = \left| \sum^3_{i=1} \left(U_{ei}\right)^{2}\xi_{i} 
{| m_{i} - \langle\bar{q} q \rangle g |} \right|.
\label{mmed}
\end{eqnarray}
The Majorana phase factor \mbox{$\xi_{i}$} is given in \cite{fmsprl}.

\begin{figure}[!t]
  \vspace*{1.2cm}
  \centerline{\psfig{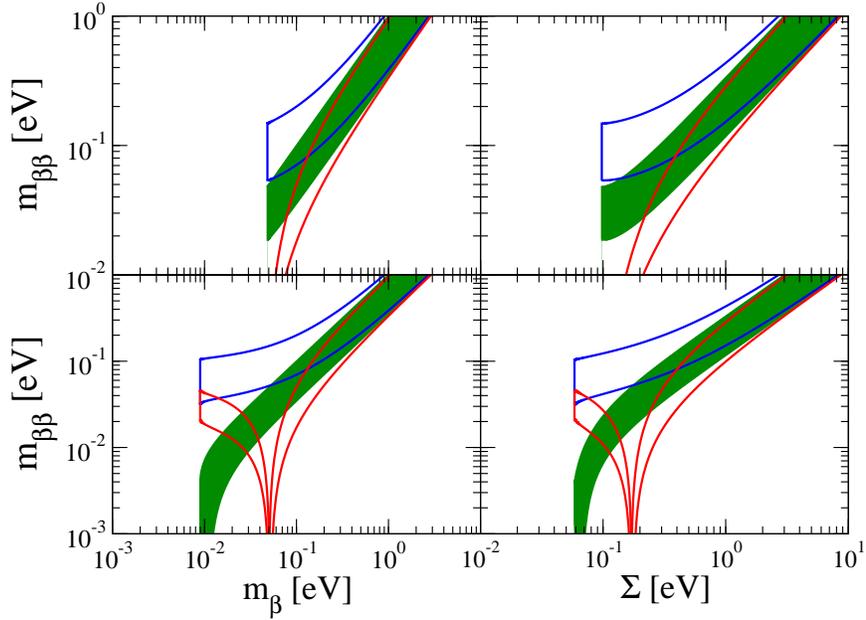}}
\caption{(Color online) 
The allowed range of values for effective Majorana mass $m_{\beta\beta}$ as a  function of
the effective electron neutrino mass $m_\beta$ (left panels) and
sum of neutrino masses $\Sigma$ (right panels). The upper and lower
panels correspond to the cases of the inverted and normal spectrum of neutrino
masses. In panels green, blue and rad  bands refer to to values 
$\langle \bar{q} q \rangle g =0$ (vacuum), $0.1$, and $-0.05$ eV,
respectively. \label{mbbnuc} }
\end{figure}

With the above simplification the quantity $m_{\beta\beta}$  in nuclear medium in comparison 
with the one in vacuum depends on the new unknown parameter $g$. 
The unknown phases in Eq. (\ref{mmed}) are varied in the interval $[0, 2\pi]$. 
In Figure \ref{mbbnuc}
 $m_{\beta\beta}$ is expressed as a function of a directly observable
parameters, namely $m_{\beta}$ and $\Sigma$. The best-fit values of vacuum mixing angles
and the neutrino mass squared differences are taken from \cite{lisiglob}. 
In upper and lower panels green, blue and red  bands refer to values 
$\langle \bar{q} q \rangle g =0$ (vacuum), $0.1$, and $-0.05$ eV, respectively.
We see that in-medium ($g\ne 0$) values of $m_{\beta \beta}$ differ significantly from those
for a vacuum ($g=0$).

If in the future the gradually improving limits on $m_\beta$ and $\Sigma$ will come
into conflict with the possible evidence of the $0\nu\beta\beta$-decay
represented by  $m_{\beta\beta}$ in vacuum, new physics would be mandatory. A possible
explanation could be a generation of in-medium Majorana neutrino mass due to nonstandard
interactions of neutrinos with nuclear matter of decaying nuclei.

The limit on Majorana neutrino mass from the $0\nu\beta\beta$-decay experiments
depends on the value of nuclear matrix elements (NMEs). Taking as an experimental
limit the  value $|m_{\beta \beta}| < 0.2$ eV and combining it with the cosmological 
and tritium  limits one finds \cite{fmsprl}
\begin{eqnarray}\label{Lim-from-g1}
\Lambda_{LNV} \geq 2.4 ~ \mbox{TeV}~~ \mbox{(Planck)}, ~~~  1.1 ~ \mbox{TeV}~~ (^3\mbox{H}).
\end{eqnarray}
For convenience the above  limits can be expressed in terms of a dimensionless 
parameter $\varepsilon_{ij}$  defined as $\varepsilon_{ij} G_{F}/\sqrt{2} = g^q_{ij}/\Lambda^{2}_{LNV}$. 
The quantity $\varepsilon_{ij}$  characterizes the relative strength of the 4-fermion 
lepton number violating operators in (\ref{EffLag}) 
with respect to the Fermi constant $G_{F}$. We find $\varepsilon_{ij} \leq 0.02$ (Planck), $0.1$ (Tritium).

%
%
%
\section[The  Majorana neutrino mechanism]{The intermediate Majorana neutrino mechanism.}
\label{sec:numec}
%

    We have seen that the determination of the scale of the neutrino mass is an urgent issue of current physics. 
    To proceed further, however, on this goal, i.e.  in our study of the neutrino mediated
		 $0\nu\beta\beta$-decay process, it is necessary to study the structure 
		 the  effective weak beta decay Hamiltonian. In general it has both left handed and right handed components. Within the $SU(2)_L\times SU(2)_R\times U(1)$ we encounter the following situations:
		
\subsection{The light neutrino mass mechanism }
\label{SecMassM}

The $0\nu\beta\beta$-decay is a process of second order in the perturbation theory of
weak interactions. In the case of the light neutrino mass mechanism of the 
$0\nu\beta\beta$-decay the weak $\beta$-decay Hamiltonian has the standard form,
\beq \label{hamilweak}
H^\beta = \frac{G_{\beta}}{\sqrt{2}} ~\left[ \left(\bar{e}\gd{\rho}(1-\gd{5})\nu_e\right)
~J^{\rho\dagger}_L + h.c.\right]
\ee
$e_L(e_R)$ and $\nu^{0}_{{e L}}(\nu^{0}_{{e R}})$ are field operators
		representing the left (right) handed electrons and  electron
		neutrinos in a  weak interaction basis in which the charged leptons are 
               diagonal.
               We have seen above the the weak neutrino eigenstates can be expressed in terms of the propagating mass eigenstates \cite {Ver86}  (see Eqs. (\ref{Eq:leftnu}) and (\ref{Eq:rightnu})). Thus omitting the subscript zero we write
\begin{equation}
\nu_{e} = \sum_{k=1}^3 U_{ek} \nu_k,
\end{equation}
with  $\nu_k$ the light neutrino mass eigenstates.
Here,  $G_{\beta}=G_F\cos{\theta_C}$, where $G_F$ and $\theta_C$ are Fermi constant and Cabbibo angle, respectively.
$\nu_k$ is the  Majorana neutrino field,   $J^{\rho}$ is the V-A hadronic  current:
\bee
\bra{p(P')}J^{\mu\dg}\ket{n(P)} &=& \bar{u}_p(P')
\left[g_V\gh{\mu}+ig_M\frac{\sigma^{\mu\nu}}{2m_N}(P'-P)_{\nu}\right.\nonumber\\
&&~~~~~~~~~-\left.g_A\gh{\mu}\gd{5}-g_p\gd{5}(P'-P)^{\mu}\phantom{\frac{i}{i}}\right]u_n(P'),
\label{Eq:lefthanded}
\eee
where the $u_p(P')$ and $u_n(P)$ are the spinors describing the proton and neutron with
corresponding four-momenta $P'^{\mu}=(E',\mathbf{p}')$ and $P^{\mu}=(E,\mathbf{p})$,
respectively. $m_N$ is the nucleon mass, $q=P'-P$ is the momentum transfer
and $q_V\equiv q_V(q^2)$, $q_M\equiv q_M(q^2)$, $q_A\equiv q_A(q^2)$ and $q_P\equiv q_P(q^2)$
are the vector, weak-magnetism, axial-vector and induced pseudoscalar form-factors, respectively.
$m_N$ is the nucleon mass.

Within the non-relativistic impulse approximation, the hadronic current 
can be written as \cite{DTK85}
\bee \label{nonhad}
J^{\rho\dg}_{L}(\mathbf{x})&=&\sum_{n}\tau_{n}^+\delta(\mathbf{x}-\mathbf{r}_n)
 \left[\left(g_{V}-g_{A}C_n\right)g^{\rho 0} \right.\nonumber\\
&&+\left.g^{\rho k}\left(g_{A}\sigma_n^k-g_{V}D_n^k-g_{P}~q_n^{k}~
\frac{\vec{\sigma}_n \cdot \mathbf{q}_n}{2m_N}
\right)\right].
\eee
Here, $\mathbf{q}_n=\mathbf{p}_n-\mathbf{p}_n^{'}$ is the momentum transfer between the nucleons.
The final proton (initial neutron) possesses  energy $E_n'$ ($E_n$)
and momentum $\mathbf{p}_n'$ ($\mathbf{p}_n$). $\vec{\sigma}_n$, $\tau^+_n$
and $\mathbf{r}_n$ are the Pauli matrix, the isospin raising operator 
and the position operator, respectively. These operators act on the $n$-th nucleon.

The nucleon recoil operators $C_n$ and $\mathbf{D}_n$ are given by
\bee
&&C_n=\frac{\vec{\sigma} \cdot \left(\mathbf{p}_n+\mathbf{p}_n^{'}\right)}{2m_N}-\frac{g_P}{g_A}\left(E_n-E_n^{'}\right)
\frac{\vec{\sigma} \cdot \mathbf{q}_n}{2m_N},\nonumber\\
&&\mathbf{D}_n=\frac{\left(\mathbf{p}_n+\mathbf{p}_n^{'}\right)}{2m_N}-i\left(1+\frac{g_M}{g_V}\right)
\frac{\vec{\sigma} \times \mathbf{q}_n}{2m_N}.
\eee
Here,   $q_M\equiv q_M(q^2)$ and $q_P\equiv q_P(q^2)$
are, respectively, the weak-magnetism and induced pseudoscalar form-factors in the case
of left-handed hadronic currents.

\subsection{The V-A mechanism with sterile neutrino(s)}
Here by the term sterile we do not mean only the usual sterile neutrino discussed in section \ref{Neutrinomass}, which is heavier than the standard neutrino but lighter than a few keV and is often included in discussions of cosmology  and neutrino oscillations. In other words here
we assume that in addition to the three conventional light neutrinos there exist other
Majorana neutrino mass eigenstates $N$ of an arbitrary mass $m_{\rm N}$, dominated by the 
sterile neutrino species $\nu_{s}$ and with some admixture of the active neutrino weak 
eigenstates, $\nu_{e,\mu,\tau}$ as
\begin{eqnarray}\label{mixing}
N = \sum\limits_{\alpha=s,e,\mu,\tau} U_{N\alpha}\  \nu_{\alpha}.
\end{eqnarray}

Massive neutrinos N have been considered in the literature in different contexts
with the masses $m_{\rm N}$ ranging from the eV to the Planck scale, in particular 
with neutrino mass at keV  (hot dark mater), Fermi \mbox{($\sim 200$ MeV)},  
TeV  (physics at LHC), GUT ($10^{16}$ GeV) or Planck ($10^{16}$ GeV) scale\cite{Abazajian:2012ys}.
Their phenomenology have been actively studied from various perspectives including their
contribution to  particle decays and production in collider experiments\cite{Helo:2010cw,Atre:2009rg}.
The corresponding searches for N have been carried out in various experiments\cite{Beringer:1900zz}.

\subsection{The mechanisms within the left-right symmetric model}
\label{sec:rightleft}
The left-right symmetric models (LRSM)  \cite{pasa74,mopa75} provide a natural framework
to understand the origin of neutrino  Majorana masses. In general one cannot
predict the scale where the left-right symmetry is realized, but it is natural
to assume that it could be  as low as $\sim$ a few TeV which can affect
the $0\nu\beta\beta$ decay rate significantly \cite{vissa11,Nemevsek,Barry13}.
In the left-right symmetric theories
in addition to the left-handed V-A weak currents also leptonic and hadronic
right-handed V+A weak currents are present.

The effective current-current interaction at low energies,
which can trigger the $0\nu\beta\beta$-decay, 
\bee \label{hamilweakrh}
H^\beta &=& \frac{G_{\beta}}{\sqrt{2}} ~\left[
j_L^{~\rho}J^{\dg}_{L\rho } - \epsilon j_L^{~\rho}J^{\dg}_{R\rho } +
~\epsilon j_R^{~\rho}J^{\dg}_{L\rho } +\kappa j_R^{~\rho}J^{\dg}_{R\rho } +  h.c.\right].
\eee
Here, $\epsilon$ is the mixing of $W_L$ and $W_R$ gauge bosons
\beq
W_L=\cos{\epsilon} W_1 - \sin{\epsilon} W_2,\quad 
W_R=\sin{\epsilon} W_1 + \cos{\epsilon} W_2
\eeq 
where $W_1$ and $W_2$ are the mass eigenstates of the gauge bosons
with masses $M_{W_1}$ and $M_{W_2}$, respectively. 
The mixing is assumed to be small, 
$\sin{\epsilon}\approx \epsilon$,  $\cos{\epsilon}\approx1$, 
and $M_{W_1}\approx m_{W_L}$, $M_{W_2}\approx m_{W_R}$.
$\kappa$ is the mass squared ratio  $\kappa=\frac{M_{W_1}^2}{M_{W_2}^2}$.
The left-handed hadron current is given Eq. (\ref{nonhad})
and right-handed hadron current takes the form
\bee \label{nonhadR}
J^{\rho\dg}_{R}(\mathbf{x}) &=& \sum_{n}\tau_{n}^+\delta(\mathbf{x}-\mathbf{r}_n)
 \left[ \left(g_{V}^{\prime}+g_{A}^{\prime}C_n\right)g^{\rho 0} \right.\nonumber\\
&&+\left.g^{\rho k}\left(-g_{A}^{\prime}\sigma_n^k-g_{V}^{\prime}D_n^k+g_{P}^{\prime}~q_n^{k}~
\frac{\vec{\sigma}_n \cdot \mathbf{q}_n}{2m_N}
\right)\right]~. \nonumber \\
\eee
As the strong and electromagnetic interactions conserves parity 
there are relations among form-factors entering the left-handed and right-handed hadronic currents
\cite{DTK85}:
\be
\frac{g_A}{g_V}=\frac{g_A^{\prime}}{g_V^{\prime}}, \qquad \frac{g_M}{g_V}=\frac{g_M^{\prime}}{g_V^{\prime}}, \qquad
 \frac{g_P}{g_V}=\frac{g_P^{\prime}}{g_V^{\prime}}.
 \ee
The left- and right-handed leptonic currents are given by 
\bee
j_L^{~\rho}=\bar{e}\gd{\rho}(1-\gd{5})\nu_{eL}, \qquad j_R^{~\rho}=\bar{e}\gd{\rho}(1+\gd{5})\nu_{eR}.\nonumber \\
\eee
The $\nu_{eL}$ and  $\nu_{eR}$ are the weak 
eigenstate electron neutrinos, which are expressed as superpositions of the light and heavy 
 mass eigenstate Majorana neutrinos   
$\nu_j$ and $N_j$, respectively. The electron neutrinos eigenstates can be expressed as
\bee\label{lambdaeta}
&&\nu_{eL}=\sum_{j=1}^3\lf U_{ej}\nu_{jL}+S_{ej} (N_{jR})^C \rh, \nonumber \\
&& \nu_{eR}=\sum_{j=1}^3 \lf T_{ej}^*(\nu_{jL})^C+V_{ej}^*N_{jR} \rh.
\eee

Before proceeding further we should mention that in the context of the above $0\nu\beta\beta$-decay is a two step process.
The neutrino is produced via the lepton current in one vertex and propagates in the other vertex. If the two current
helicities are  the same one picks out of the neutrino propagator the term:
\beq
\frac{m_j}{q^2-m_j^2}\rightarrow \left\{\begin{array}{c}m_j/q^2,\quad m^2_j \ll q^2\\-{1}/{m_j},\quad m^2_j \gg q^2\end{array} \right .
\eeq
where $q$ is the momentum transferred by the neutrino. In other words the amplitude for light neutrino becomes proportional to its mass, but for a heavy neutrino inversely proportional to its mass.

If the leptonic currents have opposite chirality the one picks out of the neutrino propagator the term:
\beq
\frac{	 \displaystyle{\not} q }{q^2-m_j^2}\rightarrow \frac{	 \displaystyle{\not} q }{q^2},\quad m^2_j \ll q^2
\label{Eq:propq}
\eeq
i.e. in the interesting case of light neutrino the amplitude does not explicitly depend on the neutrino mass. The kinematics becomes different than that for the mass term.

At this point we should note that, in general, there can be several coexisting mechanisms for the 0$\nu \beta \beta $ decay. In addition to  the light Majorana $\nu$ exchange and  heavy Majorana $\nu$ exchange, with or without right handed currents, just discussed, one can have contributions from  R parity breaking super symmetry etc.
In order to extract the most interesting information related to the light neutrino mass in the presence of two or more competing mechanisms, we need to measure the 0$\nu \beta \beta $ decay rates for several isotopes. In order to determine the relative contributions of each mechanism, we need for each one of them very precisely known  NMEs for the isotopes involved \cite{ROP12}.

Before proceeding further with theoretical issues we will review the current status and the future prospects of double beta decay searches.

%
%
%
\section[Experiments of neutrinoless double beta decay]{Experiments of neutrinoless double beta decay}
\label{sect:exp}
%

In this section we briefly 
describe experimental aspects of DBD and recent DBD experiments. Details of DBD experiments are given in reviews \cite{eji05,AEE08,ROP12,eji10}.

\subsection{Experimental methods and detectors}
 The 0$\nu \beta \beta $ decay rate is of the order of or less
 than 10$^{-27}$  and 10$^{-29}$ per year (y) in cases of inverted (IH) and normal (NH) neutrino hierarchy spectra, respectively. 
 Actually the decay rate depends quadratically on the nuclear matrix element (NME) ${M'}^{0\nu}_\nu$. 
The energy of the 0$\nu\beta\beta$ signal  is only a few MeV.  This is in the same energy region 
as backgrounds (BGs). 
The size of DBD isotope required in the target is of the order of 
multi ton (t) and multi k-ton scales for the IH and NH masses. Then BGs rates  have to be
necessarily reduced to the order of a few $\times 10^{-1}$ and a few $\times 10^{-3}$ events per year per ton of the DBD isotopes (yt).  

 The DBD nuclei are used as femto (femto m scale)
laboratories where the 0$\nu \beta \beta$ signal is enhanced and the single $\beta $ BGs are suppressed.
The luminosity of the ton-scale DBD nuclear ensemble is of the order of 
$L\approx $10$^{77}$sec$^{-1}$ cm$^{-2}$, while  
 the 0$\nu \beta \beta$ cross section for the IH $\nu $ -mass process is  of  order of $\sigma \approx $10$^{-84}$ cm$^2$. 
Thus one may expect the signal rate of the order of $L \sigma T\approx $ 3 in a year of $T$=3$\times 10^{7}$ s. 
  
DBD processes are studied by measuring the sum-energy spectrum of the two $\beta $ rays. The
 0$\nu \beta \beta $ process, which can occur  beyond the SM, is identified by the  
sharp peak of the 2 body kinematics at $Q_{\beta \beta }$, and
the neutrino-less process accompanied by a Majoron boson (0$\nu M \beta \beta $)  is characterized \cite{ROP12} by the broad peak of
 the 3 body kinematics. 
 
 The 0$\nu \beta \beta $ process may be due to the left-handed weak current (LHC) and the right-handed weak current (RHC). 
The LHC includes the light Majorana $\nu$-mass mode, the heavy Majorana $\nu$ mode, the SUSY modes, and others beyond SM. The RHC by itself includes mainly heavy neutrino mass contributions. One additional possible mechanism involves the interference of left handed and right handed leptonic currents and is due to the light neutrino component (see section \ref{sec:rightleft}), but it leads  to   neutrino mass independent lepton violating parameters $\langle\eta\rangle$  and  $\langle\lambda\rangle$  (for more details see section \ref{mechanisms of light neutrinos}). 
This  processes picks out the intermediate neutrino momentum rather than the neutrino mass (see Eq. (\ref{Eq:propq})),  and is characterized by different kinematics. It may thus be experimentally distinguished from the mass terms  by measuring the angular and energy correlations of the two $\beta $ rays. 
The modes (light $\nu$, heavy $\nu$, 
SUSY, etc) involved in the LHC are investigated by measuring the 0$\nu \beta \beta $ rates in various nuclei.

\subsection{DBD detectors and sensitivities}

We discuss  the 0$\nu \beta \beta $ process with the light Majorara-$\nu $ exchange. Then the transition rate $T^{0\nu }$ per year ton (yt) 
is expressed in terms of the nuclear sensitivity $S_N$ in units of meV$^{-2}$  and the Majorana neutrino
mass  $m_{\beta\beta}$ as\cite{eji05,ROP12}. 
\begin{equation}
(T^{0\nu})^{-1} = m_{\beta\beta}^2 S_N,~~~
S_N^{1/2}=(78)^{-1} ~|{M'}^{0\nu}_\nu| g_A^2 (G^{0\nu }/0.01 A)^{1/2},
\end{equation}
where ${M'}^{0\nu}_\nu$ is the NME, $G^{0\nu}$ is the phase space volume in units 
of 10$^{-14}$/y, and $A$ is the mass number. 

The mass sensitivity $m_m$ is defined as the minimum mass required to identify the 0$\nu \beta \beta $ signal.
 It is expressed in terms of $S_N$ and the detector sensitivity $D$ as 
\begin{equation}
m_m = S_{N}^{-1/2} D^{-1/2}, ~~~ 
D = (\epsilon NT)(\delta )^{-1},   
\end{equation}     
where $\epsilon $ is the 0$\nu \beta \beta $ peak efficiency, $N$ is the number of the DBD isotopes in units of ton, $T$ is the  
run time in unit of year and $\delta $ is the minimum number of 
counts for the peak identification with 90 $\%$ CL (confidence level).
Since $BNT$ is much larger than 1 in most DBD experiments, we use
 $\delta \approx 1.7 \times (BNT)^{1/2}$. Then $m_m$  is  
\begin{equation}
m_m\approx m^0 D^{-1/2}, 
\end{equation}
\begin{equation}
  m^0=39~G^{-1/2} (2/{M'}^{0\nu}_\nu ) 1/g^2_A, ~~~D^{-1/2}=1.3~ \epsilon^{-1/2}[B/NT]^{1/4},
\end{equation}
where $ G=  (G^{0\nu }/0.01 A)$, $m^0$ is the unit mass sensitivity in units of meV, i.e. the mass sensitivity in case of a detector 
with $D$=1 ($\epsilon $=1, $NT$=3 t y and BG rate of $B$=1/(t y)). 
 
\begin{figure}[!t]
\centerline{\psfig{file=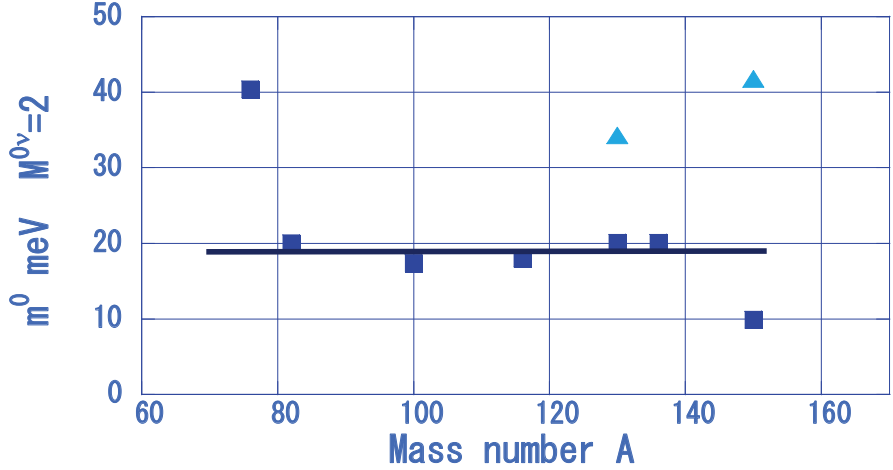,width=\textwidth}}
\caption{Unit mass sensitivities $m^0$ (squares) in case of ${M'}^{0\nu}_\nu$=2 for $^{76}$Ge, $^{82}$Se,  $^{100}$Mo, $^{116}$Cd, $^{130}$Te,  $^{136}$Xe
and $^{150}$Nd and those (triangles) for natural $^n$Te and $^n$Nd.
\label{fig:fig71}
}
\end{figure}

The nuclear sensitivity $S_n$ is proportional to the phase space factor $G^{0\nu}$ and 
$|{M'}^{0\nu}_\nu|^2$. 
The DBD isotopes used for realistic high-sensitivity experiments include $^{76}$Ge, 
 $^{82}$Se,  $^{100}$Mo, $^{116}$Cd, $^{130}$Te, and $^{136}$Xe. These isotopes have the large phase space  factor 
 around $G^{0\nu}\approx$ 1.5  in units of 10$^{-14}$ y$^{-1}$ and multi-ton scale enriched isotopes  obtained by means of the centrifugal isotope separation. 
  $^{76}$Ge detectors are used because of the high energy-resolution
although $G^{0\nu}\approx$ 0.25 is much smaller than the others. 
The unit mass sensitivities $m^0$ for typical DBD nuclei are plotted in Fig.\ref{fig:fig71}. 

Actually, the sensitivity depends on the enrichment of the isotope. If natural isotopes with the same total mass ($N$ ton) as the enriched one are used, 
the 0$\nu \beta \beta $ efficiency $\epsilon$ is reduced effectively by the abundance ratio $r$, while the BG rate $B$ remains same. Thus the mass sensitivity 
gets larger by the factor $r^{-1/2}$.  The unit mass sensitivities for $^n$Te with $r$=0.34 and $^{n}$Nd with $r$=0.056 are
as shown in Fig.\ref{fig:fig71}. 

The detector-sensitivity factor $D^{-1/2}$ is proportional to the factor $[B/NT]^{1/4}$. Then one needs low-BG large-volume detectors. 
In the case of $m^0\approx$ 20 meV, detectors with  $B \le0.3$/(t y), $\epsilon$=0.5, and $N\ge$ 3 t may be used to access the IH $\nu $ mass of 
around 15 meV for $T$=4 y, while those with, for instance, $B \le0.01$/(y t) and $N\ge $ 1 kt 
 are required to explore the NH $\nu $ mass of $m_{\beta\beta} \approx $ 1.5 meV for $T$=4 y.

The BG sources to be considered are natural RIs  as $^{208}$Tl, $^{214}$Bi and other Ur-Th chain RIs,  cosmogenic RIs, as $^{68}$Ga in the case of Ge detectors, muon and 
neutron interactions, solar-$\nu$ CC and NC interactions, the high-energy tail of the 2$\nu \beta \beta $ spectrum and others.  Then DBD experiments 
are made by using  high purity (RI-free) DBD detectors with good energy resolution at a deep underground laboratory.

The energy resolution is very important to select the 0$\nu \beta \beta $ signal at the ROI and to reduce the BG contributions since the BG spectra 
are mostly continuum.  Thus Ge detectors and cryogenic (scintillation) bolometers are very sensitive detectors.
SSSC (single site spacial correlation) analysis is used to reduce RI BGs associated with $\gamma$ rays since 
 $\gamma $ rays deposit their energies at multi-sites through Compton scatterings. PSD (pulse shape discrimination) is effective for single site time correlation (SSTC). 
SSTC analysis is used to reduce RI BGs  from $^{214}$Bi, 
$^{68}$Ga and others  by delayed anti-coincidence with the preceding $\beta $ or X rays \cite{eji05,AEE08,ROP12,eji10}.

The solar neutrinos are omnipresent, and the solar-$\nu$ CC and NC interactions with nuclei and atoms in DBD detectors are serious BG sources 
for high-sensitivity DBD experiments. The solar-$\nu$ CC interaction with  the DBD nucleus A excite  GT(1$^+$) states in the intermediate nucleus B and the  $\beta $  decay from B 
to the final nucleus C contributes to BGs at the ROI in the 0$\nu \beta \beta$ of A$\rightarrow$C. The contributions  
\cite{eji14A} are appreciable even for IH $\nu$-mass studies with medium energy-resolution ($\approx $ a few $\%$) 
experiments unless they are reduced by SSSC, SSTC and others. 

The $^8$B solar-$\nu$ CC and NC interactions with atomic electrons in DBD isotopes and those in liquid scintillators used for DBD experiments 
\cite{kamlandzen,sno14,geh10,sno11}
are also BG sources for medium energy-resolution experiments\cite{bor11,EjiZub16}  The BG contribution for the liquid 
 scintillator loaded with DBD isotopes of interest is expressed as  
\begin{equation}
B_e \approx 0.15 \times E f /(\mbox{t y}) ~~~ f=w/R, 
\end{equation}
where $E$ is the $Q$ value in units of MeV, $w$  the energy resolution in FWHM, and $R$ is the concentration of the DBD isotopes in the scintillator.
The BG rate is $B_e\approx $2-3/(t y) in a typical case of $E\approx$3 MeV and $f\approx $5, i.e. $w\approx 5$ $\%$ and $R\approx 1 \%$. 
Thus  the contribution from the solar $\nu$ interaction needs to be well considered.

\begin{figure}[!t]
\centerline{\psfig{file=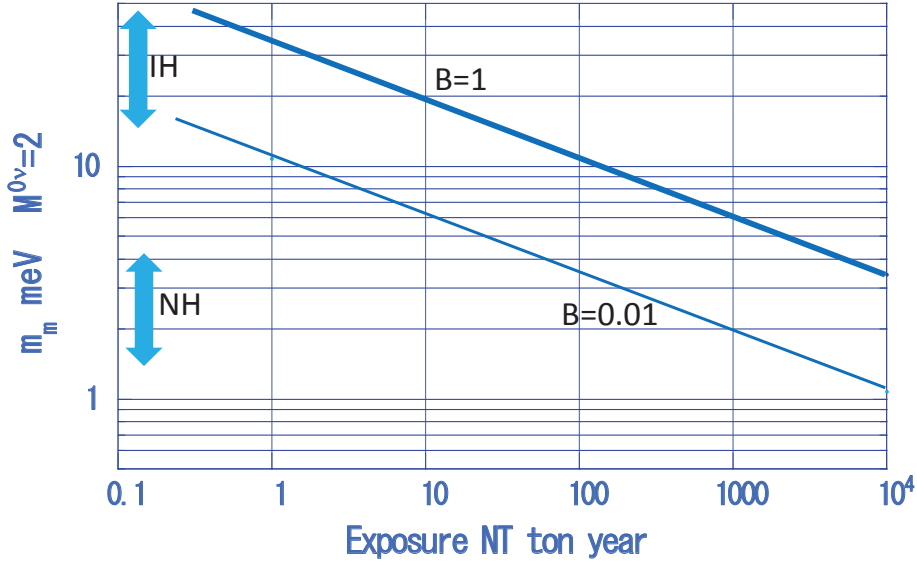,width=\textwidth}}
\caption{Neutrino-mass sensitivities $m_m$ for $^{130}$Te with ${M'}^{0\nu}_\nu$=2 (i.e.  $m_m \times {M'}^{0\nu}_\nu$/2)
 and $\epsilon$=0.5 as a function of the exposure $NT$ in cases of the BG rates of $B$=1/(t y) and 0.01/(t y), respectively.
\label{fig:72}
}
\end{figure}

The $\nu$-mass sensitivities for $^{130}$Te with ${M'}^{0\nu}_\nu$=2 and $\epsilon$=0.5 are shown as a function of the exposure $NT$ in  
cases of  $B$=1/(t y)  and $B$=0.01/(t y)   in Fig.\ref{fig:72}. Exposures required for studies of 
the IH and NH mass regions are $NT$=1-10 y t and $NT$=100-1000 y t in cases of $B$=1/(t y) and $B$=0.01/(t y), respectively. 
These are similar for $^{82}$Se,  $^{100}$Mo, $^{116}$Cd, $^{130}$Te,  $^{136}$Xe. 

\subsection[The present status of double beta decay experiments]{The present status of neutrinoless double beta decay experiments}

Experimental $0\nu\beta \beta $ studies with the QD (quasi-degenerate mass hierarchy sub eV) mass sensitivity 
have been carried out extensively as given in other  reviews\cite{eji05,AEE08,ROP12,eji10}. 
 The $^{76}$Ge experiments with $^{76}$Ge semiconductors
 (HM\cite{kla01},IGEX\cite{aal02}),  the $^{130}$Te experiment with TeO$_2$ 
cryogenic bolometers (CUORETINO\cite{arn05,arn11})  and the large volume $^{136}$Xe experiments (EXO\cite {EXO2012,EXO11},
KanLAND-Zen\cite{kamlandzen}) have been carried out to study the $\beta \beta $ decays.
Tracking detectors (ELEGANT V\cite{eji01} and  NEMO-3\cite{nemoiii05}) have been used for studying $\beta \beta $ decays from $^{82}$Se, $^{100}$Mo, 
$^{116}$Cd and other isotopes with large $Q_{\beta \beta }$ values.

Recently stringent lower limits were obtained on $T^{0\nu}_{1/2}$ 
of $^{76}$Ge, $^{100}$Mo, $^{130}$Te and $^{136}$Xe, as given  
in Table \ref{tab:eji1}. The upper limit on the effective $\nu$-mass was derived from the half-life limit by
 using the calculated ${M'}^{0\nu}_\nu$, as given in the 4th column of Table \ref{tab:eji1}. 
The mass range reflects the range of the ${M'}^{0\nu}_\nu$ values, depending on the model and the 
effective coupling constant $g_A^{\rm eff}$ used for the calculation. Accurate theoretical evaluations for
 $g_A^{\rm eff}$ are not easy. The $\nu$-mass limit  may become approximately 50 $\%$ larger 
 if one uses a 30$\%$ smaller value for $g_A^{\rm eff}$. 
  Recent experiments on neutrino-less $\beta \beta $ decays give also lower limits on the half lives $T^{0\nu M}_{1/2}$ 
for the Majoron emitting process and the Majoron neutrino coupling $\langle g_{ee}\rangle$,  e.g. in  \cite{nem3A} for $^{100}$Mo.
 Such limits exist for a number of other targets, but we do not discuss them in this work (see the earlier report \cite{ROP12}).   
 
\begin{table}[!t]
\caption{Limits on $T^{0\nu}_{1/2}$.
$Q_{\beta \beta }$ : $Q$ value for the 0$^+\rightarrow 0^+$ ground state transition. 
$G^{0\nu}$: phase volume with $g_A=1.25$ and R = 1.2 fm $A^{1/3}$.
$m_{\beta\beta}$: the range of the upper limit on the effective Majorana $\nu\nu$-mass. See text.
\label{tab:ej1}}
\begin{tabular}{lcccc} \hline
Isotope  & $Q_{\beta \beta }$ [MeV]  & $T^{0\nu }_{1/2}$ [10$^{24}$ y] &   $m_{\beta\beta}$ [meV]  & Experiment \\ 
$^{76}$Ge   & 2.039 &   52      &  160-260 & GERDA  Ge semiconductor$^a$\\
$^{100}$Mo   & 3.034 &  1      & 900 - 300  & NEMO-3  Tracking chamber$^b$\\
$^{130}$Te & 2.528  & 4    & 760 - 270 & CUORE Bolometer$^c$ \\
$^{136}$Xe & 2.459 & 11    &  450 - 190  & EXO  ionization-scintillation$^d$\\
$^{136}$Xe & 2.459 & 110    &  161-60  & KamLAND-Zen  Scintillator$^e$\\\hline
\end{tabular}

a:\cite{Gelimit}, b:\cite{nem11}, c:\cite{alf15}, d:\cite{EXO14}, e:\cite{Xelimit}.
\medskip 
\label{tab:eji1}
\end{table} 

GERDA phase I
aims at  high energy-resolution studies of $^{76}$Ge 
by using $^{76}$Ge detectors (17.7 kg +3.6 kg) immersed into  
liquid scintillator\cite{ago13}.  The BG level is 0.01 /(keV kg y) at ROI. 
The phase I data with 21.6 kg y exposure gives a  half-life limit of 
 2.1$\times$ 10$^{25}$ y with 90$\%$ CL.  The limit obtained  by combining the previous data\cite{kla01}
is 3$\times $10$^{25}$ y with 90 $\%$ CL. Thus the claim of the 0$\nu \beta \beta $ evidence\cite{kla04} 
is strongly disfavored with 99 $\%$ probability. 
The combined data set an upper limit on the effective $\nu$ mass as 200-400 meV.  Recently a new limit of 5.2 10 $^{25}$ y (160-260 meV) with 90$\%$ CL was derived \cite{Gelimit}.  

A new value for 2$\nu \beta \beta $ half life is 
 $T^{2\nu}_{1/2}$=(1.926 $\pm$ 0.095) $\times10^{21}$ y and the 0$\nu M \beta \beta $ half-life limit \cite{ago15} is
 $T^{0\nu M}$= 4.3 10$^{23}$ y, corresponding to $\left<g_{ee}\right>$=(3.4-8.7) 10$^{-5}$
 
 NEMO-3 with the tracking chambers and the PL scintillation detectors 
gives half-life limits of 1.1 10$^{24}$ y on $^{100}$Mo \cite{nem11}. 
The corresponding $\nu $-mass limit is 300-900 meV. The half-life limits on the one Majoron emitting process are $T^{0\nu M}$ =
2.7 10$^{22}$ y  for $^{100}$Mo \cite{nem3A}. The limits on the Majoron coupling 
is $\left<g_{ee}\right>$=(0.4-1.9) 10$^{-4}$.  
 
 CUORE0 is one module of CUORE, which is an expansion of CUORICINO\cite{arn11}. 
It is a high energy-resolution bolometer array to study 
$^{130}$Te $0\nu \beta \beta $ decays  at LNGS by using 39 kg TeO$_2$ (10.9 kg $^{130}$Te). 
The CUORE0  experiment reports the result of the  9.8 kg y exposure\cite{alf15} with the energy resolution of 4.9 keV in FWHM and the BG level is 
0.058 / (keV kg y). The obtained half-life limit is 2.7 $\times $10$^{24}$ y with 90$\%$ CL, 
and the limit derived by combining the previous CUORITINO experiment is 4 $\times$ 10$^{24}$ y. The corresponding 
$\nu$-mass limit is 270-760 meV.

EXO-200 is the 80$\%$-enriched $^{136}$Xe TPC experiment (150 kg in the detector proper) at WIPP.  
The energy-resolution of around 3.5  $\%$ in FWHM is achieved by measuring 
both the ionization and scintillation signals\cite{EXO2012,EXO14}.
A lower limit on $T_{1/2}^{0\nu\beta\beta}$  was derived as 1.1 $\times $10$^{25}$ y  
 ($\nu$-mass limit is 190 meV - 450 meV) with 90 $\%$ CL\cite{EXO14}.
 The BG rate at ROI is 1.7 $\times $ 10$^{-3}$/(keV kg y)\cite{EXO14}.
A lower limit of 1.2 $\times $ 10$^{24}$ with 90$\%$ CL was derived 
for $T_{1/2}^{0\nu M}$\cite{EXO15AA}. This limit corresponds to the upper limit of $\left<g_{ee}\right>$=(0.8-1.6 )$\times 10^{-5}$. 
 The precise 2$\nu \beta \beta $ half life of $T^{2\nu}_{1/2}$=2.165 $\times $10$^{21}$ y was obtained\cite{EXO14A}.

KamLAND-Zen  
studies the $^{136}$Xe DBD by means of the KamLAND detector with the 1 kt liquid scintillator 
at Kamioka\cite{kamlandzen}.  A mini balloon is set at the center for the 
$^{136}$Xe-loaded liquid scintillator with $^{136}$Xe isotopes around 320-380 kg. The energy resolution is 9.5 $\%$ in FWHM.
The 2$\nu \beta \beta $ half life of $T_{1/2}^{2\nu}$=2.38 $\times $10$^{21}$ y\cite{kamlandzen} is 
in consistent with the value by EXO\cite{EXO14A}.
A half-life limit on the 0$\nu \beta \beta $ was derived as 1.9 $\times $ 10$^{25}$ y\cite{kamlandzenA}. The limit by combining 
the EXO experiment is 3.4 $\times $ 10$^{25}$ y \cite{kamlandzenA}.
The lower limit\cite{gan12} on the $T^{0\nu M}_{1/2}$ is 2.6 10$^{24}$ y, corresponding 
to the upper limit of $\left<g_{ee}\right>$=(0.8-1.6) 10$^{-5}$.   
 Recently, a stringent limit of 1.1 $\times 10^{26}$ y was 
derived by combining the new data with
 the previous ones. It
 corresponds to the $\nu $ mass limit of 
60-161 meV \cite{Xelimit}. 

It is noted that the HM claim for the $0\nu\beta\beta$ peak\cite{evidence2,kla04} is strongly disfavored not only by the GERDA experiment\cite{ago13}, 
but also by the $^{136}$Xe 
experiments\cite{kamlandzenA,EXO2012} if one may use relative matrix elements of 
$^{76}$Ge and $^{136}$Xe given by various models.  

The 2$\nu\beta \beta $-decay rates have been observed in 12 nuclides ($^{48}$Ca, $^{76}$Ge $^{82}$Se, $^{96}$Zr, 
$^{100}$Mo, $^{116}$Cd, $^{128}$Te, $^{130}$Te, $^{136}$Xe, $^{150}$Nd, $^{130}$Ba and $^{238}$U) and in two excited states. 
The recommended half lives can be found in ref. \cite{bar15}. 
These half lives give 2$\nu\beta\beta$ 
NMEs $M^{2\nu}$.

The recent stringent limits on $T^{0\nu}_{1/2}$ give the upper limit of $m_{\beta\beta} \approx$ 60-400  meV in the QD region. 
Here the $\nu$-mass depends strongly on ${M'}^{0\nu}_\nu$. 
In order to search for the full IH-mass region of 15-45 meV, one needs to improve the mass sensitivity by one order of magnitude, 
i.e. the half-life sensitivity by 2 orders of magnitude 
The limits on $T^{0\nu M}_{1/2}$ give the upper limit of the Majoron neutrino coupling $\left<g_{ee}\right>\approx$0.8-1.6, depending on 
${M'}^{0\nu M}$. It is very crucial to get right NMEs ${M'}^{0\nu}_\nu$ and ${M'}^{0\nu M}$.

\subsection{Future high-sensitivity experiments}

Experimental DBD groups are extensively working for future high-sensitivity experiments.
Some of them are listed in Table \ref{tab:ej2}. We briefly describe them below.

\begin{table}[!t]
\caption{High-sensitivity DBD experiments in futures.   $A$:natural abundance. 
$Q_{\beta \beta }$ : $Q$ value for the 0$^+\rightarrow $0$^+$ and low BG ground state transition. 
$G^{0\nu}$: kinematic (phase space volume) factor ($g_A=1.25$ and R = 1.2 fm $A^{1/3}$).
\label{tab:ej2}}
\begin{tabular}{lccccc} \hline
isotope & $A$ & $Q_{\beta \beta }$ & $G^{0\nu}$   & Future experiments \\ 
& $[\%]$  & [MeV] & [$10^{-15}$ $y^{-1}$]  & experiments \\ \hline

$^{76}$Ge  & 7.8  & 2.039 &  2.36  & GERDA, Majorana Demonstrator  \\
 $^{82}$Se  & 9.2  & 2.992 & 10.2 & SuperNEMO, MOON  \\
$^{100}$Mo & 9.6  & 3.034 & 15.9  &   AMoRE, LUMINEU, CUPID, MOON  \\
$^{116}$Cd & 7.5  & 2.804 & 16.7  & AURORA COBRA \\ 
$^{130}$Te & 34.5 & 2.529 & 14.2  & CUORE \\
$^{136}$Xe & 8.9  & 2.467 & 14.6 & EXO, KamLAND-Zen, NEXT, Panda X-III\\
$^{150}$Nd & 5.6  & 3.368 & 63.0 & SuperNEMO, SON+, DCBA \\
\hline
\end{tabular}
\medskip 
\end{table}

GERDA Phase II studies the 0$\nu \beta \beta $ of $^{76}$Ge by introducing  additional 20 kg Ge detectors\cite{wes15}. 
The energy resolution, the PSD and others are improved to get the very low BG level around 1/(keV y t). 
The expected sensitivity is 1.4 $\times $10$^{26}$ y after 100 kg y exposure, which corresponds to 
 100-200 meV. 

MJD (MAJORANA DEMONSTRATOR)  is the 44 kg Ge (29 kg $^{76}$Ge and 15 kg $^n$Ge) experiment to study
$^{76}$Ge at SURF\cite{gue15}.  It uses PPC (p-type point contact) Ge detectors, and the expected BG rate at ROI is $B\le$3.5 /(y t). 
It is envisioned to demonstrate a path forward to achieve  
the BG rate 1/yt for the next generation ton-scale experiment to study the $\nu$-mass in the IH mass region.
 The module 1 (23 kg) and 2 (21 kg) are taking data. 
 The  Majorana and GERDA collaborations will be merged for one ton-scale future experiment 
by selecting the best techniques.
 
MOON is an extension of ELEGANT V\cite{eji01}. It is a hybrid $\beta\beta$ and solar $\nu$ 
experiment with $^{100}$Mo to study the QD/IH $\nu $-mass 
sensitivity and the low energy  solar neutrinos\cite{eji00a,nak07,eji08}.
Super-module of PL plates and wire chambers are being developed.  

SuperNEMO studies the IH mass region by large tracking chambers and PL scintillation detectors 
with 100 kg $^{82}$Se isotopes \cite{arn10,hod15,piq16}.  It uses 20 modules, each module with 5 kg $\beta \beta $ isotopes. 
The detector is based on NEMO-3, but the energy resolution 
and the BG level have been much improved.  The expected sensitivity with the BG rate 
12/(t y) is $T_{1/2}$ = 1 $\times $ 10$^{26}$ y, corresponding to the $\nu$ mass of 50-140 meV.
SuperNEMO demonstrator with 7 kg $^{82}$Se is in preparation. The sensitivity is $T_{1/2}$=6 $\times $10$^{24}$ y (200-500 meV). 
The energy resolution is around 4$\%$ at $E$=3 MeV.   

AMoRE aims to study $^{100}$Mo in 
the region of IH mass (5 $\times 10^{26}$ y) by using 200 kg $^{40}$Ca$^{100}$MoO$_4$ low-temperature detectors at Y2L\cite{par15}. 
PSD with the phonon and photon signals is used to separate $\alpha $ BGs from electron signals. The pilot experiment with a 
1.5 kg $^{48dep}$Ca$^{100}$MoO$_4$ detector is going on at Y2L.  
 
LUMINEU is developing cryogenic scintillation bolometers  of Li$_2$MoO$_4$ and ZnMoO$_4$ to study $^{100}$Mo\cite{bar14,arm15,bek16}. 
High purity crystals with PSD make it possible to get 
low BG level of around 0.3 /(keV t y) and good energy resolution of around 5 keV.

CUPID ( CUORE Upgrade with PI) is a proposed ton-scale bolometer experiment with the $\nu$-mass 
sensitivity of the order of 10 meV \cite{wan15}. The collaboration is starting  40 crystals of Li$_2$MoO$_4$ with 6 kg $^{100}$Mo  \cite{giu16}.

COBRA and AURORA study $^{116}$Cd. COBRA uses a large amount of high energy-resolution CZT(CdZnTe) semiconductors at room temperature\cite{zat15}. 
The COBRA collaboration tests 64 CZT 1 cm$^3$ detectors at LNGS. The energy resolution is $\Delta E$=1.1$\%$ at 2.6 MeV. Pixelization is  a major step forward. Recent results of the COBRA demonstrator  has been reported \cite{ebe15}.
  AURORA uses a $^{116}$CdWO$_4$ and the current half-life limit is 1.9$\times 10^{23}$ y \cite{dan16}

CUORE uses 988 natural TeO$_2$  
crystals with the natural Te isotopes (741 kg TeO$_2$, 206 kg $^{130}$Te) to 
 study $^{130}$Te at the IH $\nu $-mass region\cite{alf15,gir15}. 
 By improving the BG level to be 0.01/(keV kg y), 
the expected sensitivity is 9.5 $\times $ 10$^{25}$ y. This corresponds to the $\nu$-mass sensitivity around 50 meV-150 meV.

nEXO is an expansion of  EXO-200 by using a low BG and 2.3$\%$  energy-resolution TPC with scintillation and ionization readouts\cite{EXO15AA,lin15}. 
 The expected half-life sensitivity is around 10$^{28}$ (6-15 meV) by using 5 ton $^{136}$Xe isotopes for 5 y run.. 

KamLAND-Zen aims at  the higher-sensitivity $^{136}$Xe experiment by using 0.75 ton enriched $^{136}$Xe isotopes\cite{shi13,ino16}.
 The energy resolution and the BG rate will be same as those of the previous phases. The expected half-life sensitivity
 is around 5 $\times $10$^{26}$, which corresponds to the IH $\nu $-mass region. 
NEXT-100  uses a high pressure 
 TPC with 100 kg enriched $^{136}$Xe to study the QD $\nu $ mass ($\le$ 100 meV) at LSC\cite{NEXT12,lor14}. 
It is a low BG and good energy resolution TPC with separate readout planes for tracking and energy. 
NEXT-DEM is a demonstrator. 

SNO+ uses the 1 kt scintillation detector with 0.3 $\%$ loading of natural Te isotopes (800 kg $^{130}$Te) to 
study the $^{130}$Te decays and plan 0.5 $\%$ loading \cite{loz14,man11}. 

Current DBD experiments are mostly on high $Q_{\beta \beta }$ $\beta ^-\beta ^-$ decays
 because of the large phase volume. CANDELS  $^{48}$Ca, DCBA-MTD $^{150}$Nd, Panda X-III and others are under development as 
discussed in the reviews \cite{eji05,AEE08,ROP12} and NEUTRINO 2016. CANDLES III with 300 kg natural Ca will be enlarged to IV and V with enriched  isotopes in future \cite{kis16}.
Panda X-III is a high presure Xe TPC with 200 kg $^{136}$Xe source at CJPL \cite{gib15}. The 0$\nu \beta ^+ \beta ^+$, 0$\nu \beta+ $EC and 0$\nu $EC EC decays are studied by 
measuring the $\beta^+$ annihilation $\gamma $ 
rays and K X-rays. 

\subsection{Experimental studies of  DBD matrix elements}
\label{sect:expME}
At this point 
  we should mention the progress  towards using other experiments \cite{eji05,eji10,eji00}, mainly charge exchange reactions (CERs) and single $\beta $ decays  
  \cite{eji15,eji14} to help evaluate DBD NMEs needed in extracting the neutrino mass from $0\nu \,\beta\beta$ experiments.

One direct way to get the weak response is to use the $\nu $ CER of ($\nu _e$,e). 
Since the $\nu $ nuclear cross section is as small as $\sigma \approx 10^{-42}$ cm$^2$,
 one needs high-flux $\nu $ beams and large-volume detectors\cite{eji05,eji10}. 
 The $\nu$ beams may be obtained from pion decays and 
the pions are obtained from GeV proton beams from SNS at ORNL\cite{avi00} and J-PARC\cite{eji03a}. 

Muon ($\mu ^-$) CERs of ($\mu,\nu_{\mu}$)\cite{eji06,suh06} give $\beta ^+$ 
strengths for $J=0.1.2 $ states in the wide excitation region of $E$= 0-70 MeV. 
 Recently ($\mu ^-,$xn$\gamma $) reactions on Mo isotopes were studied by using 
DC and pulsed $\mu$ beams at RCNP and MLF J-PARC\cite{eji13a}. 
 The  $\mu$ CER of $^{100}$Mo ($\mu,x$n) $^{100-x}$Nb was studied by measuring delayed $\beta -\gamma$ rays 
from $^{100-x}$Nb isotopes. The weak strength distribution for the $\mu $ CER was deduced
from the observed Nb RI distribution by using the statistical model for the neutron emission. 
The $\mu$ capture shows a giant
 resonance around $E$=10-15 MeV. The $\mu$ capture life time gives the absolute weak strength  and  the 
NMEs with effective $g_A$.

Weak responses for $\beta ^+ $ decays are studied by using photo-nuclear ($\gamma $,x) reactions through isobaric analogue states (IAS)\cite{eji68}. 
The polarization of the photon can be used to study E1 and M1 matrix elements separately\cite{eji13}.

 Nuclear CERs with medium energy nuclear beams were used to study 
single $\beta ^{\pm}$ NMEs at IUCF, KVI, MSU, RCNP, 
Triumf and others\cite{ROP12,eji05,eji00}.   The ($^3$He,t) reactions with the high energy-resolution of $\Delta E \approx $ 
25 keV were studied extensively on DBD nuclei  by using the 420 MeV $^3$He beam at RCNP 
\cite{gue11,pup11,pup12,thi12,thi12a,thi12b,Freckers16}. Recently spin dipole (2$^-$) NMEs, which may be  relevant to 0$\nu \beta \beta $ NMEs, have been studied 
by using ($^3$He,t) reactions at RCNP \cite{eji16A}.
Double charge exchange reactions DCER provide useful information on DBD NMEs. The RCNP DCER  ($^{11}$B,$^{11}$Li) at $E/A$=80 MeV 
shows that the DCE strengths are not located at the low lying states, but mostly concentrated at the high excitation region \cite{eji16B}. Furthermore the DCER
 of ($^{18}$O,$^{18}$Ne)
 on $^{40}$Ca was measured to get the DCER strength for the ground state 0$^+\rightarrow 0^+$ transitions \cite{cap15}. Nucleon transfer reactions have been used to get nucleon vacancy and occupation probabilities of DBD initial and final nuclei \cite{SchifferGea}.

The Fermi Surface Quasi Particle model (FSQP) based on experimental single-$\beta ^{\pm}$ NMEs  
 reproduces well the 2$\nu \beta \beta $ NMEs 
$M^{2\nu}$ for the  A(Z,N) $\leftrightarrow$ C(Z+2,N-2) ground-state to ground state 
0$^+\leftrightarrow0^+$  transition\cite{eji96,eji09,eji12}. The initial, intermediate and final state nucleons involved in the ground state 
transition must necessarily be on the diffused Fermi surface. Thus 
$M^{2\nu}$ is given by the sum of the products of the single $\beta ^{\pm}$ matrix elements via the FSQP intermediate states \cite{eji05,ROP12,eji09}.
 The agreement of $M^{2\nu}(EXP)$ and $M^{2\nu}(FSQP)$ shows 
that $M^{2\nu}$ is very small due to the nuclear medium and correlation 
effects given by $k^-k^+ \approx $0.06 with $k^-\approx k^+ \approx $0.25 for the single $\beta ^{\pm}$ decay NMEs. 
FSQP may be used also for evaluating ${M'}^{0\nu}_\nu$ by using CERs and single $\beta $/EC rates to intermediate 
1$^{\pm}, 2^{\pm}$, and other states.

A recent analysis of $\beta ^{\pm}$-EC NMEs  in medium heavy nuclei shows that the axial vector 
GT(Gamow-Teller 1$^+$) and SD (spin dipole 2$^-$) NMEs 
 are reduced with respect to the single quasi-particle NMEs by the coefficient $k^{\pm}\approx $0.20-0.25 \cite{eji15,eji14}. 
The reduction may be inferred to be partly due to the nucleon
spin isospin correlation of $k_{\sigma \tau}\approx $0.4-0.5 and partly due to the non-nucleon (isobar) and nuclear medium 
effect of $k_{NM}\approx $0.5-0.6\cite{eji15,eji14}.  Here the effect of $k_{\sigma \tau}\approx $0.4-0.5 is included in QRPA with 
$\sigma \tau$ correlation, while the isobar and nuclear medium effect is a sort  of the renormalization (quenching)
of the axial-vector weak coupling. The renormalization effect may be incorporated by the effective coupling 
 of ($g_A^{\rm eff}/g_A)\approx $ 0.5-0.6. Accordingly the axial vector components of the DBD NME is reduced with respect 
to the QRPA model calculations by the coefficient $(g_A^{\rm eff}/g_A)^2 \approx $0.3.  

 So far no experimental data on 2$\nu \beta ^+$/EC NMEs are available mainly due to the small phase space. 
One possible candidate is $^{106}$Cd. 
FSQP predicts a large NME of $M^{2\nu}(FSQP)$ =0.10 in unit of $m_e^{-1}$.  The FSQP half lives  for ECEC, EC$\beta ^+$ and $\beta ^+\beta ^+$ are 
5.2$\times 10^{21}$ y,
 4.4 $\times 10^{22}$ y and 1.7$\times 10^{27}$ y, respectively, They
 are longer than experimental limits of 4.2 $\times 10^{20}$ y, 1.1 $\times 10^{21}$ y and 2.3 $\times 10^{21}$ y, respecively \cite{bel16}. 
 The decays are also being studied by Ge detectors \cite{tvg15}.

%
%
%
\section[Expression for the lifetime  of neutrinoless double beta decay]{Expression for the lifetime  of neutrinoless double beta decay}
\label{decay mechanisms}
%

We briefly review half-lives associated with $0\nu\beta\beta$-decay mechanisms
including exchange of light and heavy Majorana neutrinos. We shall pay attention
only to the $0^+ \rightarrow 0^+$ ground state to ground state $0\nu\beta\beta$-decays
transitions.

\subsection{The light Majorana neutrino mass mechanism} 

The expression for the lifetime of $0\nu\beta\beta$ decay is simplified by the fact that it is factorized into  three factors: The phase-space  factor, the nuclear matrix elements and the lepton violating neutrino parameters.

Indeed  the inverse value of the $0\nu\beta\beta$-decay half-life can be written as\cite{ROP12}
\begin{equation}
\left({T^{0\nu}_{1/2}}\right)^{-1} = \frac{m_{\beta\beta}^2}{m_e^2}~g^4_A~\left|{M'}^{0\nu}_\nu (g_A^{\rm eff})\right|^2~
G^{0\nu}(E_0,Z),
\label{halflife}
\end{equation}
where the first term $m_{\beta\beta}^2$ is the lepton violating parameter and  $G^{0\nu}{(E_0,Z)}$,  with $E_0 = E_i-E_f$ being the energy release, is the phase-space integral given by:
\bee \label{phasespace}
&& G^{0\nu}(E_0,Z) = \frac{G_{\beta}^{4}m_e^9}{32  \pi^5 R^2 \ln{(2)}} \nonumber\\ 
&& \times\frac{1}{m_e^5}\int_{m_e}^{E_i-E_f-m_e}~\left(g^2_{-1}(\varepsilon_1) + f^2_{+1}(\varepsilon_1)\right)
~\left(g^2_{-1}(\varepsilon_2) + f^2_{+1}(\varepsilon_2)\right)
~ \varepsilon_1 p_1 ~\varepsilon_1 p_1~ d\varepsilon_1\nonumber
\eee
with $\varepsilon_2 = E_i-E_f-\varepsilon_1$, $p_i=\sqrt{\varepsilon_i^2-m_e^2}$ (i=1,2). 

Unlike in previous
derivation\cite{DTK85} the exact Dirac wave functions with finite nuclear size and
electron screening of the emitted electrons in the $s_{1/2}$ and $p_{1/2}$ wave states,
\begin{eqnarray} \label{functions1}
\psi(\mathbf{r},p,s) &\simeq&  \psi_{s_{1/2}}(\mathbf{r},p,s) + \psi_{p_{1/2}}(\mathbf{r},p,s) \\
&=&\left(\begin{array}{c}
g_{-1}(\varepsilon,r)\chi_s\\
f_{+1}(\varepsilon	,r)\left(\vec{\sigma}\cdot\hat{\mathbf{p}}\right)\chi_s\\
 \end{array} \right)
+\left(\begin{array}{c}
ig_{+1}(\varepsilon,r)\left(\vec{\sigma}\cdot\hat{\mathbf{r}}\right)\left(\vec{\sigma}\cdot\hat{\mathbf{p}}\right)\chi_s\\
-if_{-1}(\varepsilon	,r)\left(\vec{\sigma}\cdot\hat{\mathbf{r}}\right)\chi_s\\
 \end{array} \right).\nonumber
\end{eqnarray}
were taken into account.
The relativistic electron wave function $\psi(\mathbf{r},p,s)$ 
in the central symmetric Coulomb field of a uniform charge distribution 
of a nucleus can be decomposed in partial waves. In the case of
the light neutrino mass mechanism of the $0\nu\beta\beta$-decay 
only the dominant $s_{1/2}$-waves of emitted electrons were taken into account: 
\begin{eqnarray} \label{functions}
  &&\psi(\mathbf{r},p,s)\simeq  \psi_{s_{1/2}}(\mathbf{r},p,s).
\end{eqnarray}
Given the Coulomb potential of the daughter nucleus and the screening potential of bound electrons in atom 
$g_{\pm 1}(\varepsilon,r)$ and $f_{\pm 1}(\varepsilon,r)$ are solutions of the radial Dirac equations.

Furthermore the electron wave functions including relativistic effect, $g_{\pm 1}(\varepsilon,r)$,  
and $f_{\pm 1}(\varepsilon,r)$ are replaced by their values at the nuclear radius R. 
We have 
\begin{equation}
g_{\pm 1}(\varepsilon,r) = g_{\pm 1}(\varepsilon,R), ~~~~~f_{\pm 1}(\varepsilon,r) = f_{\pm 1}(\varepsilon,R).
\end{equation}
The improved values of $G^{0\nu}$ calculated by taking into account the Dirac electron wave functions
with finite nuclear size and electron screening are tabulated in Ref. \cite{KotIac12}.

The nuclear matrix element ${M'}^{0\nu}_\nu$takes the form 
\begin{eqnarray}
\label{eq:0nume}
      {M'}^{0\nu}_\nu (g_A^{\rm eff}) &=& \frac{R}{2 \pi^2 g_A^2} \sum_{n} 
\int e^{i\mathbf{p}\cdot (\mathbf{x}-\mathbf{y})}
\frac{
\langle 0^+_f| {J}^{\mu\dag}_L(\mathbf{x})|n\rangle
\langle n|{J}^\dag_{L \mu} (\mathbf{y}) |0^+_i\rangle}
  {p (p+E_n-\frac{E_i-E_f}{2})}
  d^3p~ d^3x~ d^3y, \nonumber \\
\end{eqnarray}
where $\mathbf{r}_{ij}\equiv \mathbf{r}_i-\mathbf{r}_j$.
Initial and final nuclear ground states with energies 
$E_{i}$ and $E_{f}$ are denoted by $\ket{0^+_i}$ and $\ket{0^+_f}$, respectively. 
The summation index runs over intermediate nuclear states $\ket{n}$ with energies $E_{n}$.
Details on the evaluation of the NME
in Eq. (\ref{eq:0nume}) appear, e.g., in Ref. \cite{sim08} and will
be discussed in Sec. \ref{secnme}. We note that the axial-vector
$g_A^{\rm eff}(p^2)$ and induced pseudoscalar $g_P^{\rm eff}(p^2)$ 
form factors of nuclear hadron currents ${J}^{\mu\dag}$ are
``renormalized in nuclear medium''. The magnitude and origin
of this renormalization is the subject of the  analysis of many works, since it tends to increase the
$0\nu\beta\beta$-decay half-life in comparison with the case in which this
effect is absent. This issue will be addressed in Sec. \ref{secnme}.

\begin{table}[!t]
  \caption{The current experimental lower limit on the $0\nu\beta\beta$-decay half-life for 10
    isotopes with largest $Q_{\beta\beta}$-value and theoretical predictions for the hallf-life
    by assuming the cases of normal (NH) and inverted (IH) hierarchies in evaluation of
    effective Majorana neutrino mass $m_{\beta\beta}$. The averaged values of nuclear matrix elements
    with variances from Table \protect\ref{tab.nmeL} are taken into account.
   The non-quenched value of the weak-axial coupling constant is assumed. The constraints on
    $T^{0\nu - exp}_{1/2}$ are from NEMO3 ($^{48}$Ca\protect\cite{Calimit}, $^{82} \mbox{Se}$\protect\cite{nemoiii05},
    $^{100} \mbox{Mo}$\protect\cite{Molimit}, $^{150} \mbox{Nd}$\protect\cite{Ndlimit}), GERDA ($^{76} \mbox{Ge}$\protect\cite{Gelimit}),
   CAMEO ($^{116} \mbox{Cd}$\protect\cite{Cdlimit}, CUORE ($^{130}\mbox{Te}$\protect\cite{Telimit})
   and Kamlandzen ($^{136} \mbox{Xe}$\protect\cite{Xelimit})
   experiments.
\label{tab.t12}}
\begin{tabular}{lccccc}\hline\hline
  Nucl. & $G^{0\nu}$ [$10^{-15}$yr$^{-1}$]& $T^{0\nu - exp}_{1/2} $ [yr] & &  \multicolumn{2}{c}{ $T^{0\nu - theor}_{1/2}$ [yr] } \\ \cline{5-6}
  &     &                     & &   IH  &  NH   \\ \hline
${^{48}}$Ca    &   24.81  & $>2.0\times 10^{22}$  & & (0.37, 36.)$\times 10^{27}$ & (0.68, 74.)$\times 10^{29}$    \\
${^{76}}$Ge    &   2.363  & $>5.2\times 10^{25}$  & & (0.54, 9.5)$\times 10^{27}$ & (0.99, 19.)$\times 10^{29}$    \\
${^{82}}$Se    &   10.16  & $>2.5\times 10^{23}$  & & (0.17, 2.6)$\times 10^{27}$ & (0.31, 5.2)$\times 10^{29}$    \\
${^{96}}$Zr    &   20.58  &        -             & & (0.52, 74.)$\times 10^{26}$ & (0.96, 150)$\times 10^{28}$    \\
${^{100}}$Mo   &   15.92  & $>1.1\times 10^{24}$  & & (0.70, 6.4)$\times 10^{26}$ & (0.13, 1.3)$\times 10^{29}$    \\
${^{110}}$Pd   &   4.815  &        -             & & (0.14, 2.8)$\times 10^{27}$ & (0.26, 5.7)$\times 10^{29}$    \\
${^{116}}$Cd   &   16.70  & $>1.7\times 10^{23}$  & & (0.95, 12.)$\times 10^{26}$ & (0.18, 2.4)$\times 10^{29}$    \\
${^{124}}$Sn   &   9.040  &        -             & & (0.28, 6.6)$\times 10^{27}$ & (0.52, 13.)$\times 10^{29}$    \\
${^{130}}$Te   &   14.22  & $>4.0\times 10^{24}$  & & (0.11, 5.8)$\times 10^{27}$ & (0.21, 12.)$\times 10^{29}$    \\
${^{136}}$Xe   &   14.58  & $>1.1\times 10^{26}$  & & (0.21, 7.8)$\times 10^{27}$ & (0.39, 16.)$\times 10^{29}$    \\
${^{150}}$Nd   &   63.03  & $>2.0\times 10^{22}$  & & (0.35, 14.)$\times 10^{26}$ & (0.65, 29.)$\times 10^{28}$    \\
  \hline \hline
\end{tabular}
\end{table}

\begin{figure}[!t]
   \vspace*{1.2cm}
\centerline{\psfig{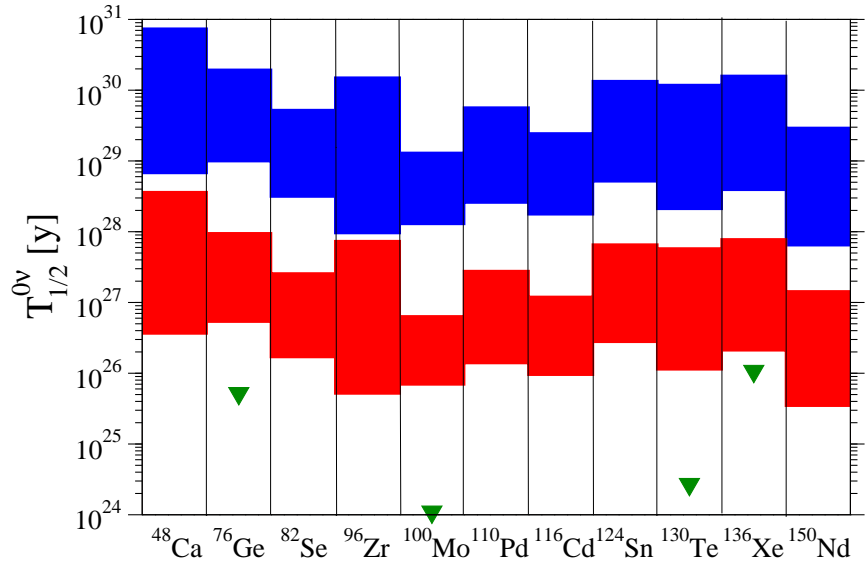}}
  \caption{(Color online.) The $0\nu\beta\beta$-decay half-lives 
    of nuclei of experimental interest calculated for the case of
    the NH (blue region) and IH (red region) (see Eqs. (\protect\ref{mbbNH})
    and (\protect\ref{mbbIH}). The averaged values with their variances
    for a given isotope from Table \protect\ref{tab.nmeL} are considered.
    The non-quenched value of weak-axial coupling constant is assumed.
    The current experimental constraints on the $0\nu\beta\beta$-decay
    half-life are shown with filled green triangles.
\label{fig.t12}}
\end{figure}

In Table \ref{tab.t12} we show predicted lifetimes of the $0\nu\beta\beta$-decay, obtained  with a Majorana neutrino mass
corresponding to the NH and IH (see Eqs. (\ref{mbbNH}) and (\ref{mbbIH})) and by asssuming averaged
values and variances of ${M'}^{0\nu}_\nu$ given in Table \ref{tab.nmeL}. They are compared with the best
experimental limits on the $0\nu\beta\beta$-decay half-life. We see that in the case of $^{136}$Xe the current
constraint is close to the range corresponding to the IH. This situation is displayed also in Fig.
\ref{fig.t12}.

\subsection{The neutrinoless double beta decay with sterile neutrinos}

 In our treatment sterile means a neutrino species, which does not participate in weak interactions. The mass of the sterile neutrinos is not known, so the could be light, but much heavier than the standard neutrinos or very heavy.

The contribution of such a neutrino to the $0\nu\beta\beta$-decay amplitude
is due to  its nonzero admixture to $\nu_e$ weak eigenstate and  as a result  is described 
by the standard neutrino exchange diagram between the two
$\beta$-decaying neutrons. Assuming  LNV  dominance
the $0\nu\beta\beta$ decay half-life for a transition to the ground state of final nucleus 
takes the form\cite{SterileN14}
\begin{equation}
[T_{1/2}^{0\nu}]^{-1} = G^{0\nu}(E_0,Z) g_{\rm A}^{4} 
\left|\sum\limits_{\rm N}\left(U^{2}_{e\rm N} m_{\rm N}\right) m_{\rm p}\, 
{M}^{\prime\, 0\nu}(m_{\rm N}, g_{\rm A}^{\rm eff}) \right|^{2},
\label{eq:1}   
\end{equation}
with  $m_{\rm p}$ the proton mass. In the above formula $g_{\rm A}$ and $g^{\rm eff}_{\rm A}$ 
stand respectively for the standard and "quenched'' values of the nucleon axial-vector coupling constants.
The relevant  nuclear matrix element  ${M}^{\prime 0\nu}$ is given by
\begin{eqnarray}\label{eq:MnuN}
&&{M}^{\prime\, 0\nu}(m_{\rm N}, g_{\rm A}^{\rm eff})
= \frac{1}{m_{\rm p}m_{\rm e}}~
\frac{R}{2 \pi^2 g^2_A} \sum_{n} \!\! 
 \int \! d^3x \, d^3y \,  d^3p \nonumber\\
&&\times e^{i\mathbf{\rm p}\cdot (\mathbf{x}-\mathbf{y})} \frac{\bra{0^+_F} 
{J}^{\mu\dag}_L(\mathbf{x})
\ket{n}\bra{n}
{J}^\dag_{L \mu} (\mathbf{y}) \ket{0^+_I}}{\sqrt{p^2+m_N^2} 
(\sqrt{p^2+m_N^2} + E_n-\frac{E_I-E_F}{2})}  \,. 
\nonumber\\
\end{eqnarray}
The dependence on $g_{\rm A}^{\rm eff}$ has been incorporated into  the weak one-body charged current
${J}^{\dag}_\mu$. 

Two conventional limiting cases are of interest, namely light, 
$m_{\rm N}\ll p_{\rm F}$, and the heavy,  $m_{\rm N} \gg p_{\rm F}$, Majorana neutrino exchange mechanisms,
where $p_{\rm F}\sim$ 200 MeV is the characteristic momentum carried via the virtual neutrino. In such cases 
 the half-life in Eq. (\ref{eq:1}) can be written as
\begin{eqnarray}
  &&[T_{1/2}^{0\nu}]^{-1} = G^{0\nu}(E_0,Z) g_{\rm A}^{4}
  \left\{\begin{array}{ll}
    \left|\frac{\langle m_{\nu}\rangle}{m_{\rm e}}\right|^{2} 
     \left|{M}^{\prime 0\nu}_{\nu}(g^{\rm eff}_{\rm A})\right|^{2},  
& \mbox{for}\  m_{\rm N} \ll p_{\rm F},    \\[3mm]
\left|\langle\frac{1}{m_{\rm N}}\rangle m_{\rm p}\right|^{2} 
   \left|{M}^{\prime 0\nu}_{\rm N}(g^{\rm eff}_{\rm A})\right|^{2}, 
    &  \mbox{for}\  m_{\rm N} \gg p_{\rm F},  
\end{array}\right.
\label{LightHeavy}   
\end{eqnarray}
with 
\begin{eqnarray}\label{meanMass}
\langle m_{\nu}\rangle = \sum\limits_{\rm N} U_{\rm eN}^{2} m_{\rm N},\ \ \ \ 
\left\langle\frac{1}{m_{\rm N}}\right\rangle =  \sum\limits_{\rm N} \frac{U_{\rm eN}^{2}}{m_{\rm N}}.
\end{eqnarray}
Here, $p_F$ is identified with  the Fermi momentum. The NMEs ${M}^{\prime 0\nu}_{\nu}, {M}^{\prime 0\nu}_{N}$ are obtained from 
${M}^{\prime 0\nu}$ appearing in Eq. (\ref{eq:MnuN}) as follows
\begin{eqnarray}\label{lim-rel-1}
{M}^{\prime 0\nu} (m_{\rm N}\rightarrow 0, g^{\rm eff}_{\rm A})
&=&  \frac{1}{m_{\rm p} m_{\rm e}} {M}^{\prime 0\nu}_{\nu}(g^{\rm eff}_{\rm A}),\\
\label{lim-rel-2}
{M}^{\prime 0\nu} (m_{\rm N}\rightarrow \infty, g^{\rm eff}_{\rm A})
 & = & \frac{1}{m_{\rm N}^{2}} {M}^{\prime 0\nu}_{\rm N}(g^{\rm eff}_{\rm A}).
\end{eqnarray}
The NMEs of ${M}^{\prime 0\nu}_{\nu}(g^{\rm eff}_{\rm A})$ and  ${M}^{\prime 0\nu}_{\rm N}(g^{\rm eff}_{\rm A})$
associated with exchange of light and very heavy neutrinos can be found in Tables
\ref{tab.nmeL} and \ref{tab.nmeH}.

It has been shown in Ref.\cite{Kovalenko:2009td} that, with the use  of the above NMEs corresponding
to the two limiting-cases, the  half-life given by  Eq. (\ref{eq:MnuN}) can be obtained  with a reasonably good
accuracy using  an  "interpolating formula'' given by 
\begin{eqnarray}\label{interpol}
[T_{1/2}^{0\nu}]^{-1} = {\cal A}~
\left|m_{\rm p} \sum\limits_{\rm N}U^{2}_{e\rm N} \frac{m_{\rm N}} {\langle p^{2}\rangle + m_{\rm N}^{2}} \right|^{2},
\end{eqnarray}
where 
\begin{eqnarray}
{\cal A} \ \ &=& G^{0\nu}(E_0,Z) g_{\rm A}^{4}  \left|M^{\prime 0\nu}_{\rm N}(g^{\rm eff}_{\rm A})  \right|^{2}, \\   
\langle p^{2}\rangle &=& m_{\rm p} m_{\rm e} \left|\frac{M^{\prime 0\nu}_{\rm N}(g^{\rm eff}_{\rm A}) }
{M^{\prime 0\nu}_{\nu}(g^{\rm eff}_{\rm A}) } \right|   \label{coeff-2}
\end{eqnarray}

The values parameters $\langle p^{2}\rangle$ obtained from
${M}^{\prime 0\nu}_{\nu}(g^{\rm eff}_{\rm A})$ and  ${M}^{\prime 0\nu}_{\rm N}(g^{\rm eff}_{\rm A})$, 
 in the context of various   nuclear structure methods, are given Table \ref{tab.ster}.
A numerical comparison of the  "exact'' results, obtained within the QRPA with isospin
restoration\cite{SRFV13}, with those calculated with the interpolating formula (\ref{interpol})
showed a rather good agreement, with the possible exception of the transition region where the accuracy is about 
20\% - 25\% \cite{SterileN14}. By glancing the Table  \ref{tab.ster} one can  see that there is
a significant difference in the  values of $\sqrt{\langle p^{2}\rangle}$ calculated by
different methods, using different short range two nucleon  correlations. 
We note that the parameter  $\sqrt{\langle p^{2}\rangle}$with  typical values 
$\sim$ 150-200 MeV can be interpreted as the mean Fermi momentum
 $p_{\rm F}$ of the nucleons in the nucleus, as suggested by the structure of the NME
in Eq. (\ref{eq:MnuN}).

\begin{table}[!t]
  \caption{The values of the parameter $\langle p^2\rangle$ of the interpolating formula
    specified in Eq. (\protect\ref{coeff-2}) calculated within different nuclear structure
    approaches: interacting shell model (ISM) (Strasbourg-Madrid (StMa)\protect\cite{StMa09}
and Central Michigan University (CMU)\protect\cite{CMU16} groups),
interacting boson model (IBM)\protect\cite{IBM15},
quasiparticle random phase approximation (QRPA) (Tuebingen-Bratislava-Caltech 
(TBC)\protect\cite{TBC13,Fang15} and Jyv\"askyla (Jy)\protect\cite{QJy15} groups), 
projected Hartree-Fock Bogoliubov approach (PHFB)\protect\cite{phfb12H}.
The Argonne, CD-Bonn and UCOM two-nucleon short-range correlations  are taken into account.
The non-quenched value of weak axial-vector coupling $g_A$ is assumed.
\label{tab.ster}}
\begin{tabular}{lcccccccc}\hline\hline
  &       &         & \multicolumn{6}{c}{$\sqrt{\langle p^{2}\rangle}$ [MeV]}  \\ \cline{4-9}
  Method & $g_A$ & src     & ${^{48}}$Ca  & ${^{76}}$Ge  & ${^{82}}$Se  & ${^{96}}$Zr & ${^{100}}$Mo & ${^{110}}$Pd
  \\ \hline
ISM-StMa   & 1.25 &  UCOM   &   178    &   150       &     149      &             &            &           \\
ISM-CMU    & 1.27 & Argonne &   178    &   134       &     138      &             &            &           \\
           &      & CD-Bonn &   203    &   165       &     162      &             &            &           \\
IBM        & 1.27 & Argonne &   113    &   103       &     103      &    129      &   136      &   135     \\
QRPA-TBC   & 1.27 & Argonne &   189    &   163       &     164      &    180      &   174      &   166     \\
           &      & CD-Bonn &   231    &   193       &     194      &    211      &   204      &   194     \\
QRPA-Jy    & 1.26 & CD-Bonn &          &   191      &      192      &    217     &    207      &   187     \\
PHFB       & 1.25 & Argonne &          &             &              &    130      &   127      &   124     \\
           &      & CD-Bonn &          &             &              &    150      &   145      &   143     \\ \hline\hline
         &       &         & \multicolumn{6}{c}{$\sqrt{\langle p^{2}\rangle}$ [MeV]}  \\ \cline{4-9}
Method & $g_A$ & src       & ${^{116}}$Cd & ${^{124}}$Sn & ${^{128}}$Te & ${^{130}}$Te & ${^{136}}$Xe & ${^{150}}$Nd
  \\ \hline 
ISM-StMa    & 1.25 &  UCOM   &            &    160      &            &    161      &     159    &            \\
ISM-CMU     & 1.27 & Argonne &            &    153      &            &    159      &     170   &           \\
            &      & CD-Bonn &            &    177      &            &    184      &     197    &           \\
IBM         & 1.27 & Argonne &   130      &    109      &     109    &    109      &     107     &     155  \\
QRPA-TBC    & 1.27 & Argonne &   157      &    186      &     178    &    180      &     183    &           \\
            &      & CD-Bonn &   182      &    214      &     207    &    209      &     211    &           \\
QRPA-Jy     & 1.26 & CD-Bonn &   177      &    202      &     196    &    201      &     175    &           \\
PHFB        & 1.27 & Argonne &            &             &    131     &    132      &            &     121  \\
            &      & CD-Bonn &            &             &    150     &    150      &            &     139   
\\ \hline \hline
\end{tabular}
\end{table}

The apparent advantage of the formula (\ref{interpol}) is the fact  that it exhibits explicitly
the $m_{\rm N}$ dependence of the $0\nu\beta\beta$ amplitude or the half-life. It can be conveniently used for an analysis  of the neutrino sector
without relying on  the sophisticated machinery of the nuclear structure calculations.
Also any upgrade in the  nuclear structure approach typically brings out asymptotic
NMEs for $m_{\rm N}\ll p_{\rm F}$ and $m_{\rm N}\gg p_{\rm F}$. This method allows one to immediately
reconstruct with a good accuracy upgraded NMEs  for any value of  $m_{\rm N}$.   

\begin{figure}[!t]
   \vspace*{1.2cm}  
\centerline{\psfig{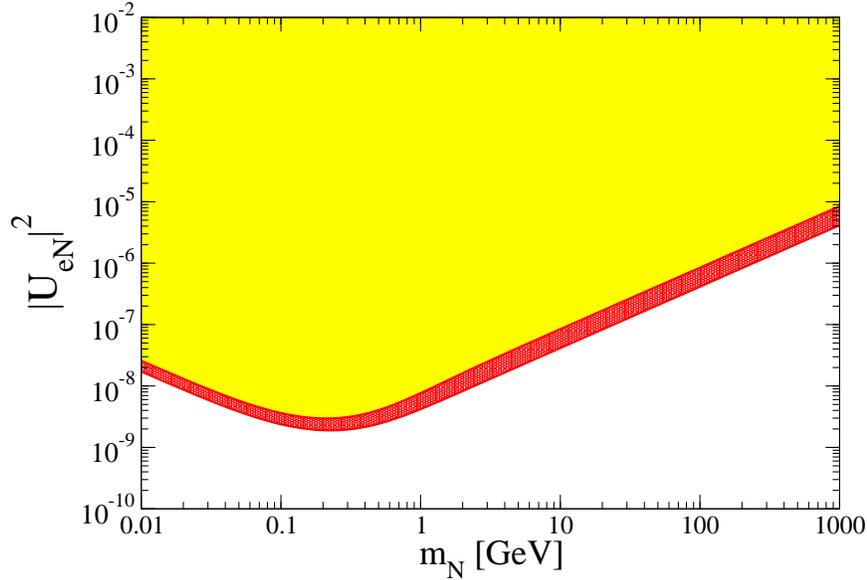}}
\caption{The exclusion plot in the $|U_{e\rm N}|^2-m_{\rm N}$ plane (yellow region) derived from the
  lower limit on the $0\nu\beta\beta$-decay half-life of $^{136}$Xe
  ($T_{1/2}^{0\nu-exp}({^{136}Xe}) > 3.4~ 10^{25}~ \mbox{yr}$,
  combined EXO+KamlandZEN) \protect\cite{kamlandzenA}.
  The weakest (strongest) limit is obtained for $M^{0\nu}(m_{\rm N})$ calculated with Argonne potential (CD-Bonn potential)
  within the QRPA with isospin restoration \protect\cite{SRFV13} and assuming $g_A^{\rm eff}=1.00$ ($g_A^{\rm eff}=1.269$).
  \label{fig.ster}}
\end{figure}

The role of the intermediate mass sterile neutrinos N in various LNV processes
has been extensively  studied in the literature (for a recent review, c.f. \cite{Helo:2010cw,Atre:2009rg})
and suitable  limits in the $|U_{\alpha N}|^{2}-m_{\rm N}$-plane have been derived. In fact it has been
shown that $0\nu\beta\beta$-decay limits for $|U_{\rm eN}|^{2}-m_{\rm N}$ are the most
stringent compared to those derived from  other LNV processes, with the possible  exception of  
a narrow region of the parametric plane \cite{Helo:2010cw,benes05,Mitra:2011qr}.
In Fig. \ref{fig.ster} we present the exclusion plot  $|U_{e N}|^{2}$ vs $m_{\rm N}$. The most stringent
half-life limit $T^{0\nu-exp}_{1/2} > 1.07\times 10^{26}$ yr \cite{Xelimit} has been used in the
analysis.

In Ref.\cite{SterileN14} the $0\nu\beta\beta$-decay half-life formula for a generic neutrino spectrum
has been  presented. It incorporates a popular scenario $\nu$MSM \cite{Asaka:2005pn,Asaka:2005an} and offers
a solution of the DM and baryon asymmetry problems utilizing massive Majorana neutrinos.
It contains the following ingredients:\\
i) three light neutrinos $\nu_{k=1,2,3}$
with the masses $m_{\nu (k)}\ll p_{\rm F}\sim$200 MeV dominated by $\nu_{e,\mu,\tau}$;\\
ii) a number of  neutrinos $\nu^{DM}_{i}$  as  dark matter (DM) candidates  with
the masses $m^{DM}_{i}$ at the keV scale;\\
iii) a number of heavy neutrinos $N$ with the masses $m_{\rm N}\gg p_{\rm F}$, \\
(iv) several
intermediate mass $m_{h}$ neutrinos $h$ including  a pair highly degenerate
in mass, which is  needed for the generation of the baryon asymmetry via leptogenesis \cite{Asaka:2005an}.\\
Taking the advantage  of the  "interpolating'' formula  (\ref{interpol})
the $0\nu\beta\beta$-decay half-life is given by\cite{SterileN14}
\begin{eqnarray}\label{Example}
&&[T_{1/2}^{0\nu}]^{-1} =  {\cal A} 
\left|\frac{m_{\rm p}}{\langle p^{2}\rangle}  
\sum\limits_{k=1}^{3}U^{2}_{ek}m_{k}  +
\frac{m_{\rm p}}{\langle p^{2}\rangle}  \sum\limits_{i}\left(U^{DM}_{ei}\right)^{2}m^{DM}_{i}  
\right.\nonumber\\
&& \left.~~~~~~~~~~~~~~~~~~ + m_{\rm p}
\sum\limits_{\rm N}\frac{U^{2}_{e\rm N} }{ m_{\rm N}} + m_{\rm p}
\sum\limits_{h}\frac{U^{2}_{eh}  m_{h}} {\langle p^{2}\rangle + m_{h}^{2}} \right|^{2}. 
\end{eqnarray}
Here, since the expected mixing $|U^{DM}_{ei}|, |U_{\rm eN}|, |U_{eh}|\ll |U_{ek}|$ between the sterile and the standard neutrinos is very small,  the
mixing between the light neutrinos $\nu_{k}$ to a good accuracy can be identified with 
the PMNS mixing matrix \mbox{$U_{ek} \approx U^{PMNS}$}.

\subsection{The right-handed current mechanisms of light neutrinos}
\label{mechanisms of light neutrinos}

Recently, the description of $0\nu\beta\beta$-decay mechanisms due to the interference of the the left handed and right handed leptonic 
currents (see section \ref{sec:rightleft}), restricted to  light neutrino exchange, has been revived and  improved \cite{DusanLR}.
Furthermore  the effect of the induced pseudoscalar term
of nucleon current was considered. Then, within the standard approximations
the $0\nu\beta\beta$-decay half-life takes the form
\bee \label{halflifem}
&&\left[T_{1/2}^{0\nu}\right]^{-1} = g_A^4\left|M_{GT}\right|^2\left\{C_{mm}\lf
\frac{\left| m_{\beta\beta}  \right|}{m_e}\rh^2 
\right.\nonumber\\ 
&& + C_{m\lambda}\frac{\left| m_{\beta\beta}\right|}{m_e}\left\langle \lambda\right\rangle\cos{\psi_1}
+ C_{m\eta}\frac{\left| m_{\beta\beta}\right|}
{m_e}\left\langle \eta\right\rangle\cos{\psi_2}\nonumber\\  
&& + \left. C_{\lambda\lambda}\left\langle \lambda\right\rangle^2 +C_{\eta\eta}\left\langle \eta\right\rangle^2
+C_{\lambda\eta}\left\langle \lambda\right\rangle\left\langle \eta\right\rangle \cos{(\psi_1-\psi_2)}\phantom{\frac{!}{!}} \right\}.\nonumber\\
\eee
The effective lepton number violating parameters 
and the relative phases appearing in the previous equation are given by
\bee\label{effective}
\left\langle \lambda\right\rangle&=&\kappa|\sum_{j=1}^3 U_{ej}T_{ej}^*(g_V'/g_V)|,~~~~~~
\left\langle \eta\right\rangle = \epsilon|\sum_{j=1}^3 U_{ej}T_{ej}^*|, \nonumber\\
\psi_1&=&\textrm{arg}[(\sum_{j=1}^3 m_j U_{ej}^2 )(\sum_{j=1}^3 U_{ej}T_{ej}^*(g_V'/g_V))^{*}],\nonumber\\
\psi_2&=&\textrm{arg}[(\sum_{j=1}^3 m_j U_{ej}^2 )(\sum_{j=1}^3 U_{ej}T_{ej}^*)^{*}].\nonumber\\
\eee
With help of (\ref{effective}) and by assuming (\ref{whynot}), $U_0 \simeq V_0$ and $(g_V'/g_V)\simeq 1$
we get  
\begin{eqnarray}
  \left\langle \lambda\right\rangle \approx (M_{W_1}/M_{W_2})^2 \frac{m_D}{m_{LNV}} \left|\xi\right|,
  ~~~~\left\langle \eta\right\rangle    \approx \epsilon \frac{m_D}{m_{LNV}} \left|\xi\right|,
\end{eqnarray}  
with 
\begin{eqnarray}
  |\xi | &=& |c_{23} c_{12}^2 c_{13} s_{13}^2 - c_{12}^3 c_{13}^3 - c_{13}c_{23} c_{12}^2 s_{13}^2
  -c_{12}c_{13}\lf c_{13}^2 s_{12}^2+s_{13}^2 \rh | \nonumber\\
  &\simeq& 0.82 
\end{eqnarray}
The experimental upper bound on the mixing angle  of left and right vector bosons is $\epsilon < 0.013$
and, provided that  the CP-violating phases in the mixing matrix for right-handed quarks are small, one gets
$\epsilon < 0.0025$. Flavor and CP-violating processes of kaons and B-mesons allow one to  to deduce
lower bound on the mass of the heavy vector boson $M_{W_2} > 2.9$ TeV \cite{Dev}. In left right symmetric models (LRSM) there could
be additional contributions to $0\nu\beta\beta$-decay due to the doubly charged Higgs triplet.  This triplet, also  considered in the case of type-I seesaw mechanism,  as pointed in Ref. \cite{Dev}, was found to make a  negligible contribution in this case.  

The coefficients $C_I$ (I=$mm$, $m\lambda$, $m\eta$, $\lambda\lambda$, $\eta\eta$ and $\lambda\eta$)
can be expressed as appropriate combinations of nuclear matrix elements and phase-space factors as follows:
\bee\label{Ccoef}
C_{mm} &=& (1-\chi_F+\chi_T)^2G^{0\nu}_{01}, \nonumber\\
C_{m\lambda} &=&-(1-\chi_F+\chi_T)\left[\chi_{2-}G^{0\nu}_{03}-\chi_{1+}G^{0\nu}_{04}\right], \nonumber\\
C_{m\eta} &=&(1-\chi_F+\chi_T)\nonumber\\
&\times&\left[\chi_{2+}G^{0\nu}_{03}-\chi_{1-}G^{0\nu}_{04}-\chi_{P}G^{0\nu}_{05}+\chi_{R}G^{0\nu}_{06}\right], \nonumber\\
C_{\lambda\lambda} &=&\chi_{2-}^2G^{0\nu}_{02}+\frac{1}{9}\chi_{1+}^2G^{0\nu}_{011}-\frac{2}{9}\chi_{1+}\chi_{2-}G^{0\nu}_{010}, \nonumber\\
C_{\eta\eta} &=&\chi_{2+}^2G^{0\nu}_{02}+\frac{1}{9}\chi_{1-}^2G^{0\nu}_{011}-\frac{2}{9}\chi_{1-}\chi_{2+}G^{0\nu}_{010}+\chi_{P}^2G^{0\nu}_{08}\nonumber\\
&-&\chi_{P}\chi_{R}G^{0\nu}_{07}+\chi_{R}^2G^{0\nu}_{09}, \nonumber\\
C_{\lambda\eta} &=&-2[\chi_{2-}\chi_{2+}G^{0\nu}_{02}-\frac{1}{9}\lf\chi_{1+}\chi_{2+}+\chi_{2-}\chi_{1-}\rh G^{0\nu}_{010}\nonumber\\
&+&\frac{1}{9}\chi_{1+}\chi_{1-}G^{0\nu}_{011}].  \\\nonumber
\eee
The explicit form of nuclear matrix elements $M_{GT}$ and the  ratios $\chi_I$ of the other matrix elements over this one
is given in\cite{DusanLR}. The integrated kinematical factors as recently improved  are given in Table
\ref{tab:phas}.

\begin{table*}[!t]
  \begin{center}
    \caption{Phase-space factors $G^{0\nu}_{0j}$ (j=1, $\cdots$, 11) in units $yr^{-1}$ 
obtained using screened exact finite-size Coulomb wave functions for $s_{1/2}$ and $p_{1/2}$
states of electron.\label{tab:phas}}
\centering 
\renewcommand{\arraystretch}{1.1}  
      \begin{tabular}{lrrrrrr}\hline\hline
  & $^{48}$Ca & $^{76}$Ge & $^{82}$Se & $^{96}$Zr & $^{100}$Mo & $^{110}$Pd \\ \hline
$10^{14}$$G^{0\nu}_{01}$   & $ 2.483$ & $0.237$ & $1.018$ & $ 2.062$ & $ 1.595$ & $0.483$  \\  
$10^{14}$$G^{0\nu}_{02}$   & $16.229$ & $0.391$ & $3.529$ & $ 8.959$ & $ 5.787$ & $0.814$  \\  
$10^{15}$$G^{0\nu}_{03}$   & $18.907$ & $1.305$ & $6.913$ & $14.777$ & $10.974$ & $2.672$  \\  
$10^{15}$$G^{0\nu}_{04}$   & $ 5.327$ & $0.470$ & $2.141$ & $ 4.429$ & $ 3.400$ & $0.978$  \\  
$10^{13}$$G^{0\nu}_{05}$   & $ 3.007$ & $0.566$ & $2.004$ & $ 4.120$ & $ 3.484$ & $1.400$  \\  
$10^{12}$$G^{0\nu}_{06}$   & $ 3.984$ & $0.531$ & $1.733$ & $ 3.043$ & $ 2.478$ & $0.934$  \\ 
$10^{10}$$G^{0\nu}_{07}$   & $ 2.682$ & $0.270$ & $1.163$ & $ 2.459$ & $ 1.927$ & $0.599$  \\  
$10^{11}$$G^{0\nu}_{08}$   & $ 1.109$ & $0.149$ & $0.708$ & $ 1.755$ & $ 1.420$ & $0.462$  \\  
$10^{10}$$G^{0\nu}_{09}$   & $16.246$ & $1.223$ & $4.779$ & $ 8.619$ & $ 6.540$ & $1.939$  \\ 
$10^{14}$$G^{0\nu}_{010}$  & $ 2.116$ & $0.141$ & $0.801$ & $ 1.855$ & $ 1.359$ & $0.309$ \\  
$10^{15}$$G^{0\nu}_{011}$  & $ 5.376$ & $0.476$ & $2.183$ & $ 4.557$ & $ 3.502$ & $1.010$ \\ \hline\hline 
 & $^{116}$Cd & $^{124}$Sn & $^{130}$Te & $^{136}$Xe & $^{150}$Nd  & \\ \hline
$10^{14}$$G^{0\nu}_{01}$      & $ 1.673$ & $0.906$ & $1.425$ & $1.462$ & $ 6.316$ & \\
$10^{14}$$G^{0\nu}_{02}$      & $ 5.349$ & $1.967$ & $3.761$ & $3.679$ & $29.187$ &  \\
$10^{15}$$G^{0\nu}_{03}$      & $11.128$ & $5.403$ & $8.967$ & $9.047$ & $45.130$ & \\
$10^{15}$$G^{0\nu}_{04}$      & $ 3.569$ & $1.886$ & $3.021$ & $3.099$ & $14.066$ & \\
$10^{13}$$G^{0\nu}_{05}$      & $ 4.060$ & $2.517$ & $3.790$ & $4.015$ & $14.873$ & \\
$10^{12}$$G^{0\nu}_{06}$      & $ 2.563$ & $1.543$ & $2.227$ & $2.275$ & $ 7.497$ & \\
$10^{10}$$G^{0\nu}_{07}$      & $ 2.062$ & $1.113$ & $1.755$ & $1.812$ & $ 8.085$ & \\
$10^{11}$$G^{0\nu}_{08}$      & $ 1.703$ & $0.939$ & $1.549$ & $1.657$ & $ 8.405$ & \\
$10^{10}$$G^{0\nu}_{09}$      & $ 6.243$ & $3.301$ & $4.972$ & $4.956$ & $19.454$ & \\
$10^{14}$$G^{0\nu}_{010}$     & $ 1.418$ & $0.660$ & $1.146$ & $1.165$ & $ 7.115$ & \\
$10^{15}$$G^{0\nu}_{011}$     & $ 3.704$ & $1.955$ & $3.148$ & $3.238$ & $15.055$ & \\
\hline\hline
      \end{tabular}
  \end{center}
\end{table*}     

\begin{figure}[!t]
   \vspace*{1.3cm}  
\centerline{\psfig{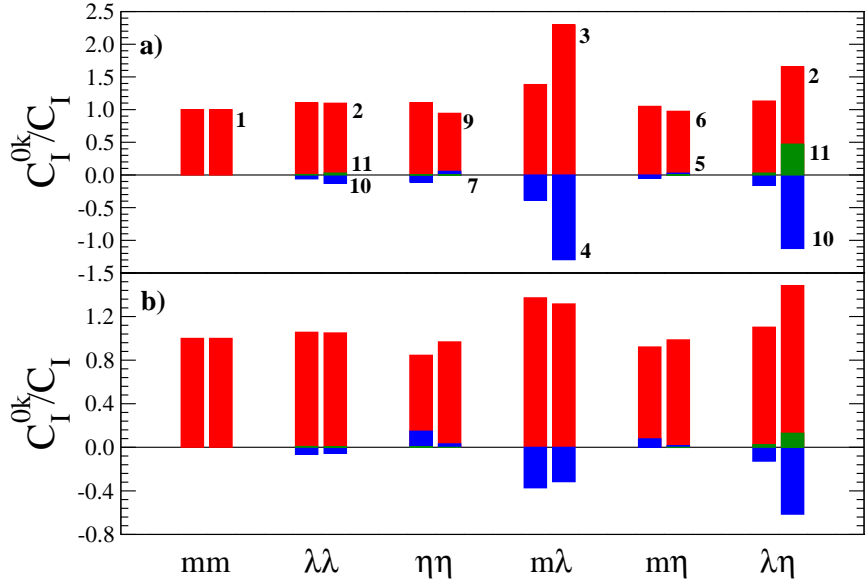}}
  \caption{(Color online) The decomposition of coefficients 
    $C_I$ (I=$mm$, $m\lambda$, $m\eta$, $\lambda\lambda$, $\eta\eta$ and $\lambda\eta$,
    see Eqs. (\ref{halflifem}) and (\ref{Ccoef}) ) on partial contributions $C^{0k}_I$ associated with
    phase space factors $G^{0\nu}_{0k}$ (k=1, $\cdots$,11). The symbols standing for index I are shown
    on the x-axis. The partial contributions are identified by index k, whose value is shown by
    the corresponding bar. The contributions from largest to the third largest are  displayed
    in red, blue and orange colors, respectively.
    Ratios $C^{0k}_I/C_I$ calculated with the ISM and QRPA matrix elements
    are presented with left and right bars for each value of index I, respectively. Results for $^{76}$Ge
    and $^{136}$Xe are presented in the lower  b) and upper a) panels, respectively.
\label{fig:cfac}  }
\end{figure}

In this case the  $0\nu\beta\beta$-decay rate (\ref{halflifem})  depends a a number of parameters, $\langle\lambda\rangle$, $\langle\eta\rangle$ $\psi_1$ and $\psi_2$, whose values are unknown. Their importance also depends  on the value of the  coefficients 
$C_I$ (I=$mm$, $m\lambda$, $m\eta$, $\lambda\lambda$, $\eta\eta$ and $\lambda\eta$,
which can be calculated.  The corresponding  quantity is a superposition of contributions $C^{0k}_I$ associated with
phase space factors $G^{0\nu}_{0k}$ (k=1, $\cdots$,11). In Fig. \ref{fig:cfac} ratios
$C^{0k}_I/C_I$ for the $0\nu\beta\beta$-decay $^{76}$Ge and $^{136}$Xe are displayed.
They were obtained by using the quasiparticle random phase approximation (QRPA) \cite{MBK88b}
and the interacting shell-model (ISM) \cite{CNPR96}nuclear matrix elements.
We should mention that the coefficients $C_{mm}$, $C_{\lambda\lambda}$, $C_{\eta\eta}$,
and $C_{m\eta}$ are dominated by a single contribution.
In the case of $C_{m\lambda}$ and  $C_{\lambda\eta}$, however,  there  may exist  a competition between two contributions.

Using these nuclear matrix elements  \cite{MBK88b,CNPR96} and the phase-space factors from Table \ref{tab:phas}
one can deduce from the experimental data $T^{0\nu-exp}_{1/2} > 1.07\times 10^{26}$ yr \cite{Xelimit}
for $^{136}$Xe decay \cite{Xelimit}
 in the case of  left-right symmetric theories constraints on the parameters  as follows\cite{DusanLR}:
\begin{eqnarray}
  \langle\eta\rangle   &\le&  1.2\times 10^{-9}~~(\rm QRPA),~~~1.7\times 10^{-9}~~(\rm ISM),\nonumber\\
  \langle\lambda\rangle &\le& 2.4\times 10^{-7}~~(\rm QRPA),~~~1.9\times 10^{-7}~~(\rm ISM). 
\end{eqnarray}  
Furthermore by assuming  the values $\epsilon$ = 0.013 and  0.0025 mentioned earlier as well as the current limit
-
$\langle\eta\rangle \le 1.7~ 10^{-9}$ ($^{136}$Xe, ISM)
we end up with $m_D/m_{LNV} < 1.6~ 10^{-7}$ and $<8.3~10^{-7}$, respectively.
For $M_{W_2} = 2.9$ TeV and $\langle\lambda\rangle \le 1.9~ 10^{-7}$
($^{136}$Xe, ISM) we get $m_D/m_{LNV} = 3.0~ 10^{-4}$.
Thus, from the more stringent limits on $\langle\eta\rangle$ we obtain
$m_{LNV}$/TeV $>$ (1.2 - 6.3) $m_D$/MeV, in agreement with the assumption that the basic
scale of LRSM is O(TeV). It is therefore obvious that already the present
limits of $0\nu\beta\beta$-decay half-lives can be used to constrain 
the allowed parameter space of LRSM. Furthermore the present  mechanism associated with
right-handed currents, somewhat forgotten in recent years,  can, in principle,  compete with the one based on $m_{\beta\beta}$ that
is commonly used.

Before concluding this section we should mention again that the contributions discussed in this section can involves different kinematics, since due o the chiralities involved the neutrino momentum is picked  in the propagator (see Eq. (\ref{Eq:propq})) rather than the neutrino mass. It can thus, in principle, be  distinguished from the left handed $\nu$ - mass contribution  by measuring energy and angular correlations of the two $\beta $ rays as discussed in more detail in earlier reviews \cite{eji05,ROP12}. To this end dedicated tracking detectors such as ELEGANT/MOON and NEMO-3/SuperNEMO  can be  used for the correlation studies (see section \ref{sect:exp}).

\subsection{Mechanism with heavy neutrinos within LR symmetric models}

In connection with LHC facility at CERN there is a frequently addressed
question whether the origin of neutrino masses is due to physics
at the TeV scale. A brief overview of a class of TeV scale theories for neutrino
masses based on the LR extension of the standard model can be find
e.g. in Ref. \cite{Mohapatra16}. 

By considering only those $0\nu\beta\beta$-decay mechanisms within the LR theories, which amplitudes are
proportional to light light neutrino masses $m_i$ or inverse proportional to heavy
neutrino masses $M_i$, for the half-life we get
\begin{equation}
  \left({T^{0\nu}_{1/2}} G^{0\nu} g_A^4\right)^{-1} =
  \left|\eta_\nu {M'}^{0\nu}_\nu  (g_A^{\rm eff}) + \eta_N^L {M'}^{0\nu}_N (g_A^{\rm eff})\right|^2~
  + \left|\eta_N^R {M'}^{0\nu}_N (g_A^{\rm eff})\right|^2~\label{halflife1}
\end{equation}
with 
\begin{eqnarray}\label{analysLNV}
  \eta_{\nu} &=& \frac{m_{\beta\beta}^2}{m_e^2} = \sum_{i} ({U}^{(11)}_{ei})^2 \frac{m_i}{m_e}
  \approx  \frac{m_p}{m_{LNV}} \frac{m^2_D}{m_e m_p}
  \sum_{i} {(U_0)}_{ei}^2 \frac{m_i~m_{LNV}}{m^2_D},\\
 \eta_{N}^L &=& \frac{m_p}{m_{LNV}} \sum_{i} (U^{(12)}_{ei})^2 \frac{m_{LNV}}{M_i}
 \approx \frac{m_p}{m_{LNV}} \left(\frac{m_D}{m_{LNV}}\right)^2 \sum_{i} \frac{m_{LNV}}{M_i},\nonumber\\
  \eta_{N}^R &=& \frac{m_p}{m_{LNV}} \left( \frac{M_{W_{1}}}{M_{W_{2}}} \right)^2
  \sum_{i} (U^{22}_{ei})^2 \frac{m_{LNV}}{M_i}
  \approx \frac{m_p}{m_{LNV}} \left(\frac{M_{W_1}}{M_{W_2}}\right)^2
  \sum_{i} {(V_0)}_{ei}^2 \frac{m_{LNV}}{M_i}.  \nonumber     
\end{eqnarray}

We note that the negligible interference term between left and right handed contributions
is not taken into account. From the qualitative analysis in Eq. (\ref{analysLNV} it follows
that if ${M_{W_{1}}}/{M_{W_{2}}} \gg {m_D}/{m_{LNV}}$ the heavy neutrino mass contribution due
to right-handed currents to the $0\nu\beta\beta$-decay
($\eta^R_N$ mechanism) dominates over the heavy neutrino mass contribution determined by only
left handed currents ($\eta^L_N$ mechanism) unless there is some strong suppression due to
mixing among heavy neutrinos. Further, the $\eta_\nu$ and $\eta^R_N$ mechanisms could be of
a comparable importance, if we could write
\begin{eqnarray}
  \sum_{i} {(U_0)}_{ei}^2 \frac{m_i~m_{LNV}}{m^2_D} \simeq \sum_{i} {(V_0)}_{ei}^2 \frac{m_{LNV}}{M_i}
  \nonumber\\
  \frac{m^2_D}{m_e m_p} {M'}^{0\nu}_\nu \simeq \left( \frac{M_{W_{1}}}{M_{W_{2}}} \right)^2 {M'}^{0\nu}_N.\end{eqnarray}  
In such a case a competion between these two mechanism can play an important role. 

There is a general consensus that a measurement of the $0\nu\beta\beta$-decay 
in one isotope does not allow us to determine the underlying physics mechanism. 
Complementary measurements in different isotopes is very important especially
for the case there are competing mechanisms of the $0\nu\beta\beta$-decay.
Different cases of competing $0\nu\beta\beta$-decay mechanisms were discussed in  Ref. \cite{ROP12} and more recently in Refs
\cite{SPVF,FMPSV11,Meroni13}.

If the $0\nu\beta\beta$-decay is induced by two ``non-interfering'' mechanisms
$\eta_\nu$ and $\eta_N^R$ one can determine the absolute values of these
two fundamental parameters from data on the half-lives
of two nuclear isotopes \cite{FMPSV11,Meroni13}. Given a pair of nuclei, 
$|\eta_\nu|^2$ and $|\eta^R_N|^2$ are solutions of a
system of two linear equations
\begin{eqnarray}
\left(T^{0\nu}_{1/2}(i) G^{0\nu}(i) g_A^4 \right)^{-1} ~= ~
|\eta_{\nu} ~{M'}^{0\nu}_{\nu}(i)|^{2} + |\eta^R_{N} ~{M'}^{0\nu}_{N}(i)|^{2},
\end{eqnarray}
where the index i denotes the isotope. From the ``positivity'' conditions 
($|\eta_\nu|^2 > 0$ and $|\eta^R_N|^2 > 0$)  it follows that 
the ratio of  half-lives is within the range \cite{FMPSV11,Meroni13}
\begin{eqnarray}\label{nonintm}
\frac{ G^{0\nu}(i) |{M}^{0\nu}_{N}(i)|^{2}} 
{ G^{0\nu}(j) |{M}^{0\nu}_{N}(j)|^{2}} 
  \le \frac{T^{0\nu}_{1/2}(j)}{T^{0\nu}_{1/2}(i)} \le 
\frac{ G^{0\nu}(i) |{M}^{0\nu}_{\nu}(i)|^{2}} 
{ G^{0\nu}(j) |{M}^{0\nu}_{\nu}(j)|^{2}}. 
\end{eqnarray}

Surprisingly, the physical solutions are possible only 
if the ratio of the half-lives takes values in narrow 
intervals \cite{FMPSV11}. This also is true in the case of other competing mechanisms  \cite{ROP12}. Often the non physical solutions can be eliminated by already existing data \cite{ROP12}.

\begin{figure}[!t]
   \vspace*{1.3cm}  
\centerline{\psfig{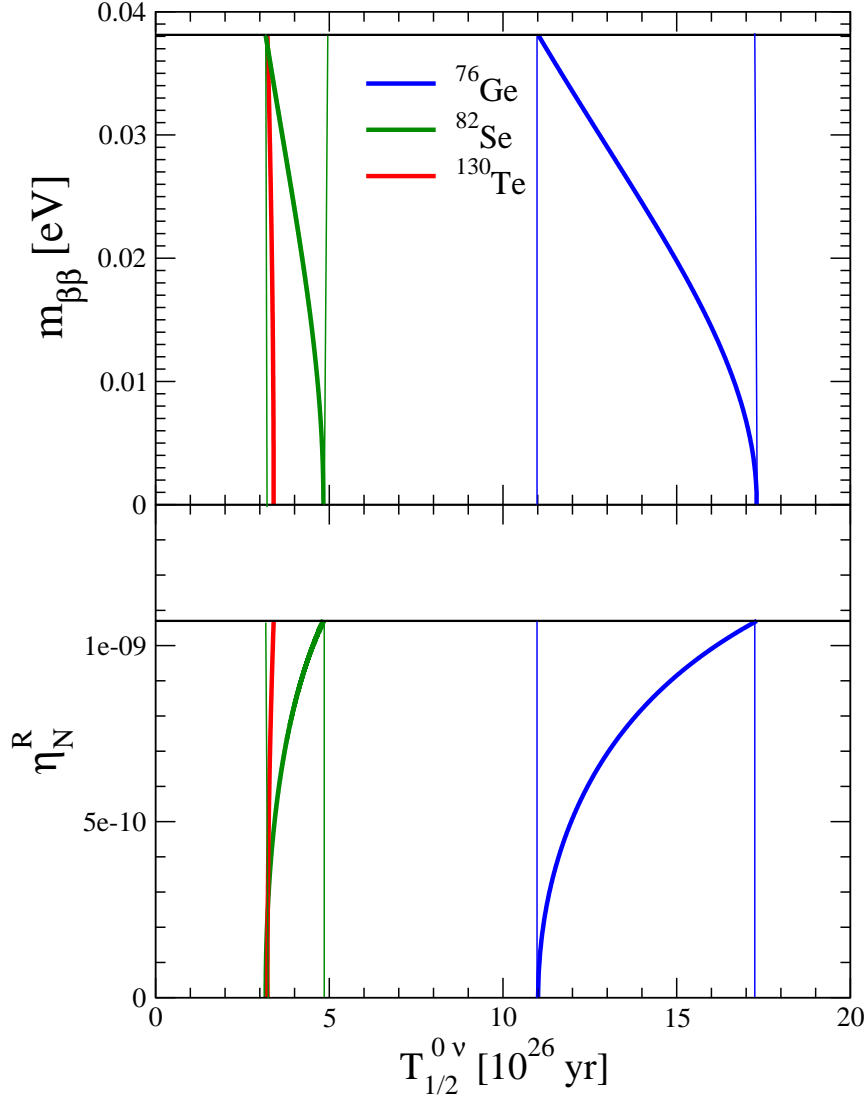}}
\caption{(Color online)
  The parameters $m_{\beta\beta}$ and $\eta^R_N$ versus the $0\nu\beta\beta$-decay
  half-life of $^{76}$Ge (blue), $^{82}$Se(green) and $^{130}$Te(red)
  for $T^{0\nu-exp}_{1/2}(^{136}Xe) = 1.0~ 10^{27}$  yr. The horizontal
  lines correspond to values of $m_{\beta\beta}$ and $\eta^R_N$, if there
  is a dominance of single mechanism. The QRPA NMEs (TBC, Argonne src) and
  the non-quenched value of $g_A^{\rm eff}$ are considered (see Tables 
  \ref{tab.nmeL} and \ref{tab.nmeH}).
  \label{etambb}
}
\end{figure}

In Fig. \ref{etambb} the allowed ranges of $m_{\beta\beta}$ and $\eta^R_N$
parameters obtained as solution of Eq. (\ref{nonintm}) are presented
as function of half-life of $^{76}$Ge, $^{76}$Se and $^{76}$Te by assuming
$T^{0\nu-exp}_{1/2}(^{136}Xe) = 1.0~ 10^{27}$  y, what is compatible
with inverted hierarchy of neutrino masses. In the case of
a dominance of a single mechanism we have $m_{\beta\beta}$ = 38 meV and
$\eta^R_N = 1.1~ 10^{-9}$. We notice that
linear set of equations in Eq. (\ref{nonintm}) has solution for a
very narrow interval for $T^{0\nu-exp}_{1/2}(^{130}Te)$ and that by assuming
co-existance of both mechanisms the values of parameters
$m_{\beta\beta}$ and $\eta^R_N$ are smaller.

%
%
%
\section[The neutrinoless double-beta decay nuclear matrix elements]{The neutrinoless double-beta decay nuclear matrix elements }
\label{secnme}
%

The nuclear matrix elements for the $0\nu\beta\beta$ decay must be 
evaluated using nuclear structure methods. There are no 
observables that could be directly related to the magnitude of 
$0\nu\beta\beta$ nuclear matrix elements and that could be used to determine 
them in a model independent way.

The nuclear matrix elements ${M'}^{0\nu}_\nu$ and ${M'}^{0\nu}_N$ associated
with exchange of light (sub eV scale) and heavy (TeV scale) neutrinos, respectively,
can be written as 
\begin{equation}
{M'}^{0\nu}_{\nu, N} =  \left(\frac{g^{\rm eff}_A}{g_A}\right)^2 {M}^{0\nu}_{\nu, N} (g^{\rm eff}_A),
\label{primeNME}
\end{equation}
where ${M}^{0\nu}_{\nu, N}$ consists of Fermi, Gamow-Teller and tensor parts as
\begin{equation}
  {M}^{0\nu}_{\nu, N}(g^{\rm eff}_A)  =  - \frac{M^{(\nu, N)}_{F}}{(g^{\rm eff}_A)^2} +
  M^{(\nu, N)}_{GT}(g^{\rm eff}_A) + M^{(\nu, N)}_T(g^{\rm eff}_A).
\end{equation}
This definition of ${M'}^{0\nu}_{\nu, N}$ allows to display the effects of uncertainties
in $g^{\rm eff}_A$. 

There is a common practice to calculate $0\nu\beta\beta$-decay NMEs by taking advantage
of relative coordinates of two decaying nucleons. For $0^+ \rightarrow 0^+$ transition
we have 
\begin{eqnarray}
M^{(\nu, N)}_K  =  \sum_{J^{\pi}_m,\mathcal{J}} \sum_{pnp'n'}
(-1)^{j_n + j_{p'} + J + {\mathcal J}}
\sqrt{ 2 {\mathcal J} + 1}
\left\{
\begin{array}{c c c}
j_p & j_n & J  \\
 j_{n'} & j_{p'} & {\mathcal J}
\end{array}
\right\}  \times~~~~~~~~~~~~~~~~\\
\langle p(1), p'(2); {\mathcal J} \parallel 
{\cal O}^{(\nu, N)}_K 
\parallel n(1), n'(2); {\mathcal J} \rangle 
~\langle 0_f^+ \parallel
[ \widetilde{c_{p'}^+ \tilde{c}_{n'}}]_J \parallel J^{\pi}_m \rangle
 \langle  J^{\pi}_m\parallel [c_p^+ \tilde{c}_n]_J \parallel 0_i^+ \rangle ~.
\nonumber
\label{eq:long}
\end{eqnarray}
The reduced matrix elements of the one-body proton-neutron operators
$c_p^+ \tilde{c}_n$ ($\tilde{c}_n$ denotes the time-reversed state)
in the Eq. (\ref{eq:long})

The two-body operators $O^{(\nu, N)}_K,~ K$ = Fermi (F), Gamow-Teller (GT), and Tensor
(T) in (\ref{eq:long}) contain neutrino potentials and spin and isospin 
operators, and energies of excited states $E^{m}_{J^\pi}$: 
\begin{eqnarray}
O^{(\nu, N)}_F(r_{12},E^k_{J^\pi}) &=& \tau^+(1)\tau^+(2)~ H^{(\nu, N)}_F(r_{12},E^m_{J^\pi})~,
\nonumber\\
O^{(\nu, N)}_{GT}(r_{12},E^k_{J^\pi}) &=& \tau^+(1)\tau^+(2)~ H^{(\nu, N)}_{GT}(r_{12},E^m_{J^\pi})
~\sigma_{12}~,
\nonumber\\
O^{(\nu, N)}_{T}(r_{12},E^k_{J^\pi}) &=& \tau^+(1)\tau^+(2)~ H^{(\nu, N)}_{T}(r_{12},E^m_{J^\pi})
~S_{12}.
\label{a12}
\end{eqnarray}
with
\begin{eqnarray}
{\bf r}_{12} &= &{\bf r}_1-{\bf r}_2, ~~~ r_{12} = |{\bf r}_{12}|,
~~~
\hat{{\bf r}}_{12} = \frac{{\bf r}_{12}}{r_{12}},~\nonumber \\
\sigma_{12} &=& {\vec{ \sigma}}_1\cdot {\vec{ \sigma}}_2,~~~
S_{12} = 3({\vec{ \sigma}}_1\cdot \hat{{\bf r}}_{12})
       ({\vec{\sigma}}_2 \cdot \hat{{\bf r}}_{12})
      - \sigma_{12}.
			\label{Eq:tensor}
\end{eqnarray}
Here, ${\bf r}_1$ and ${\bf r}_2$ are the coordinates of the
nucleons undergoing beta decay.

The neutrino potentials are integrals over the exchanged momentum $q$,
\begin{eqnarray}
H^{(\nu, N)}_K (r_{12},E^k_{J^\pi}) =
\frac{2}{\pi} {R} \int_0^{\infty}~ f_K(qr_{12})~F^{(\nu, N)}(q^2)~h_K (q^2) q^2 dq 
\label{eq:pot}
\end{eqnarray}
The functions $f_{F,GT}(qr_{12}) = j_0(qr_{12})$ and $f_{T}(qr_{12})= - j_2(qr_{12})$
are spherical Bessel functions. The functions $h_K(q^2)$ that enter the $H_K$'s 
through the integrals over $q$ in Eq.\ (\ref{eq:pot}) are 
\begin{eqnarray}
h_{F} ({ q}^{2})  & = & g^2_V({ q}^{2}) \\
h_{GT} ({ q}^{2}) & = & \frac{g^2_A({ q}^{2})}{g^2_A} 
[ 1 - \frac{2}{3} \frac{ { q}^{2}}{ { q}^{2} + m^2_\pi } + 
\frac{1}{3} ( \frac{ { q}^{2}}{ { q}^{2} + m^2_\pi } )^2 ]
+ \frac{2}{3} \frac{g^2_M({ q}^{2} )}{(g^{\rm eff}_A)^2} \frac{{ q}^{2} }{4 m^2_p }, 
\nonumber \\
h_T ({ q}^{2}) & = & \frac{g^2_A({ q}^{2})}{g^2_A} [ 
\frac{2}{3} \frac{ { q}^{2}}{ { q}^{2} + m^2_\pi } -
\frac{1}{3} ( \frac{ { q}^{2}}{ { q}^{2} + m^2_\pi } )^2 ]
+ \frac{1}{3} \frac{g^2_M ({ q}^{2} )}{(g^{\rm eff}_A)^2} \frac{{ q}^{2} }{4 m^2_p }  \nonumber  
\end{eqnarray}
Here, the Partially Conserved Axial Current (PCAC) hypothesis has been employed.
The finite nucleon size  is taken into account via momentum dependence 
of the nucleon form-factors. 
For the vector, weak-magnetism and axial-vector form factors we adopt 
the usual dipole approximation as follows: 
\begin{eqnarray}
g_V({ q}^{2}) = \frac{g_V}{(1+{ q}^{2}/{M^2_V})^2},~~
g_M({ q}^{2}) = (\mu_p-\mu_n) g_V({ q}^{2}),~~
g_A({ q}^{2}) = \frac{g^{\rm eff}_A}{(1+{ q}^{2}/{M^2_A})^2},\nonumber\\
\end{eqnarray}
where $g_V = 1$, $(\mu_p - \mu_n) = 3.70$. The parameters $M_V = 850 ~MeV$ and 
$M_A = 1~086 ~MeV$ come from electron scattering and 
neutrino charged-current scattering experiments.

The difference in the calculation of ${M}^{0\nu}_{\nu} (g^{\rm eff}_A)$ and
${M}^{0\nu}_{N} (g^{\rm eff}_A)$ has origin in factors $F^{(\nu, N)}(q^2)$.
We have 
\begin{eqnarray}
  F^{(\nu)}(q^2)
  = \frac{1} {q~\left(q + E^m_{J^\pi} - (E_i + E_f)/2\right)} ~,~~~~
  F^{(\nu, N)}(q^2) = \frac{1} {m_e~m_p}.
\end{eqnarray}
From the form of $F^{(\nu)}(q^2)$ it follows that corresponding potentials 
(\ref{eq:pot}) depend weakly on the energies of the virtual
intermediate states, $E^m_{J^\pi}$ as the mean neutrino momentum is large
about 100-200 MeV. As   $F^{(\nu, N)}(q^2)$ does not depend on the energy of intermediate states,
the summation over these states by using completeness relation
\begin{equation}
1 = \sum_{J^\pi_m} |J^\pi_m\rangle\langle J^\pi_m\rangle. 
\end{equation}  
Then the calculation of ${M}^{0\nu}_{N} (g^{\rm eff}_A)$ requires only the knowledge
of initial and final ground state wave functions. The same is valid also for  
calculation of ${M}^{0\nu}_{\nu} (g^{\rm eff}_A)$ once the closure approximation
for intermediate nuclear states is considered by replacing energies of intermediate
states $[E^m_{J^\pi} - (E_i + E_f)/2]$ by an average value ${\overline E} \approx 10~MeV$.\\
However, we note that a construction of reliable ground state wave functions requires
diagonalization procedure by which all states of the intermediate nucleus are calculated. 
The problem with the averaged energy ${\overline E}$ is that its value is unknown
and there is no good way to calculate it. 

The nuclear structure approaches do not allow one to introduce the short-range 
correlations into the two-nucleon relative wave function. The 
traditional way to take them into account  is to introduce an explicit Jastrow-type correlation 
function $f(r_{12})$ into the involved two-body transition 
matrix elements
\begin{eqnarray}
 \langle {{\Psi}}_{\mathcal J} 
\parallel f(r_{12}) 
{\cal O}_K (r_{12}) f(r_{12}) 
\parallel {{\Psi}}_{\mathcal J}  \rangle.
\label{correl}
\end{eqnarray}
Here, 
\begin{eqnarray}
|{\overline{\Psi}}_{\mathcal J}  \rangle
=  f(r_{12}) ~| {{\Psi}}_{\mathcal J} \rangle,~~~~   
| {{\Psi}}_{\mathcal J}  \rangle \equiv
| n(1), n'(2); {\mathcal J} \rangle 
\label{cwf}
\end{eqnarray}
are the relative wave function with and without the short-range
correlations, respectively. Currently, for purpose of numerical calculation of
the $0\nu\beta\beta$-decay NMEs the coupled cluster method short-range correlation functions 
in an analytic form of Jastrow-like function are used,
\begin{equation}
f(r_{12}) = 1 ~-~ c~ e^{-a r^2}(1-b r^2). 
\end{equation}
The set of parameters for Argonne  and CD-Bonn NN interactions 
is given by
\begin{eqnarray}
\mbox{set A}:~~a &=& 1.59~fm^{-2},~~b = 1.45~fm^{-2},~~c = 0.92,\nonumber\\
\mbox{set B}:~~a &=& 1.52~fm^{-2},~~b = 1.88~fm^{-2},~~c = 0.46.\nonumber\\
\end{eqnarray}
Another option is to adopt the the unitary correlation operator method (UCOM)
approach for description  of the two-body correlated wave function \cite{roth04}.

\subsection{Nuclear matrix elements for non-quenched value of axial vector coupling constant}

There is a variety of nuclear structure methods used for the calculation of the
$0\nu\beta\beta$-decay NMEs. The differences among them are due to
construction of the mean field,  the ways residual interaction is fixed, 
the size of the considered single nucleon model space, the 
many-body approximations are used in the diagonalization of the nuclear Hamiltonian
and the type considered short-range correlations.

In the last few years there is a lot of activity in the field of calculation 
of the $0\nu\beta\beta$-decay NMEs and significant progress has been
achieved. Five different many-body approximate methods have been applied
for the calculation as follows:\\
\begin{enumerate}
\item The Interacting Shell Model (ISM) (Strasbourg-Madrid (StMa)\cite{StMa09}
and Central Michigan University (CMU)\cite{CMU16} groups).\\
The disadvantage of the ISM is that only a limited number
of orbits close to the Fermi level is considered The advantage is that
all possible correlations within the space are included. As a result 
a good spectroscopy of low-lying excited states for parent and daughter nuclei
is achieved. This approach has been applied  only for six double beta decay
systems with A=48, 76, 82, 124, 130 and 136. The CMU group managed
to perform calculation without consideration of the closure approximation 
\cite{CMU16,HoroiSe14,HoroiTe15,HoroiGe16}. In addition, useful
multipole decomposition of the $0\nu\beta\beta$-decay nuclear matrix element
was presented. \\
For a given choice of harmonic single particle states it can allow  a large number of nucleons to be put in these orbitals leading to large configuration mixing. Unfortunately, however, even with the most advanced computers, in order to get  a manageable size of the resulting shell model matrices,  the number of orbitals is restricted, usually $\le 4$.

 This limitation can be somewhat overcome by considering doorway states in perturbation theory. Suppose that the double beta decay operator for light neutrinos  in the case of the closure approximation is given by
\beq
T_{0}=\sum_{i< j}\tau^+_{i}\tau^+_{j}\left( f^2_V-g^2_A\bfs_i .\bfs_j \right ),\,T_{1}=\sum_{i< j}\tau^+_{i}\tau^+_{j}\left( f^2_V-g^2_A\bfs_i .\bfs_j \right )\frac{1}{r_{ij}}
\eeq
Suppose now that the ground state of the initial nucleus $(N,Z)$ obtained in the ISM is $|i\rangle>$ and that of the final $(N-2,Z+2)$ nucleus is $|f\rangle$. Consider now the door way states $d_i$ and $d_f$ built on the initial and final states respectively, namely 
\beq
d_i=T_0|i\rangle>,\,d_f=T^\dagger_0|f\rangle>.
\eeq
Suppose further  that the states $d_i$ and $d_f$ are normalized and orthogonalized to the final and and initial ISM states respectively, i.e. :
\beq
\langle d_i|f\rangle=0,\,\langle d_f|i\rangle=0,\,\langle d_i|d_i\rangle=1,\,\langle d_f|d_f\rangle=1.
\eeq
Then using the employed  nucleon-nucleon interaction one computes the mixing to the final and initial states as follows:
\beq
C_f=\frac{\langle d_i|V|f\rangle}{E_{d_i}-E_f},\,C_i=\frac{\langle d_f|V|i\rangle}{E_{d_f}-E_i}
\eeq
where $E_{d_i}$,  $E_{d_f}$ are the energies of the doorway states and $E_i$ and $E_f$ are the energies of the initial and final states obtained in the ISM.

The the neutrinoless double beta decay  nuclear matrix element can be given as:
\beq
M^{0\nu}_{\nu}=\langle f|T|i\rangle+C_f\langle d_i|T|i\rangle+C_i\langle f|T|d_f\rangle
\eeq
where $T$ is the transition operator, as described in section \ref{decay mechanisms}, with the first ME as obtained in the lSM. In this approach one will manage to include any missing  the spin-orbit partner in the ISM, which
guarantees that the Ikeda sum rule is fulfilled. \\If one wants to be more ambitious one may use the operator$T_1$ instead of $T_0$.

It is clear that this treatment does not avoid the number of operations involved in the full Hilbert space. The only great advantage is that  it does not lead to an  increase of the already large matrix dimensions involved in the ISM. 

\item   Quasiparticle Random Phase Approximation (QRPA) (Tuebingen-Bratislava-Caltech 
(TBC)\cite{TBC13,Fang15}, Jyv\"askyla (Jy)\cite{QJy15}, 
and Noth-Caroline University (NC)\cite{QSky13} groups).\\
The formalism of the spherical\cite{TBC13}  and deformed\cite{Fang15} QRPA has been 
improved by the TBC group by achieving partial restoration of the isospin
symmetry by the requirement that $2\nu\beta\beta$-decay Fermi matrix element
vanishes, as it should, unlike in previous version of the method. It was
achieved by separating the renormalization parameter $g_{pp}$ of the particle-particle
proton-neutron interaction into isovector and isoscalar parts. The isovector
parameter $g^{T=1}_{pp}$ was chosen to be essentially equal to the constant
$g_{pair}$ used to renormalize the nucleon pairing interactions. So,
no new parameter was needed. As a result in  the case of $0\nu\beta\beta$-decay the Fermi
matrix element is substantially reduced, while the full matrix element
${M}^{0\nu}_{\nu}$  is reduced by $\approx$ 10\%.  The Jy group improved their
calculation \cite{QJy15} by following the calculation procedure of the TBC group apart from 
the isospin restoration.  Unlike in their  previous calculations\cite{kor07,korte1,korte3}
the tensor contribution to ${M}^{0\nu}_{\nu}$ has been found to be non-negligible in agreement
with the TBC results \cite{TBC13}.
In both, TBC and Jy calculations the pairing and residual interactions of the nuclear 
Hamiltonian and the two-nucleon short-range correlations (SRC) are
derived from the same realistic nucleon-nucleon interaction
(CD-Bonn and Argonne potentials) by exploiting the Brueckner-Hartree-Fock
and coupled cluster methods. Contrary, calculation of the NC group are performed
within the deformed QRPa approach by exploiting the Skyrme interaction. 
We note that there is a new QRPA approached developed by Terasaki in which 
the $0\nu\beta\beta$-decay NME is calculated with particle-particle 
QRPA by two particle transfer to the intermediate nucleus (A+2,Z+2) nucleus
under the closure approximation, instead the true double-beta path through
(A,Z+1) nucleus \cite{jun16}. The obtained result for $^{150}$Nd are in a rather good
agreement with that of the TBC group within deformed QRPA approach\cite{Fang15}.
\item  Interacting Boson Model (IBM)\cite{IBM15}.\\
  In the IBM the low lying states of the nucleus are modeled in terms of
  either L=0 (s boson) or L=2 (d boson) states. Recently, the IBM-2 approach has
  been also improved by considering isospin restoration in the calculation
  of the $0\nu\beta\beta$-decay NMEs \cite{IBM15}.  It affected slightly the
  Fermi contribution to  ${M}^{0\nu}_{\nu}$, which become smaller. 
\item The Projected Hartree-Fock-Bogoliubov Method (PHFB) \cite{phfb13}
  and non-relativistic\cite{NREDF10} and relativistic \cite{REDF15}
  the Energy Density Functional Method (EDF) methods.\\
  Recently, within the PHFB approach a systematic study of the uncertainties in calculated
  $0\nu\beta\beta$-decay matrix elements has been performed by considering different
  sets of schematic residual interaction and Jastrow-like short-range correlations\cite{phfb13}.
  The advantage  of the PHFB is that  the PHFB wave functions posses
  good particle number and angular momentum obtained by projection on the axially
  symmetric intrinsic HFB states. The EDF is considered to be an improvement
  with respect to the PHFB. Compared with the PHFB, the EDF includes
  additional correlations connected with particle number projection,
  as well as fluctuations in quadrupole shapes and pairing gaps. Recently,
  a systematic study of the $0\nu\beta\beta$-decay matrix elements was
  presented in the framework of the relativistic EDF (REDF) method \cite{REDF15}.
  For sake of simplicity often calculations are performed without taking into account
  two-body short-range correlations.
\item The ab initio methods \cite{schwenk16b,schwenk16a,navratil14,navratil16}. 
  The issue of ab initio methods is the description of nuclei starting from the
  constituent nucleons and the realistic interactions among them. The nuclear
  forces include two-,  three- and possibly higher  many-nucleon  components.
  This approach has so far  been applied successfully for nuclear structure properties
  of light nuclei. Thus up to now there have been no results for double beta transitions
  apart for some toy model calculations. It might be that the ab initio $0\nu\beta\beta$-decay
  results for $^{48}$Ca will appear soon.
\end{enumerate}
It goes without saying that each of the applied methods has some advantages and drawbacks
and that there is  space for further improvement of each of them. 

\begin{table}[!t]
\caption{The NME of the $0\nu\beta\beta$-decay $M^{0\nu}_{\nu}$ calculated in the framework of
different approaches:
interacting shell model (ISM) (Strasbourg-Madrid (StMa)\protect\cite{StMa09}
and Central Michigan University (CMU)\protect\cite{StMa09,CMU16} groups),
interacting boson model (IBM)\protect\cite{IBM15},
quasiparticle random phase approximation (QRPA) (Tuebingen-Bratislava-Caltech 
(TBC)\protect\cite{TBC13,Fang15}, Jyv\"askyla (Jy)\protect\cite{QJy15}, 
and Noth-Caroline University \protect\cite{QSky13} groups), 
projected Hartree-Fock Bogoliubov approach (PHFB)\protect\cite{phfb13},
non-relativistic\protect\cite{NREDF10}
energy density functional method\protect\cite{REDF15}. 
The Argonne, CD-Bonn and UCOM two-nucleon short-range correlations  are taken into account.
Averaged nuclear matrix element of different approaches and its variances for a given isotope
are calculated following Refs.\protect\cite{qrpa1,qrpa2}.
The non-quenched value of weak axial-vector coupling $g_A$ and R = 1.2 fm $A^{1/3}$ are assumed.
\label{tab.nmeL}}
\begin{tabular}{lcccccccc}\hline\hline
  Method & $g_A$ & src &  \multicolumn{6}{c}{ $M^{0\nu}_{\nu}$} \\ \cline{4-9}
         &       &          & ${^{48}}$Ca  & ${^{76}}$Ge  & ${^{82}}$Se  & ${^{96}}$Zr & ${^{100}}$Mo & ${^{110}}$Pd
  \\ \hline
ISM-StMa  & 1.25  &  UCOM   &   0.85   &   2.81      &     2.64     &             &            &           \\
ISM-CMU   & 1.27  & Argonne &   0.80   &   3.37      &     3.19     &             &            &           \\
          &       & CD-Bonn &   0.88   &   3.57      &     3.39     &             &            &           \\
IBM       & 1.27 & Argonne &   1.75   &   4.68      &     3.73     &    2.83     &   4.22     &   4.05    \\
QRPA-TBC  & 1.27 & Argonne &   0.54   &   5.16      &     4.64     &    2.72     &   5.40     &   5.76    \\
          &       & CD-Bonn &   0.59   &   5.57      &     5.02     &    2.96     &   5.85     &   6.26    \\
QRPA-Jy   & 1.26  & CD-Bonn &          &   5.26      &     3.73     &    3.14     &   3.90     &   6.52    \\
dQRPA-NC  & 1.25  & without &          &   5.09      &              &             &            &           \\
PHFB      & 1.25 & Argonne &          &             &              &    2.84     &   5.82     &   7.12    \\
          &       & CD-Bonn &          &             &              &    2.98     &   6.07     &   7.42    \\
  NREDF   & 1.25  &  UCOM   &   2.37   &   4.60      &     4.22     &    5.65     &     5.08   &           \\
  REDF    & 1.25 & without &   2.94   &   6.13      &     5.40     &    6.47     &     6.58   &         \\ \hline
 Mean value &      &         &   1.34   &   4.55      &     4.02     &    3.78     &    5.57    &   6.12  \\
 variance   &      &         &   0.81   &   1.20      &     0.91     &    2.49     &    0.58    &   1.78   \\ \hline\hline 
  Method & $g_A$ & src        & \multicolumn{6}{c}{ $M^{0\nu}_{\nu}$} \\ \cline{4-9}
          &       &         & ${^{116}}$Cd & ${^{124}}$Sn & ${^{128}}$Te & ${^{130}}$Te & ${^{136}}$Xe & ${^{150}}$Nd
  \\ \hline \hline
ISM-StMa  & 1.25  &  UCOM   &            &    2.62     &            &    2.65     &     2.19   &            \\
ISM-CMU   & 1.27  & Argonne &            &    2.00     &            &    1.79     &     1.63   &           \\
          &       & CD-Bonn &            &    2.15     &            &    1.93     &     1.76   &           \\
IBM       & 1.27 & Argonne &   3.10     &    3.19     &     4.10   &    3.70     &     3.05   &     2.67  \\
QRPA-TBC  & 1.27 & Argonne &   4.04     &    2.56     &     4.56   &    3.89     &     2.18   &           \\
          &       & CD-Bonn &   4.34     &    2.91     &     5.08   &    4.37     &     2.46   &     3.37  \\
QRPA-Jy   & 1.26  & CD-Bonn &   4.26     &    5.30     &     4.92   &    4.00     &     2.91   &           \\
dQRPA-NC  & 1.25  & without &            &             &            &    1.37     &     1.55   &     2.71  \\
PHFB      & 1.27 & Argonne &            &             &    3.90    &    3.81     &            &     2.58  \\
          &       & CD-Bonn &            &             &    4.08    &    3.98     &            &     2.68  \\
  NREDF   & 1.25  &  UCOM   &   4.72     &    4.81     &     4.11   &    5.13     &     4.20   &     1.71  \\
  REDF    & 1.25 & without &   5.52     &    4.33     &            &    4.98     &     4.32   &     5.60 \\ \hline
 Mean value &      &         &   4.34     &    3.07     &    4.34    &    3.42     &    2.59    &   3.01         \\
 variance   &      &         &   0.79     &    1.01     &    0.23    &    1.67     &    1.10    &   1.34
\\ \hline \hline
\end{tabular}
\end{table}

In Table \ref{tab.nmeL}, recent results of the different methods for ${M}^{0\nu}_\nu$ 
are summarized. 
The presented numbers have been obtained with the non-quenched value of the axial
coupling constant ($g_A^{\rm eff}=g_A$)\footnote{
A modern value of the axial-vector coupling  constant is $g_A =1.269$. We note 
that in the referred calculations of the $0\nu\beta\beta$-decay NMEs
the previously accepted value $g_A^{\rm eff}=g_A=1.25$ was assumed.}, 
Thus, the discrepancies among the results of different 
approaches are mostly related to the approximations on which a given 
nuclear many-body method is based. From Table  \ref{tab.nmeL} it follows that there
exists significant difference between the results of the ISM and other approaches.
The UCOM  and Brueckner (CD-Bonn and Argonne) short-range correlations were considered. 
The impact of the choice of the short-range correlations on the NME
is not large, about 10\%. The spread of values for a given isotope is mostly
given by factor 2 and for $^{48}$Ca about factor 5. The spread for a given
isotope is affected by the presence of results calculated with the ISM and
the REDF approaches, which usually offer the smallest and largest value
of calculated NME. In order to estimate the current uncertainty in NMEs
for a given isotope we take advantage of calculation of mean
values and variances following Refs.\cite{qrpa1,qrpa2}. We see that the smallest
uncertainty is reported by $^{128}$Te, $^{100}$Mo (not calculated within the ISM) and $^{82}$Se
and the largest by $^{48}$Ca. 

In Table \ref{tab.nmeH}, recent results of different methods for ${M}^{0\nu}_N$ 
are presented.  They have been achieved within the ISM, IBM, QRPA and PHFB
methods but not in the EDF approach. This fact is probably a reason of small
spread of results for $^{48}$Ca. The results with the CD-Bonn short-range correlations
are significantly larger in comparison with those using Argonne short-range
correlations. The largest values are reported by the QRPA. The explanation
might be that there is larger momentum transfer between nucleons 
by the heavy neutrino in comparison with light neutrino exchange. Therefore,
the role of transitions through higher multipoles is expected to be
important. But, these states are not described within the ISM and the IBM and as result
the value of ${M}^{0\nu}_N$ is reduced.

The range of results produced within different nuclear models for a given isotope 
(see Tables \ref{tab.nmeL} and \ref{tab.nmeH}) means that some of them,
or generally all of them, are deficient in the way they incorporate some important
physics. It is not clear which nuclear physics observables and/or nuclear models
need be reproduced to give a reliable prediction for the $0\nu\beta\beta$-decay
NME, apart from the $2\nu\beta\beta$-decay NMEs  deduced from the measured
half-lives. But, none of the discussed models can do it reliably
now. Complementary experimental information from processes  like charge-exchange
and particle transfer reactions, muon capture and charged current (anti)neutrino-nucleus
reactions is assumed to be also relevant (see section \ref{sect:expME}). The occupancies of valence neutron and proton orbits have been
extracted by accurate measurements of one nucleon
adding and removing transfer reactions by J. Schiffer and collaborators\cite{Freeman12}
for double beta decay systems with A= 76 (protons and neutrons\cite{SchifferGea,SchifferGeb}),
100 (neutrons\cite{SchifferMo}), 130  (neutrons\cite{SchifferTe} and protons\cite{SchifferTeXe})
and 136 (proton\cite{SchifferTeXe}). The Gamow-Teller strengths to intermediate
nucleus have been investigated via charge-exchange reactions for
A=48 ($\beta^-$\cite{Yako09} and $\beta^+$\cite{Yako09}),
76 ($\beta^-$\cite{thi12}), 96 ($\beta^-$\cite{thi12b}), 
100 ($\beta^-$\cite{thi12a}), 128 ($\beta^-$\cite{pup12}),
130 ($\beta^-$\cite{pup12}), 130 ($\beta^-$\cite{pup11}) and 
150 ($\beta^-$\cite{gue11} and $\beta^+$\cite{gue11}). Unfortunately, there is
no available or useful information about the $\beta^+$ strengths connecting
intermediate and final nuclei for many double beta decaying systems. These transitions  might be probed not only by charge-exchange
reactions, but  with muon capture as well. 

There are two important questions in the context of the calculation of the $0\nu\beta\beta$-decay
NME: i) What is behind the smallness of both $2\nu\beta\beta$- and
$0\nu\beta\beta$-decay matrix elements? ii) Is the sensitivity the QRPA to the renormalization
of particle-particle interaction of nuclear Hamiltonian an artifact of the QRPA? 
Vogel and Zirnbauer discussed a possibility that the underlying symmetry is the
spin-isospin Wigner SU(4) symmetry\cite{VZ86,VZ88}. The main argument against it was the fact that
in medium-heavy and heavy nuclei the SU(4) symmetry is badly broken by spin-orbit
splitting. Recently, the $2\nu\beta\beta$-decay Gamow-Teller and Fermi transitions
were studied within an exactly model, which allows a violation of both spin-isospin SU(4)
and isospin SU(2) symmetries\cite{Dusan15}. It was found that the model reproduces the main features of realistic
calculation within the QRPA with isospin symmetry restoration, in particular 
the dependence of the $2\nu\beta\beta$-decay decay matrix elements on ithe sovector and
isoscalar particle-particle interactions. By using perturbation theory an explicit
dependence of the $2\nu\beta\beta$-decay matrix elements on the like-nucleon pairing,
particle-particle proton-neutron interaction was obtained. It was found that these matrix
elements do not depend on the mean field part of Hamiltonian and that they are governed
by a weak violation of both SU(2) and SU(4) symmetries by the particle-particle interaction
of Hamiltonian\cite{Dusan15}. The fact that mean field, which breaks the SU(4) symmetry, plays only
secondary role in the evaluation of double beta decay NMEs might be an explanation
of both of the above posed questions. In order to prove it is desired to perform
beyond closure calculation of $2\nu\beta\beta$-decay NMEs within  nuclear
structure models of interest (ISM, (R)EDF, IBM etc)
and to study sensitivity of obtained results to proton-neutron residual interaction.

\begin{table}[!t]
  \caption{The NME of the $0\nu\beta\beta$-decay $M^{0\nu}_{N}$ calculated in the framework of
different approaches:
interacting shell model (ISM) (Strasbourg-Madrid (StMa)\protect\cite{StMa09}
and Central Michigan University (CMU)\protect\cite{CMU16} groups),
interacting boson model (IBM)\protect\cite{IBM15},
quasiparticle random phase approximation (QRPA) (Tuebingen-Bratislava-Caltech 
(TBC)\protect\cite{TBC13,Fang15} and Jyv\"askyla (Jy)\protect\cite{QJy15} groups), 
projected Hartree-Fock Bogoliubov approach (PHFB)\protect\cite{phfb12H}.
The Argonne, CD-Bonn and UCOM two-nucleon short-range correlations  are taken into account.
The non-quenched value of weak axial-vector coupling $g_A$ and R = 1.2 fm $A^{1/3}$ are assumed.
\label{tab.nmeH}}
\begin{tabular}{lcccccccc}\hline\hline
         &       &         & \multicolumn{6}{c}{ $M^{0\nu}_{N}$} \\ \cline{4-9}
  Method & $g_A$ & src     & ${^{48}}$Ca  & ${^{76}}$Ge  & ${^{82}}$Se  & ${^{96}}$Zr & ${^{100}}$Mo & ${^{110}}$Pd
  \\ \hline
ISM-StMa   & 1.25 &  UCOM   &   56.5   &   133       &     122      &             &            &           \\
ISM-CMU    & 1.27 & Argonne &   52.9   &   126       &     127      &             &            &           \\
           &      & CD-Bonn &   75.5   &   202       &     187      &             &            &           \\
IBM        & 1.27 & Argonne &   47     &   104       &      83      &     99      &   164      &   154     \\
QRPA-TBC   & 1.27 & Argonne &   40.3   &   287       &     262      &    184      &   342      &   333     \\
           &      & CD-Bonn &   66.3   &   433       &     394      &    276      &   508      &   492     \\
QRPA-Jy    & 1.26 & CD-Bonn &          &   401      &      287     &     308     &    350      &   476     \\
PHFB       & 1.25 & Argonne &          &             &              &    101      &   195      &   230     \\
           &      & CD-Bonn &          &             &              &    141      &   267      &   316     \\ \hline\hline
         &       &         & \multicolumn{6}{c}{ $M^{0\nu}_{N}$} \\ \cline{4-9}
Method & $g_A$ & src       & ${^{116}}$Cd & ${^{124}}$Sn & ${^{128}}$Te & ${^{130}}$Te & ${^{136}}$Xe & ${^{150}}$Nd
  \\ \hline 
ISM-StMa    & 1.25 &  UCOM   &            &    141      &            &    144      &     115    &            \\
ISM-CMU     & 1.27 & Argonne &            &    97.4     &            &    94.5     &     98.8   &           \\
            &      & CD-Bonn &            &    141      &            &    136      &     143    &           \\
IBM         & 1.27 & Argonne &   110      &    79       &     101    &     92      &     73     &     116  \\
QRPA-TBC    & 1.27 & Argonne &   209      &    184      &     302    &    264      &     152    &           \\
            &      & CD-Bonn &   302      &    279      &     454    &    400      &     228    &           \\
QRPA-Jy     & 1.26 & CD-Bonn &   278      &    453      &     396    &    338     &      186    &           \\
PHFB        & 1.27 & Argonne &            &             &    139     &    138      &            &     78.5  \\
            &      & CD-Bonn &            &             &    191     &    188      &            &     108  
\\ \hline \hline
\end{tabular}
\end{table}

\subsection{Impact of quenching of weak axial-vector coupling constant on the NMEs}

One important source of uncertainty in ${M'}^{0\nu}_{\nu, N}(g^{\rm eff}_A)$ (see Eq. \ref{primeNME}) comes from the fact that 
the effective value of the axial-vector coupling constant  $g^{\rm eff}_A$ is not well known. The axial-vector coupling constant is renormalized due to   nuclear medium effects and, unfortunately, it is reduced, i.e. quenched. The vector coupling constant, which is not expected to be renormalized due to conserved vector current (CVC) hypothesis, yields a smaller contribution to the nuclear matrix element. To a good accuracy
${M'}^{0\nu}_{\nu, N}(g^{\rm eff}_A)$ is proportional to the squared value of $g^{\rm eff}_A$
and correspondingly the decay-rate
to its the fourth power. Thus, quenching of axial-vector coupling constant is very important for 
 double beta decay. The origin of the quenching is not completely known. 
This effect is usually attributed to the $\Delta$-isobar admixture in 
the nuclear wave function or to the shift of the GT strength to higher 
excitation energies due to the short-range tensor correlations. \\
It has been shown that the axial vector single $\beta$-decay NMEs for GT 1$^+$, SD 2$^-$ and others,
which may affect the  DBD NMEs, are reduced with respect to pnQRPA calculations due
to the non-nuclear and nuclear medium effects which are not included in
pnQRPA \cite{eji14,eji15}, but it is not clear what is the most important part,
i.e. the fraction to nuclear medium effects, which  will affect the NME obtained
in other nuclear models, e.g. in the context of ISM. 

Different observations and nuclear structure calculations predict various values
of $g^{\rm eff}_A$:
\begin{itemize}
\item $(g_A)^4=(1.269)^4=2.59$. A modern value of axial-vector coupling constant of
  a free nucleon is $g_A$=1.269 (previously, $g_A = 1.254$ was considered). This
  non-quenched value ($g^{\rm eff}_A=g_A$) is often adopted in the calculation of
  ${M'}^{0\nu}_{\nu, N}(g^{\rm eff}_A)$ and offers its largest value.\\ 
\item $(g^{\rm eff}_A)^4 \simeq 1.00$ (Experimental prediction). It is well known that sum 
  of Gamow???Teller $\beta^-$-strengths to individual final states estimated
  by the IKEDA sum rule is significantly larger than the experimental ones.
  That effect is  known as the axial-vector current matrix elements quenching.
  In order to account for this, it is customary to quench the calculated GT matrix elements
  up to 70\%. Formally, this is accomplished by replacing the true value of the
  coupling constant $g_A = 1.269$ by a quenched  value $g^{\rm eff}_A$= 1.0. \\
\item $(g^{\rm eff}_A)^4 \simeq $ 0.66 ($^{48}$Ca), 0.66 ($^{76}$Ge),  0.30 ($^{76}$Se),  
0.20 ($^{130}$Te) and   0.11 ($^{136}$Xe) (The ISM calculation\cite{poves12,Barea13}).
The shell model, which describes qualitatively well energy spectra, does 
reproduce experimental values of $M^{2\nu}$ only by consideration of 
significant quenching of the Gamow-Teller operator, typically by 0.45 to 70\%.\\ 
\item $(g^{\rm eff}_A)^4\simeq (1.269~A^{-0.18})^4 = 0.063$ (The IBM prediction \cite{Barea13}).  This is an incredible result.  
The quenching of the axial-vector coupling within the  IBM-2 is more like 60\%
\cite{Barea13}. It has been determined by theoretical prediction for the
$2\nu\beta\beta$-decay half-lives, which were based on within closure
approximation calculated corresponding NMEs, with the measured half-lives.\\
\item $(g^{\rm eff}_A)^4 \simeq 0.30$  and $0.50$ for $^{100}$Mo and  $^{116}$Cd,
  respectively (The QRPA prediction).  In Ref. \cite{lisi08}
  $g^{\rm eff}_A$ was treated as a completely free parameter alongside $g_{pp}$
  (used to renormalize particl-particle interaction) by performing calculations
  within the QRPA and RQRPA. It was found that a least-squares
  fit of $g^{\rm eff}_A$ and $g_{pp}$, where possible, to the $\beta$-decay rate and 
  $\beta^+$/EC rate of the $J^\pi$ = $1^+$ ground state in the intermediate nuclei 
   involved in double-beta decay in addition to the $2\nu\beta\beta$ rates of the initial nuclei, 
   leads to an effective $g^{\rm eff}_A$ of about 0.7 or 0.8. This value, which is comparable to 
   that needed in the shell model to reproduce $2\nu\beta\beta$-decay rates is significantly 
   smaller than 1.0 - 1.27, the range usually used in the QRPA \cite{SRFV13}.
   The above statistical approach has been extended also for analysis of 47 isobaric triplets 
   28 more extended isobaric chains of nuclei to extract values and uncertainties for
   $g^{\rm eff}_A$ in calculations performed within the QRPA. A comparatively small value of 
   $g^{\rm eff}_A$ was found \cite{DeppischGA}.
\end{itemize}
We see that the uncertainty in the calculated $0\nu\beta\beta$-decay half-life
due to quenching is quite large.

The quenching of axial-vector coupling constant is assumed to have different sources
like the truncation of the many-nucleon Hilbert space and the many-body
currents, which  reduce matrix elements by amounts that are still in
dispute. The effect of two-body currents can be isolated by defining
new \emph{effective} one-body current with the factor $g_A(p^2)$
replaced by an effective coupling $g_A^{\rm eff}(p^2)$, given by
\begin{eqnarray}
\label{eq:gaef}
g_A^{\rm eff}(p^2) &=& g_A(p^2) \left( 1-\frac{\rho}{F_\pi^2} \left[ \frac{c_D}{g_a
\Lambda_\chi} + \frac{2}{3}c_3\frac{p^2}{4m_\pi^2+p^2} \right. \right. \nonumber \\
 &&\left. \left. + I(\rho,0) \left( \frac{1}{3}(2c_4-c_3) + \frac{1}{6m}
 \right) \right] \right) \,, 
\end{eqnarray}
with
\begin{eqnarray}
\label{eq:integral}
I(\rho,&P)& = 1 -\frac{3m_\pi^2}{2k_F^2}+\frac{3m_\pi^3}{2k_F^3}\textrm{acot}
\left[ \frac{m_\pi^2+\frac{P^2}{4}-k_F^2}{2m_\pi k_F} \right] \\
&+& \frac{3m_\pi^2}{4k_F^3P}\left( k_F^2+m_\pi^2-\frac{P^2}{4} \right)
\textrm{ln}\left[ \frac{m_\pi^2+(k_F-\frac{P}{2})^2}{m_\pi^2+(k_F+
\frac{P}{2})^2} \right] \,. \nonumber
\end{eqnarray}
Here, $k_F$ is the Fermi momentum and $P$ is the center-of-mass
momentum of the decaying nucleons, which can be set to zero without altering
$I(\rho,P)$ significantly \cite{gansm}.  The constants $c_3$, $c_4$, and $c_D$
are the effective field theory parameters, fit to data in light nuclei \cite{gansm}. 

\begin{figure}[!t]
   \vspace*{1.3cm}  
\centerline{\psfig{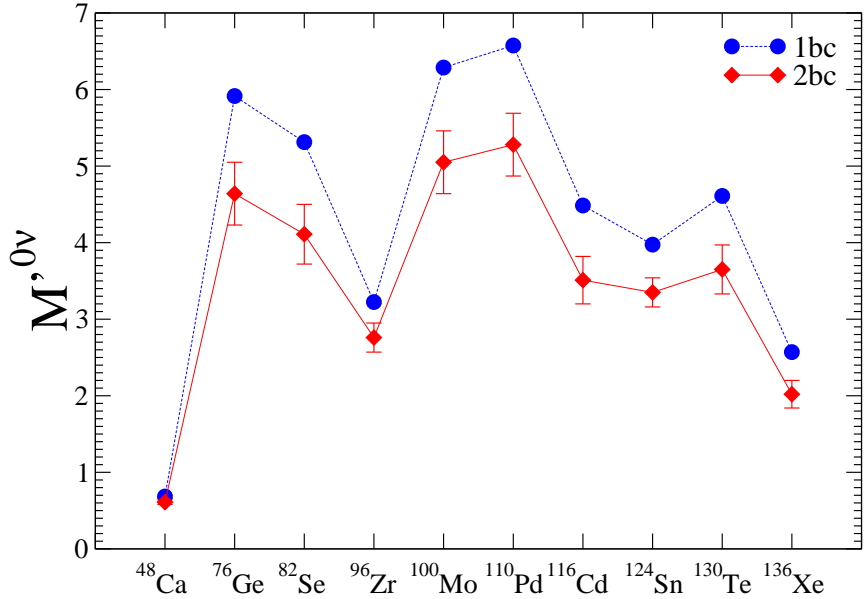}}
\caption{(Color online) 
Nuclear matrix element ${M'}^{0\nu}_\nu$.
The blue circles represent the results with the one-body current (1bc) only,
and the red diamonds the average of the results with two-body currents (2bc)
included. The error bars represent the dispersion in those values.
\label{fig:quench}}
\end{figure}

The $0\nu\beta\beta$-decay matrix element has a different form than does
the $2\nu\beta\beta$ matrix element and it might be that the
quenching $2\nu\beta\beta$ decay has a smaller  effect on $0\nu\beta\beta$-decay.
First, this issue was discussed in the context of the two-body currents
and the ISM in Ref. \cite{gansm}. It was found that 
the effect of the two-body currents decreases as the momentum transfer increases, and so such
currents will quench $2\nu\beta\beta$-decay NME, for which the momentum transfer is essentially zero, 
more than the $0\nu\beta\beta$-decay NME, since in the latter case the intermediate neutrino can transfer several 
hundred MeV of momentum from one decaying nucleon to the other. Later this finding was
also confirmed within the QRPA\cite{gaqrpa}. The conclusion is that, if most of the quenching is 
due to two-body currents, as effective field theory suggests,  the
$0\nu\beta\beta$-matrix elements may  be quenched by a factor on the order
of 30\% \cite{gansm,gaqrpa}. \\The nuclear matrix element ${M'}^{0\nu}$ calculated for the nuclei
of experimental interest, obtained by consideration both one- and two-body currents within the
QRPA\cite{gaqrpa}, are displayed in Fig. \ref{fig:quench}.

Currently, it is difficult to assign both systematic and statistical
uncertainties to calculated NMEs\cite{EngelReview}. The most sophisticated way
to do it has been proposed in the framework of the QRPA\cite{qrpa1} and applied later
within the PHFB\cite{phfb13}. Further, within the QRPA for a given set of nuclei,
it has been shown that the correlations among NME errors are as important as their size
\cite{lisi09,Lisi15}. It is desired that also other groups  present ``statistical samples''
of NME calculations  as well, in order to provide independent estimates of (co)variances
for their NME estimates. A covariance analysis proposed \cite{lisi09,Lisi15}
is the way to estimate correctly current or prospective sensitivities
to effective Majorana neutrino mass $m_{\beta\beta}$.

The investigation of the quenching of the axial current in double beta decay will be illuminated by the relevant experimental searches (see section \ref{sect:expME}).

Accurate determination of the NMEs, and a realistic estimate of their uncertainty,
remains of great importance. Methods are developed and improved. Increasing computer
power should allow them to include all important nuclear physics aspects and to
reproduce relevant nuclear physics observables. The ultimate goal is to evaluate 
Nuclear matrix elements with uncertainty of less than 30\% to establish the unknown neutrino
mass scale. 
To this end, it is crucial that the theory groups consider all factors outlined above in order  to evaluate NMEs within 15 $\%$ uncertainty and the experimental groups
provide relevant data to be used in checking  these models.

%
%
%
\section[Concluding remarks]{Concluding remarks}
%

In this review we analyzed the $0\nu \beta\beta$-decay process, which is the oldest and perhaps the best to test lepton number violation and at the same time settle the issue of the nature of the neutrino, i.e. whether  the neutrino mass eigenstates are of the Majorana or  Dirac type. We also have seen that it is the best process to settle the neutrino mass scale in the few meV scale. To achieve these goals first, and  most important, some serious experimental problems must be overcome in order to be able  measure life times of the order of $\geq10^{26}$y. We have discussed the ongoing, planned and future experiments. We have witnessed a great progress in  tackling the various background problems, improving the energy resolution and preparing large  masses of the needed isotopes.\\ To encourage and guide the experiments we considered a variety of particle models, which allow the double beta decay to occur at a reasonable level. The most important mechanism involves the light Majorana  neutrino, but we considered other competing mechanisms as well. \\In order to extract useful information from the data i) the phase space integrals must be reliably calculated. There has been great progress in this direction, but we are haunted by the quenching of the axial current. This is crucial in determining the expected life time. There appears to be hope towards  determining its effective value by using other related experimental information. ii)  The precise evaluation of the nuclear matrix elements themselves. It is encouraging that all available nuclear models (the Interacting Shell Model, the Quasiparticle Random Phase Approximation, the Interacting Boson Model, the Projected Hartree-Fock-Bogoliubov Method and the Energy Density Factional Method) are being used in the evaluation of all the needed $0\nu\beta\beta$-decay nuclear matrix elements.  The fact that these vastly different  models agree in the predicted values of these ME for various mechanisms and a number of different targets as well as the fact that they  reproduce  data on other related experiments, such  as single beta decay rates and charge exchange reactions, gives us confidence that they are perhaps estimated better than a factor of 2.  Admittedly  we have not witnessed  any great progress during the last few years and this situation must be improved.  It is hoped that some constraints may arise by utilizing experimental input from other processes. \\It is clear  that the mere observation of  double beta decay will be a triumph of physics. The next day, however, the extraction of the most useful parameter, namely the determination of  the scale of the neutrino mass, will begin.  It has been shown that  \cite{ROP12}, given  adequate information on a number of judiciously chosen targets, the neutrino mass can, in principle, be extracted from the data even in the case that other competing mechanisms  may significantly contribute to this process.

%
%
%
\section*{Acknowledgements}

F. \v{S} acknowledges the support in part by the VEGA Grant Agency of the Slovak Republic under Contract No. 1/0922/16, by
Slovak Research and Development Agency under Contract No. APVV-14-0524, and by the Ministry of Education,
Youth and Sports of the Czech Republic under Contract No. LM2011027. J.D.V acknowledges that part of this work was supported
by IBS-R017-D1-2016-a00 in the Republic of Korea and  by the ARC Centre of Excellence in Particle Physics (CoEPP), University of Adelaide, Australia. He is indebted to Professors A. Thomas and Y. Semertzidis for support and hospitality.

\section*{References}

\begin{thebibliography}{100}

\bibitem{GOEPMAY}
M.~Goeppert-Mayer {\em Phys. Rev.}, vol.~48, p.~512, 1035.

\bibitem{emajorana}
E.~Majorana {\em Nuovo Cim.}, vol.~14, p.~171, 1937.

\bibitem{Pr52}
H.~Primakoff {\em Phys. Rev.}, vol.~85, p.~888, 1952.

\bibitem{davisno}
J.~R.~Davis {\em Phys. Rev.}, vol.~97, p.~766, 1955.

\bibitem{ing}
M.~G. Ingram and J.~H. Reynolds {\em Phys. Rev.}, vol.~78, p.~822, 1950.

\bibitem{kir67}
T.~Kirsten, W.~Gentner, and O.~Schaeffer {\em Z. Physik}, vol.~202, p.~273,
  1967.

\bibitem{kir67b}
T.~Kirsten, W.~Gentner, and O.~Miller {\em Z. Naturf. A}, vol.~22, p.~1783,
  1967.

\bibitem{tak66}
N.~Takaota and K.~Ogata {\em Z. Naturf. A}, vol.~21, p.~84, 1966.

\bibitem{sir72}
B.~Srinvasan, O.~E.C.~Alexander, and Manuel {\em J. Inorg. Nucl. Chem.},
  vol.~34, p.~2381, 1972.

\bibitem{DTNOT}
M.~Doi, T.~Kotani, N.~Nishiura, K.~Okuda, and E.~Takasugi {\em Phys. Lett B.},
  vol.~103, p.~219, 1981.

\bibitem{SVa82}
J.~Schechter and J.~W.~F. Valle {\em Phys. Rev. D}, vol.~25, p.~2951, 1982.

\bibitem{klko96}
M. Hirsch, H.V. Klapdor-Kleingrothaus, and S.G. Kovalenko {\it Phys. Lett. B},
  vol. {372}, p. 181, 1996.

\bibitem{FKSS97}
A.~Faessler, S.~Kovalenko, F.~{\v S}imkovic, and J.~Schwieger {\em Phys. Rev.
  Lett.}, vol.~78, p.~183, 1997.

\bibitem{FKS98}
A.~Faessler, S.~Kovalenko, and F.~{\v S}imkovic {\em Phys. Rev. D}, vol.~58,
  p.~055004, 1998.

\bibitem{WKS99}
A.~Wodecki, W.~A. Kami{\' n}ski, and F.~{\v S}imkovic {\em Phys. Rev. D},
  vol.~60, p.~115007, 1999.

\bibitem{zdes02}
V.~I. Tretyak and Y.~G. Zdesenko {\em At. Dat. Nucl. Dat. Tabl.}, vol.~80,
  p.~83, 2002.

\bibitem{eji05}
H.~Ejiri {\em J. Phys. Soc. Jap.}, vol.~74, p.~2101, 2005.

\bibitem{AEE08}
F.~Avignone, S.~Elliott, and J.Engel {\em Rev. Mod. Phys.}, vol.~80, p.~481,
  2008.

\bibitem{RMP08}
L.~Camilleri, E.~Lisi, and J.~F. Wilkerson {\em Ann. Rev. Nucl. Part. Sci.},
  vol.~58, p.~343, 2008.

\bibitem{ell87}
S.~R. Elliott, A.~A. Hahn, and M.~Moe {\em Phys. Rev. Lett.}, vol.~59, p.~2020,
  1987.

\bibitem{SUPERKAMIOKANDE}
Y. Fukuda {\it et al.} (The Super-Kamiokande Collaboration) {\it Phys. Rev.
  Lett.}, {\it ibid} {vol. 81}, p. 1562 $\&$ 1158, 1998; {\it ibid} {vol. 82},
  p. 1810, 1999; {\it ibid} {vol. 85}, p. 3999, 2000; {vol. 86}, p. 5651, 2001.

\bibitem{SOLAROSC}
Q.R. Ahmad {\it et al.} (The SNO Collaboration) {\it Phys. Rev.Lett.}, {vol.
  89}, p. 011302; {\it ibid} {vol. 89}, p. 011301, 2002; {\it ibid} {vol. 87},
  p. 071301, 2001;\\ K. Lande {\it et al.} (The Homestake Collaboration) {\it
  Astrophys. J.}, {vol. 496}, p. 505, 1998;\\ W. Hampel {\it et al.} (The
  Gallex Collaboration) {\it Phys. Lett.B}, {vol. 447}, p. 127, 1999;\\ J.N.
  Abdurashitov {\it et al.} (The Sage Collaboration) {\it Phys. Rev. C}, {vol.
  80}, p. 056801, 1999;\\ G.L Fogli {\it et al.}, {\it Phys. Rev. D}, {vol.
  66}, p. 053010, 2002.

\bibitem{KAMLAND}
K. Eguchi {\it et al.} (The KamLAND Collaboration) {\it Phys. Rev. Lett.},
  {vol. 90}, p. {021802}, 2003.

\bibitem{ROP12}
J.~D. Vergados, H.~Ejiri, and F.~{\v S}imkovic {\em Rep. Prog. Phys.}, vol.~75,
  2012.

\bibitem{Xelimit}
A. Gando {\it et al.} (The KamLAND-Zen Collaboration) {\it Phys. Rev. Lett.},
  vol. 117, p. 082503, 2016.

\bibitem{WINTER}
R.~Winter {\em Phys. Rev.}, vol.~100, p.~142, 1955.

\bibitem{Ver83}
J.~Vergados {\em Nuc. Phys. B}, vol.~218, p.~109, 1983.

\bibitem{BeRuJar83}
J.~Bernabeu, A.~de~Rujula, and C.~Jarlskog {\em Phys. Rev. C}, vol.~223, p.~15,
  1983.

\bibitem{DK93}
M.~Doi and T.~Kotani {\em Prog. Theor. Phys.}, vol.~89, p.~139, 1993.

\bibitem{SujWy04}
Z.~Sujkowski and S.~Wycech {\em Phys. Rev. C}, vol.~70, p.~052501, 2004.

\bibitem{lukas}
L.~Lukaszuk, Z.~Sujkowski, and S.~Wycech {\em Eur. Phys. J. A}, vol.~27, p.~63,
  2006.

\bibitem{SIMKO11}
M.~Krivoruchenko, F.~{\v S}imkovic, D.~Frekers, and A.~Faessler {\em Nucl.
  Phys. A}, vol.~859, p.~140, 2011.

\bibitem{SimKriv09}
F.~{\v S}imkovic and M.~Krivoruchenko {\em Phys. Part. Nuc.}, vol.~6, p.~298,
  2009.

\bibitem{verg11}
J.~Vergados {\em Phys. Rev. C}, vol.~84, p.~044328, 2011.

\bibitem{AUDI03}
G.~Audi {\em et~al.} {\em Nucl. Phys. A}, vol.~729, p.~3, 2003.

\bibitem{BLAUM06}
K.~Blaum {\em Phys. Rep.}, vol.~425, p.~1, 2006.

\bibitem{penning1}
G.~Douysset {\em et~al.} {\em Phys. Rev. Lett.}, vol.~86, p.~4259, 2001.

\bibitem{BNW10}
K.~Blaum, Y.~Novikov, and G.~Werth {\em Contemp. Phys.}, vol.~51, p.~149, 2010.

\bibitem{DEC11}
S.~Eliseev {\em et~al.} {\em Phys. Rev. Lett.}, vol.~106, p.~052504, 2011.

\bibitem{redshaw07}
M.~Redshaw {\em et~al.} {\em Phys. Rev. Lett.}, vol.~98, p.~053003, 2007.

\bibitem{redshaw09}
M.~Redshaw {\em et~al.} {\em Phys. Rev. Lett.}, vol.~102, p.~212502, 2009.

\bibitem{SCIE09}
N.~Scielzo {\em et~al.} {\em Phys. Rev. C}, vol.~80, p.~0225501, 2009.

\bibitem{RAKH09}
S.~Rahaman {\em et~al.} {\em Phys. Rev. Lett.}, vol.~103, p.~042501, 2009.

\bibitem{kolhinen}
V.~Kolhinen {\em et~al.} {\em Phys. Lett. B}, vol.~684, p.~17, 2010.

\bibitem{mount10}
B.~Mount, M.~Redshaw, , and E.~G. Myers {\em Phys. Rev. C}, vol.~81, p.~032501,
  2010.

\bibitem{elis2}
S.~Eliseev {\em et~al.} {\em Phys. Rev. C}, vol.~83, p.~038501, 2011.

\bibitem{EliNov}
S.~Eliseev {\em et~al.} {\em Phys. Rev. C}, vol.~84, p.~012501, 2011.

\bibitem{elis4}
M.~Goncharov {\em et~al.} {\em Phys. Rev. C}, vol.~84, p.~028501, 2011.

\bibitem{elis5}
S.~Eliseev {\em et~al.} {\em Phys. Rev. Lett.}, vol.~107, p.~152501, 2011.

\bibitem{droese11}
C.~Droese {\em et~al.} {\em Nucl. Phys. A}, vol.~875, p.~1, 2012.

\bibitem{BELLI11}
P.~Belli {\em et~al.} {\em Eur. Phys. J. A}, vol.~47, p.~91, 2011.

\bibitem{barab1}
A.~Barabash, P.~Hubert, A.~Nachab, and V.~Umatov {\em Nucl. Phys. A}, vol.~785,
  p.~371, 2007.

\bibitem{barab2}
A.~Barabash, P.~Hubert, A.~Nachab, S.~Konovalov, I.~Vanyushin, and V.~Umatov
  {\em Nucl. Phys. A}, vol.~807, p.~269, 2008.

\bibitem{belli09}
P.~Belli {\em et~al.} {\em Nucl. Phys. A}, vol.~824, p.~101, 2009.

\bibitem{rukh11}

\newblock N.I. Rukhadze {\it et al.} (TGV Collaboration) {\it Nucl. Phys. A},
  {vol. 852}, {p. 197}, 2011.

\bibitem{FPS11}
D.~Frekers, P.~Puppe, J.~Thies, P.~Povinec, F.~{\v S}imkovic, J.~Stani{\v c}ek,
  and I.~S{\' y}kora {\em Nucl. Phys. A}, vol.~860, p.~1, 2011.

\bibitem{BELLI11b}
P.~Belli {\em et~al.} {\em Nucl. Phys. A}, vol.~859, p.~126, 2011.

\bibitem{Ver86}
J.~Vergados {\em Phys. Rep.}, vol.~133, p.~1, 1986.

\bibitem{HS84}
{W.C. Haxton and G.S. Stephenson, Jr.} {\it Prog. Part. Nucl. Phys.}, {vol.
  12}, {p. 409}, 1984.

\bibitem{DTK85}
M.~Doi, T.~Kotani, and E.~Tagasugi {\em Prog. Theor. Phys. (Supp.)}, vol.~83,
  p.~1, 1985.

\bibitem{Tom91}
T.~Tomoda {\em Rep. Prog. Phys.}, vol.~54, p.~53, 1991.

\bibitem{SC98}
J.~Suhonen and O.~Civitarese {\em Phys. Rep.}, vol.~300, p.~123, 1998.

\bibitem{FS98}
A.~Faessler and F.~{\v S}imkovic {\em J. Phys. G}, vol.~24, p.~2139, 1998.

\bibitem{Ver02}
J.~Vergados {\em Phys. Rep.}, vol.~361, p.~1, 2002.

\bibitem{RODEJ11}
W.~Rodejohann {\em Int. J. Mod. Phys. E}, vol.~20, p.~1833, 2011.

\bibitem{SDSJ97}
J.~Suhonen, P.~Divari, L.~Skouras, and I.~D. Johnstone {\em Phys. Rev. C},
  vol.~55, p.~714, 1997.

\bibitem{RCN95}
J.~Retamosa, E.~Caurier, and F.~Novacki {\em Phys. Rev. C}, vol.~51, p.~371,
  1995.

\bibitem{CNPR96}
E.~Caurier, F.~Novacki, A.~Poves, and J.~Retamosa {\em Phys. Lett. B}, vol.~77,
  p.~1954, 1996.

\bibitem{SSDV92}
J.~Sinatkas, L.~Skouras, D.~Strottman, and J.~Vergados {\em J. Phys. G},
  vol.~18, p.~1377, 1992.

\bibitem{CPZ90}
E.~Caurier, A.~Poves, and A.~Zucker {\em Phys. Lett B}, vol.~252, p.~13, 1990.

\bibitem{AHARMIN}
B.~Aharmin {\em et~al.} {\em Phys. Rev. C}, vol.~72, p.~055502, 2005.

\bibitem{CHOOZ}
M. Apollonio {\it et al.} (The CHOOZ Collaboration) {\it Phys. Lett. B}, {vol.
  446}, {p. 415}, {1999}.

\bibitem{ARAKI}
T.~Araki {\em et~al.} {\em Phys. Rev. Lett.}, vol.~94, p.~081801, 2005.

\bibitem{dayabay}
F.P. An {\it et al.} (The Daya Bay Collaboration) {\it Phys.Rev.Lett.}, vol.
  108, p. 171803, 2012, arXiv:1203.1669[hep-ex].

\bibitem{reno12}
J.K. Ahn {\it et al.} (The RENO Collaboration),{\it Phys. Rev. Lett.}, vol.
  108, p. 191802, 2012, arXiv:1204.0626[hep-ex].

\bibitem{SCHWETZ}
T.~Schwetz, M.~T\'ortola, and J.~Valle {\em New J. Phys.}, vol.~10, p.~113011,
  2008.

\bibitem{Capozzi14}
F.~Capozzi, G.~Fogli, E.~Lisi, A.~Marrone, D.~Montanino, and A.~Palazzo {\em
  Phys. Rev. D}, vol.~89, p.~093018, 2014.

\bibitem{G-GM08}
M.~Gonzales-Garcia and M.~Maltoni {\em Phys. Rep.}, vol.~460, p.~1, 2008.

\bibitem{PaesRod15}
H.~P{\"a}s and W.~Rodejohann {\em New J. Phys.}, vol.~17, p.~115010, 2015.

\bibitem{HHOPS15}
J.~Helo, M.~Hirsch, T.~Ota, and F.~P. dos Santos {\em JHEP}, vol.~1505, p.~092,
  2015.

\bibitem{baysian15}
M.~Gerbino, M.~Lattanzi, and A.~Melchiorri {\em Phys. Rev. D}, vol.~93,
  p.~033001, 2016.

\bibitem{MCMC02}
A.~Lewis and S.~Bridle {\em Phys. Rev. D}, vol.~66, p.~103511, 2002.

\bibitem{Ver76}
J.~Vergados {\em Phys. Rev. C}, vol.~13, p.~865, 1976.

\bibitem{HSS82}
W.~C. Haxton, G.~S. Stephenson, and D.~Strottman {\em Phys. Rev. D}, vol.~25,
  p.~2360, 1982.

\bibitem{SV83}
L.~Skouras and J.~Vergados {\em Phys. Rev. C}, vol.~28, p.~2122, 1983.

\bibitem{ZBR90}
L.~Zhao, B.~Brown, and W.~Richter {\em Phys. Rev. C}, vol.~42, p.~1120, 1990.

\bibitem{ZB93}
L.~Zhao and B.~Brown {\em Phys. Rev. C}, vol.~47, p.~2641, 1993.

\bibitem{Retal96}
R.~Radha {\em et~al.} {\em Phys. Rev. Lett.}, vol.~76, p.~2642, 1996.

\bibitem{NSM96}
T.~S. H.~Nakada and K.~Muto {\em Nucl. Phys. A}, vol.~607, p.~235, 1996.

\bibitem{KDL97}
S.~Koonin, D.~Dean, and K.~Langanke {\em Phys. Rep.}, vol.~278, p.~1, 1997.

\bibitem{edf}
T.~Rodrigez and G.~Martinez-Pinedo {\em Phys. Rev. Lett.}, vol.~105, p.~252503,
  2010.

\bibitem{VZ86}
P.~Vogel and M.~Zirnbauer {\em Phys. Rev. Lett}, vol.~57, p.~3148, 1986.

\bibitem{CAT87}
O.~Civitarese, A.~Faessler, and T.~Tomoda {\em Phys. Lett. B}, vol.~194, p.~11,
  1987.

\bibitem{MBK88}
K.~Muto and H.~Klapdor {\em Phys. Lett. B}, vol.~208, p.~53, 1988.

\bibitem{EVJP91}
J.~Engel, P.~Vogel, X.~Ji, and S.~Pittel {\em Phys. Lett. B}, vol.~225, p.~5,
  1989.

\bibitem{RFSK91}
A.~A. Raduta, A.~Faessler, S.~Stoica, and W.~A. Kami{\' n}ski {\em Phys. Lett.
  B}, vol.~254, p.~7, 1991.

\bibitem{GV92}
A.~Griffiths and P.~Vogel {\em Phys. Rev. C}, vol.~46, p.~181, 1992.

\bibitem{SC93}
J.~Suhonen and O.~Civitarese {\em Phys. Lett. B}, vol.~308, p.~212, 1993.

\bibitem{CS94}
O.~Civitarese and J.~Suhonen {\em Nuc. Phys. A}, vol.~575, p.~251, 1994.

\bibitem{SSVP97}
F.~{\v S}imkovic, J., Schwieger, M.~Veselsk\'y, G.~Pantis, and A.~Faessler {\em
  Phys. Lett. B}, vol.~393, p.~267, 1997.

\bibitem{SPF98}
F.~{\v S}imkovic, G.~Pantis, and A.~Faessler {\em Prog Part. Phys.}, vol.~40,
  p.~285, 1998.

\bibitem{cheoun}
M.~Cheoun, A.~Bobyk, A.~Faessler, F.~{\v S}imkovic, and G.~Teneva {\em Nucl.
  Phys. A}, vol.~561, p.~74, 1993.

\bibitem{MUT97}
K.~Muto {\em Phys. Lett. B}, vol.~391, p.~243, 1997.

\bibitem{SRFV13}
F.~{\v S}imkovic, V.~Rodin, A.~Faessler, and P.~Vogel {\em Phys. Rev. C},
  vol.~87, p.~045501, 2013.

\bibitem{TS95}
J.~Toivanen and J.~Suhonen {\em Phys. Rev. Lett.}, vol.~75, p.~410, 1995.

\bibitem{SSF96}
J.~Schwieger, F.~{\v S}imkovic, and A.~Faessler {\em Nuc. Phys. A}, vol.~600,
  p.~179, 1996.

\bibitem{TBC13}
F.~{\v S}imkovic, V.~Rodin, A.~Faessler, and P.~Vogel {\em Phys. Rev. C},
  vol.~87, p.~045501, 2013.

\bibitem{phfb}
P.~Rath, R.~Chandra, K.~Chaturvedi, P.~Raina, and J.~Hirsch {\em Phys. Rev. C},
  vol.~82, p.~064310, 2010.

\bibitem{IBM09}
J.~Barea and F.~Iachello {\em Phys. Rev. C}, vol.~79, p.~044301, 2009.

\bibitem{BaKotIac15}
J.~Barea, J.~Kotila, and F.~Iachello {\em Phys. Rev. C}, vol.~91, p.~034304,
  2015.

\bibitem{REDF15}
J.~Yao, L.~Song, K.~Hagino, P.~Ring, and J.~Meng {\em Phys. Rev. C}, vol.~91,
  p.~024316, 2015.

\bibitem{SMIRNOV04}
A.Yu. Smirnov, arXiv: hep-ph/0411194.

\bibitem{WEINBERG79}
S.~Weinberg {\em Phys. Rev. Lett.}, vol.~43, p.~1566, 1979.

\bibitem{SeeSaw07}
A.~Abada, C.Biggio, F.~Bonnet, M.~Gavela, and T.~Hambye {\em JHEP}, vol.~0712,
  p.~061, 2007.
\newblock arXiv:0707.4058 (hep-ph).

\bibitem{L-PMP15}
J. Lopez-Pavon, E. Molinaro, S. T. Petcov, Radiative Corrections to Light
  Neutrino Masses in Low Scale Type I Seesaw Scenarios and Neutrinoless Double
  Beta Decay, arXiv:1506.05296 (hep-ph).

\bibitem{HuLiSm15}
P. Humbert, M. Lindner and J. Smirnov, arXiv:1503.03066 (hep-ph); P. Humbert,
  M. Lindner, S. Patra and J. Smirnov, arXiv:1505.07453 (hep-ph).

\bibitem{BonHirOtaWint}
F.~Bonnet, M.~Hirsch, T.~Ota, and W.~Winter {\em JHEP}, vol.~1303, p.~055,
  2013.
\newblock [Erratum: J. High Energy Phys. 1404, 090 (2014)].

\bibitem{Pascoli08}
C.~Boehm, Y.~Farzan, T.~Hambye, S.~Palomares-Ruiz, and S.~Pascoli {\em Phys.
  Rev. D}, vol.~77, p.~043516, 2008.

\bibitem{GKSA06}
M.~G{\' o}{\' z}d{\' z}, W.~Kami{\' n}ski, F.~{\v S}imkovic, and A.~Faessler
  {\em Phys. Rev. D}, vol.~74, p.~055007, 2006.

\bibitem{Twoloop80}
E.~Witten {\em Phys. Lett. B}, vol.~91, p.~81, 1980.

\bibitem{Twoloop15}
C-Q Geng and L-H Tsai, {\it Annals of Physics}, {vol. 365}, p. 210, 2016; C-Q
  Geng, D. Huang and L-H Tsai {\it Phys. Rev. D}, {vol. 90}, p. 113005, 2014.

\bibitem{Twoloop91}
G.~K. Leontaris and J.~D. Vergados {\em Phys. Lett. B}, vol.~258, p.~111, 1991.

\bibitem{Xing}
Z.-Z. Xing {\em Phys. Rev. D}, vol.~85, p.~013008, 2012.

\bibitem{Ma01}
E.~Ma and G.~Rajasekaran {\em Phys. Rev. D}, vol.~64, p.~113012, 2001.

\bibitem{BAM-VK15}
F.~Bjoerkeroth, F.~de~Anda, I.~de~Medeiros~Varzielas, and S.~F. King {\em
  JHEP}, vol.~06, p.~141, 2015.
\newblock arXiv:1503.03306 (hep-h).

\bibitem{DKN14}
G.J. Ding, S.F. King and T. Neder {\it JHEP}, {vol. 12}, {p. 007}, 2014,
  arXiv:1409.8005[hep-ph]; Y. Shimizu and M. Tanimoto {\it JHEP} {vol. 12}, p.
  132, 2015, arXiv:1507.06221 [hep-ph].

\bibitem{AltFer10}
G.~Altarelli and F.~Feruglio {\em Rev. Mod. Phys.}, vol.~82, p.~2701, 2010.
\newblock arXiv:1002.0211[hep-h].

\bibitem{OSVV16}
S.~Dell'Oro, S.~Marcocci, M.~Viel, and F.~Vissani {\em Adv.High Energy Phys.},
  vol.~2016, p.~2162659, 2016.
\newblock arXiv:1601.07512v2 [hep-ph].

\bibitem{DisSym15}
C Arbela'ez, A. E. Ca'rcamo Hern'andez, S, Kovalenko, and I. Schmidt {\it Phys.
  Rev. D}, vol. 92, p. 115015, 2015, arXiv:1507.03852[hep-ph]; P. Ballett, S.
  Pascoli and J. Turner {\it Phys. Rev. D}, vol. 92, p. 093008, 2015,
  arXiv:1503.07543 [hep-ph].

\bibitem{FonHir15}
R.~Fonseca and M.~Hirsch {\em Phys. Rev. D}, vol.~92, p.~015014, 2015.
\newblock arXiv:1505.06121[hep-ph].

\bibitem{VienNgocKhoi15}
V.~Vien, H.~Long, and D.~P. Khoi {\em Int. J. Mod. Phys. A}, vol.~30,
  p.~1550102, 2015.
\newblock arXiv:1506.06063[hep-h].

\bibitem{Vergados16}
J. D. Vergados, Reduction of $SU_f(3)\supset SO(3)\supset A_4$-The scalar
  potential, arXiv:1604.00678 [phys.gen-ph].

\bibitem{MohEtal07}
R.~Mohapatra {\em et~al.} {\em Rep. Prog. Phys.}, vol.~70, p.~1757, 2007.

\bibitem{lisiglob}
F.~Capozzi, E.~Lisi, A.~Maronne, D.~Montanino, and A.~Palazzo {\em Nucl. Phys.
  B}, vol.~908, p.~218, 2016.

\bibitem{globfit16}
M.~Gonzalez-Garcia, M.~Maltoni, and T.~Schwetz {\em Nucl. Phys. B}, vol.~908,
  p.~199, 2016.

\bibitem{Pascoli05}
S.~Pascoli, S.~Petcov, and T.~Schwetz {\em Nuc. Phys. B}, vol.~734, p.~24,
  2006.

\bibitem{Katrin}
A. Osipowicz A {\it et al.} (The KATRIN Collaboration) 2001, hep-ex/0109033;\\
  J. Angrik {\it et al.} (The KATRIN Collaboration) 2004, KATRIN Design Report
  http://bibliothek.fzk.de/zb/berichte/FZKA7090.pdf.

\bibitem{otten}
E.~Otten and C.~Weinheimer {\em Rep. Prog. Phys.}, vol.~71, p.~086201, 2008.

\bibitem{Mare}
E. Andreotti (The MARE Collaboration) {\it Nucl. Instrum. Meth.}, {vol. 572},
  {p. 208}, 2007.

\bibitem{Abarajan11}
K.~N. Abazajian {\em et~al.} {\em Astropart. Phys.}, vol.~35, p.~177, 2011.

\bibitem{ThAbdaLah10}
S.~A. Thomas, F.~B. Abdalla, and O.~Lahav {\em Phys.Rev.Let.}, vol.~105,
  p.~031301, 2010.

\bibitem{SDSS05}
U. Seljak {\it et al.} (The SDSS Collaboration) {\it Phys. Rev. D}, {vol. 71},
  {p. 103515}, 2005.

\bibitem{RSPD14}
S.~Riemer, S.~Sorensen, D.~Parkinson, and T.~Davis {\em Phys. Rev. D}, vol.~89,
  p.~103505, 2014.

\bibitem{CSVB14}
M.~Costanzi, B.~Sartoris, M.~Viel, and S.~Borgani {\em JCAP}, vol.~1410,
  p.~081, 2014.

\bibitem{Planck15}
P. Ade {\it et al.} (The Planck Collaboration) {\it Astron. Astrophys.}, vol.
  594, p. A13, 2016, arXiv:1502.01589[astroph.CO]).

\bibitem{DMVV15}
S.~Dell'Oro, S.~Marcocci, M.~Viel, and F.~Vissani {\em JCAP}, vol.~1512,
  p.~023, 2015.
\newblock arXiv:1505.02722[astroph.CO].

\bibitem{Vissani16}
S.~D. Oro, S.~Marcocci, M.~Viel, and F.~Vissani {\em Adv. High Energy Phys.},
  vol.~2016, p.~2162659, 2016.

\bibitem{fmsprl}
S.~Kovalenko, M.~Krivoruchenko, and F.~{\v S}imkovic {\em Phys. Rev. C},
  vol.~112, p.~142503, 2014.

\bibitem{Abazajian:2012ys}
K.N. Abazajian {\it et al.} arXiv:1204.5379[hep-ph].

\bibitem{Helo:2010cw}
J.~Helo, S.~Kovalenko, and I.~Schmidt {\em Nucl. Phys. B}, vol.~853, p.~80,
  2011.

\bibitem{Atre:2009rg}
A.~Atre, T.~Han, S.~Pascoli, and B.~Zhang {\em JHEP}, vol.~0905, p.~030, 2009.

\bibitem{Beringer:1900zz}
J. Beringer {\it et~al.} (The Particle Data Group) {\it Phys. Rev. D}, vol. 86,
  p. 010001, 2012.

\bibitem{pasa74}
J.~C. Pati and A.~Salam {\em Phys. Rev. D}, vol.~10, p.~275, 1974.

\bibitem{mopa75}
R.~N. Mohapatra and J.~C. Pati {\em Phys. Rev. D}, vol.~11, p.~2558, 1975.

\bibitem{vissa11}
V.~Tello, M.~Nemev\v{s}ek, F.~Nesti, G.~Senjanovi\'{c}, and F.~Vissani {\em
  Phys. Rev. Lett.}, vol.~106, p.~151801, 2011.

\bibitem{Nemevsek}
M. Nemev{\v s}ek, F. Nesti, G. Senjanovi{\' c} and V. Tello, arXiv:1112.3061
  [hep-ph].

\bibitem{Barry13}
J.~Barry and W.~Reodejohann {\em JHEP}, vol.~1309, p.~153, 2013.

\bibitem{eji10}
H.~Ejiri {\em Prog. Part. Nucl. Phys.}, vol.~64, p.~249, 2010.

\bibitem{eji14A}
H.~Ejiri and S.~Elliott {\em Phys. Rev. C}, vol.~89, p.~055501, 2014.

\bibitem{kamlandzen}
A. Gando {\it et al.} (The KamLAND-Zen Collaboration) {\it Phys. Rev. C}, {vol.
  85}, p. 045504, 2012, arXiv:1201.4664 [hep-ex].

\bibitem{sno14}
V. Lozza {\it et al.} (The SNO-collaboration) {\it EPJ Web. Conf.}, {vol. 65},
  p. 010003, 2014.

\bibitem{geh10}
V.~Gehman, P.~Doe, R.~Robertson, D.~Will, H.~Ejiri, and R.~Hazama {\em Nuc.
  Phys. Proc. Suppl}, vol.~622, p.~602, 2010.

\bibitem{sno11}
J. Maneira {\it et al.} (The SNO Collaboration) {Nuc. Phys. Proc. Suppl.}, vol
  217, p. 50, 2011.

\bibitem{bor11}
N.~Borros and K.~Zuber {\em J. Phys.}, vol.~38, p.~10521, 2011.

\bibitem{EjiZub16}
H.~Ejiri and K.~Zuber {\em J. Phys. G: Nucl. Part. Phys.}, vol.~43, p.~045201,
  2016.

\bibitem{kla01}
H.~V. Klapdor-Kleingrothaus {\em et~al.} {\em Phys. Rev. D}, vol.~63,
  p.~073005, 2001.

\bibitem{aal02}
C.~E. Aalseth {\em et~al.} {\em Phys. Rev}, vol.~D 65, p.~092007, 2002.

\bibitem{arn05}
C.~Arnaboldi {\em et~al.} {\em Phys. Rev. Lett.}, vol.~95, p.~142501, 2005.

\bibitem{arn11}
C.~Arnaboldi {\em et~al.} {\em Astropart. Phys.}, vol.~34, p.~822, 2011.

\bibitem{EXO2012}
M. Auger {\it et al.} (The EXO Collaboration) {\it Phys. Rev. Lett.}, vol.109,
  p. 032505, 2012, arXiv:1205.5608v1 [hep-ex].

\bibitem{EXO11}
N.~Ackermann {\em et~al.} {\em Phys. Rev. Lett.}, vol.~107, p.~212501, 2011.

\bibitem{eji01}
E.~Ejiri {\em et~al.} {\em Phys. Rev. C}, vol.~63, p.~065501, 2001.

\bibitem{nemoiii05}
R. Arnold {\it et al.} (The NEMO-3 Collaboration) {\it Phys. Rev. Lett.}, {vol.
  95}, {p. 182302}, {2005}.

\bibitem{nem3A}
R. Arnold {\it et al.} (The NEMO-3 Collaboration), {\it Nucl. Phys.}, {vol.
  765}, {p. 483}, 2006.

\bibitem{Gelimit}
M. Agostini (The GERDA Collaboration), NEUTRINO 2016 conf., http:
  http://neutrino2016.iopconfs.org.

\bibitem{nem11}
R. Arnold {\it et al.} (The NEMO-3 Collaboration) {\it Phys. Rev. D}, {vol.
  89}, {p. 111101}, {2014}.

\bibitem{alf15}
K. Alfonso {\it et al.} (The CUORE Collaboration) {\it Phys. Rev. Lett.}, vol.
  102502, p. 115, 2015, arXiv:1504.02454v1[nucl-exp].

\bibitem{EXO14}
J.B. Albert {\it et al.} (The EXO Collaboration), {\it Nature}, {vol. 510}, {p.
  229}, {2014}.

\bibitem{ago13}
M.~Agostini {\em et~al.} {\em Phys. Rev. Lett.}, vol.~111, p.~122503, 2013.

\bibitem{kla04}
H.~Klapdor-Kleingrothause {\em et~al.} {\em Phys. Lett. B}, vol.~586, p.~198,
  2004.

\bibitem{ago15}
M. Agostini {\it et al.} (The GERDA Collaboration) {\it Eur. Phys. J. C}, {vol.
  75}, {p. 416}, 2015.

\bibitem{EXO15AA}
The EXO Collab. (J.B. Albert et al.) Phys. Rev. D 90 (2014) 092004.

\bibitem{EXO14A}
J. Albert {\it et al.} (The EXO collaboration) {\it Phys. Rev. C}, vol. 89, p.
  015502, 2014.

\bibitem{kamlandzenA}
A. Gando {\it et al.} (The KamLAND-Zen Collaboration) {\it Phys. Rev. Lett.},
  {vol. 110}, {p. 002502}, {2013}.

\bibitem{gan12}
A. Gando {\it et al.} (The KamLAND-Zen Collaboration) {\it Phys. Rev. C}, vol.
  86, 021601R, 2012.

\bibitem{evidence2}
H.~V. Klapdor-Kleingrothaus and I.~Krivosheina {\em Mod. Phys. Lett. A},
  vol.~21, p.~1547, 2006.

\bibitem{bar15}
A. Gando {\it et al.} (The KamLAND-Zen Collaboration) {\it Nucl. Phys. A},
  {vol. 935}, {p. 52}, {2015}.

\bibitem{wes15}
T. Wester for GERDA collaboration {\it MEDEX15 Proc. AIP conference proc.}
  (2015).

\bibitem{gue15}
{C. Cuesta {\it et al.} (The Majorana Collaboration) {\it AIP Conf. Proc.},
  vol. 1686, p. 020005, 2015, arXiv:1507.07612[physics.ins-det]}.

\bibitem{eji00a}
H.~Ejiri {\em et~al.} {\em Phys. Rev. Lett.}, vol.~85, p.~2917, 2000.

\bibitem{nak07}
H.~Nakamura {\em et~al.} {\em J. Phys. Soc. Japan}, vol.~76, p.~114201, 2007.

\bibitem{eji08}
H.~Ejiri {\em et~al.} {\em Eur. Phys. J. ST}, vol.~162, p.~239, 2008.

\bibitem{arn10}
R. Arnold {\it et al.} (The SuperNEMO Collaboration) {\it Eur. Phys. J. C},
  {vol. 70}, {p. 927}, {2010}.

\bibitem{hod15}
R. Hod{\' a}k for SuperNEMO collaboration {\it AIP Conf.Proc.}, vol. 1686, p.
  020012, 2015.

\bibitem{piq16}
F.~Piquemal Private communication 2016.

\bibitem{par15}
H. Park for AMoRE Collaboration {\it AIP Conf.Proc.}, vol. 1686, p. 020016,
  2015.

\bibitem{bar14}
A.~Barabash {\em et~al.} {\em Eur. Phys. J. C}, vol.~74, p.~3133, 2014.

\bibitem{arm15}
E.~Armengaud {\em et~al.} {\em JINST}, vol.~10, p.~PO5007, 2015.

\bibitem{bek16}
T.~B. Becker {\em et~al.} {\em Astropart. Phys.}, vol.~72, p.~38, 2016.

\bibitem{wan15}
G. Wang {\it et al.} (The CUPID Collaboration) arXiv:1504.03599
  [physics.ins-det]; arXiv:1504.03612 [physics.ins-det].

\bibitem{giu16}
A. Giuliani, Private communication, 2016; L. Canonica, NEUTRINO 2016 conf. ,
  http: http://neutrino2016.iopconfs.org.

\bibitem{zat15}
S. Zatschler for COBRA Collaboration {\it AIP Conf. Proc.}, vol. 1686, p.
  020027, 2015.

\bibitem{ebe15}
J. Ebert {\it et al.} (The COBRA collaboration) {\it Phys. Rev. C}, vol. 94, p.
  024603, 2016, arXiv 1509.04113[nucl-ex].

\bibitem{dan16}
F.~A. Danevich {\em et~al.} {\em J. Phys. Conf. Series}, vol.~718, p.~062009,
  2016.

\bibitem{gir15}
L. Gironi {\it et al.} {\it AIP Conf. Proc.}, vol. 1686, p. 020011, 2015.

\bibitem{lin15}
Y. Lin, for nEXO collaboration, {\it APR15 meeting of APS}.

\bibitem{shi13}

\newblock J. Shirai for KamLAND-Zen Collaboration {\it Nucl. Phys. Proc.
  Suppl.}, vol. 28, p. 237, 2013.

\bibitem{ino16}
K. Inoue, private communication 2016.

\bibitem{NEXT12}
V. \'{A}lvarez {\it et al.} (The NEXT Collaboration) {\it JINST}, vol. 7, p.
  T06001, arXiv:1202.0721v2 [physics.ins-det].

\bibitem{lor14}
D. Lorca for NEXT Collaboration, Proc. 20th Int. Conf. on Part. and Nucl.
  (PANIC 14): Hamburg, Germany, August 24-29, 2014,
  DOI:10.3204/DESY-PROC-2014-04/65, arXiv:1411.0475[physics.ins-det].

\bibitem{loz14}
V. Lozza {\it et al.} (The SNO+ Collaboration) {\it EPJ Web Conf.}, vol. 65, p.
  01003, 2014.

\bibitem{man11}
J. Maneira {\it et al.} (The SNO+ Collaboration) {\it Nucl. Phys. Proc.
  Suppl.}, vol. 217, p. 50, 2011.

\bibitem{kis16}

\newblock T. Iida {\it et al.} (The CANDLES Collaboration) {\it J. Phys. Conf.
  Series}, vol. 718, p. 062026, 2016.

\bibitem{gib15}
K.L.Giboni, Panda X-III collaboration, KEK seminar, Dec. 2015.

\bibitem{eji00}
H.~Ejiri {\em Phys. Rep.}, vol.~338, p.~265, 2000.
\newblock and refs. therein.

\bibitem{eji15}
H.~Ejiri and J.~Suhonen {\em J. Phys. G}, vol.~42, p.~055201, 2015.

\bibitem{eji14}
H.~Ejiri, N.~Soukouti, and J.~Suhonen {\em Phys. Lett. B}, vol.~729, p.~27,
  2014.

\bibitem{avi00}
F. Avignone {\it Workshop Neutr. Nucl. Phys. Stopped $\pi \mu $ Facility (Oak
  Ridge)}, 2000.

\bibitem{eji03a}
H.~Ejiri {\em Nucl. Instr. Meth. Phys. Research}, vol.~503, p.~276, 2003.

\bibitem{eji06}
H.~Ejiri {\em Czech. J. Phys.}, vol.~56, p.~459, 2006.

\bibitem{suh06}
J.~Suhonen and M.~Kortelainen {\em Czech J. Phys.}, vol.~56, p.~519, 2006.

\bibitem{eji13a}
I. Hashim, H. Ejiri {\it et al.} {Proc. Neutrino nuclear response workshop}
  Osaka, 2016.

\bibitem{eji68}
H.~Ejiri {\em et~al.} {\em Phys. Rev. Lett.}, vol.~21, p.~373, 1968.

\bibitem{eji13}
H.~Ejiri, S.~Titov, M.~Boswell, and A.~Young {\em Phys. Rev.}, vol.~C 88,
  p.~045610, 2013.

\bibitem{gue11}
C.~Guess {\em et~al.} {\em Phys. Rev. C}, vol.~83, p.~064318, 2011.

\bibitem{pup11}
P.~Puppe {\em et~al.} {\em Phys. Rev. C}, vol.~84, p.~051305, 2011.

\bibitem{pup12}
P.~Puppe {\em et~al.} {\em Phys. Rev. C}, vol.~86, p.~044603, 2012.

\bibitem{thi12}
J.~H. Thies {\em et~al.} {\em Phys. Rev. C}, vol.~86, p.~014304, 2012.

\bibitem{thi12a}
J.~H. Thies {\em et~al.} {\em Phys. Rev. C}, vol.~86, p.~044309, 2012.

\bibitem{thi12b}
J.~H. Thies {\em et~al.} {\em Phys. Rev. C}, vol.~86, p.~054323, 2012.

\bibitem{Freckers16}
D.~Freckers {\em et~al.} {\em Phys. Rev. C}, vol.~94, p.~014614, 2016.

\bibitem{eji16A}
H. Ejiri and D. Frekers, {\it J. Phys. G}, vol. 43, p. 11LT01, 2016.

\bibitem{eji16B}
H.~Ejiri and K.~Takahisa Private communication.

\bibitem{cap15}
F.~Cappuzzello {\em et~al.} {\em Eur. Phys. J}, vol.~A 51, p.~145, 2015.

\bibitem{SchifferGea}
J.~Schiffer {\em et~al.} {\em Phys. Rev. Lett.}, vol.~100, p.~112501, 2008.

\bibitem{eji96}
H.~Ejiri and H.~Toki {\em J. Phys. Soc. Japan}, vol.~65, p.~7, 1996.

\bibitem{eji09}
H.~Ejiri {\em J. Phys. Soc. Japan}, vol.~78, p.~074201, 2009.

\bibitem{eji12}
H.~Ejiri {\em J. Phys. Soc. Japan}, vol.~81, p.~033201, 2012.

\bibitem{bel16}
P.~Belli {\em et~al.} {\em Phys. Rev.}, vol.~C 93, p.~045502, 2016.

\bibitem{tvg15}
N.J. Rukhadze {\it et al.} {\it AIP Conf. Proc.}, vol. 1686, p. 020020, 2015.

\bibitem{KotIac12}
J.~Kotila and F.~Iachello {\em Phys. Rev. C}, vol.~85, p.~034316, 2012.

\bibitem{sim08}
F.~{\v S}imkovic, A.~Faessler, V.~Rodin, P.~Vogel, and J.~Engel {\em Phys. Rev.
  C}, vol.~77, p.~045503, 2008.

\bibitem{Calimit}
R. Arnold {\it et al.} (The NEMO-3 Collaboration) {\it Phys. Rev. D}, vol. 93,
  p. 112008, 2016.

\bibitem{Molimit}
R. Arnold {\it et al.} (The NEMO-3 Collaboration) {\it Phys. Rev. D}, vol. 92,
  p. 072011, 2015, arXiv:1506.05825[nucl-ex].

\bibitem{Ndlimit}
R. Arnold {\it et al.} (The NEMO-3 Collaboration) {\it Phys. Rev. D}, vol. 94,
  p. 072003, 2016, arXiv:1606.08494 [hep-ex].

\bibitem{Cdlimit}
F.A. Danevich {\it et al.} (The CAMEO Collaboration) {\it Nucl. Phys. Proc.
  Suppl.}, vol. 138, p. 230, 2005.

\bibitem{Telimit}
K. Alfonso {\it et al.} (The CUORE Collaboration) {\it Phys. Rev. Lett.}, vol.
  115, p. 102502, 2015.

\bibitem{SterileN14}
A.~Faessler, M.~Gonz{\' a}lez, S.~Kovalenko, and F.~{\v S}imkovic {\em Phys.
  Rev. D}, vol.~90, p.~096010, 2014.

\bibitem{Kovalenko:2009td}
S.~Kovalenko, Z.~Lu, and I.~Schmidt {\em Phys. Rev. D}, vol.~80, p.~073014,
  2009.

\bibitem{StMa09}
J.~Menendez, A.~Poves, E.~Caurier, and F.~Nowacki {\em Nucl. Phys. A},
  vol.~818, p.~139, 2009.

\bibitem{CMU16}
M.~Horoi and A.~Neacsu {\em Phys. Rev. C}, vol.~93, p.~024308, 2016.

\bibitem{IBM15}
J.~Barea, J.~Kotila, and F.~Iachello {\em Phys. Rev. C}, vol.~91, p.~034304,
  2015.

\bibitem{Fang15}
D.~Fang, A.~Faessler, and F.~{\v S}imkovic {\em Phys. Rev. C}, vol.~92,
  p.~044301, 2015.

\bibitem{QJy15}
J.~Hyv{\" a}rinen and J.~Suhonen {\em Phys. Rev. C}, vol.~91, p.~024613, 2015.

\bibitem{phfb12H}
P.~Rath, R.~Chandra, P.~Raina, K.~Chaturvedi, and J.~Hirsch {\em Phys. Rev. C},
  vol.~85, p.~014308, 2012.

\bibitem{benes05}
P.~Bene\v{s}, A.~Faessler, S.~Kovalenko, and F.~{\v S}imkovic {\em Phys. Rev.
  D}, vol.~71, p.~077901, 2005.

\bibitem{Mitra:2011qr}
M.~Mitra, G.~Senjanovic, and F.~Vissani {\em Nucl. Phys. B}, vol.~856, p.~26,
  2012.

\bibitem{Asaka:2005pn}
T.~Asaka and M.~Shaposhnikov {\em Phy. Lett. B}, vol.~620, p.~17, 2005.

\bibitem{Asaka:2005an}
T.~Asaka, S.~Blanchet, and M.~Shaposhnikov {\em Phys. Lett. B}, vol.~631,
  p.~151, 2005.

\bibitem{DusanLR}
D.~{\v S}tef{\' a}nik, R.~Dvornick{\' y}, F.~{\v S}imkovic, and P.~Vogel {\em
  Phys. Rev. C}, vol.~92, p.~055502, 2015.

\bibitem{Dev}
P.~B. Dev, S.~Goswami, and M.~Mitra {\em Phys. Rev. D}, vol.~91, p.~113004,
  2015.

\bibitem{MBK88b}
K.~Muto, E.~Bender, and H.~Klapdor {\em Z. Phys. A}, vol.~334, p.~177, 1989.

\bibitem{Mohapatra16}
R.~Mohapatra {\em Nucl. Phys. B}, vol.~908, p.~423, 2016.

\bibitem{SPVF}
F.~{\v S}imkovic, G.~Pantis, J.~Vergados, and A.~Faessler {\em Phys. Rev. C},
  vol.~60, p.~055502, 1999.

\bibitem{FMPSV11}
A.~Faessler, A.~Meroni, S.~T. Petcov, F.~{\v S}imkovic, and J.~D. Vergados {\em
  Phys. Rev. D}, vol.~83, p.~113003, 2011.

\bibitem{Meroni13}
A.~Meroni, S.~Petcov, and F.~{\v S}imkovic {\em JHEP}, vol.~02, p.~025, 2013.

\bibitem{roth04}
R.~Roth, T.~Neff, H.~Hergert, and H.~Feldmeier {\em Nucl. Phys. A}, vol.~745,
  p.~3, 2004.

\bibitem{HoroiSe14}
R.~Sen{'}kov, M.~Horoi, and B.~Brown {\em Phys. Rev. C}, vol.~89, p.~054304,
  2014.

\bibitem{HoroiTe15}
A.~Neascu and M.~Horoi {\em Phys. Rev. C}, vol.~91, p.~024309, 2015.

\bibitem{HoroiGe16}
R.~Sen{'}kov and M.~Horoi {\em Phys. Rev. C}, vol.~93, p.~044334, 2016.

\bibitem{QSky13}
M.~Mustonen and J.~Engel {\em Phys. Rev. C}, vol.~87, p.~064302, 2013.

\bibitem{kor07}
M.~Kortelainen and J.~Suhonen {\em Phys. Rev. C}, vol.~75, p.~051303(R), 2007.

\bibitem{korte1}
M.~Kortelainen, O.~Civitarese, J.~Suhonen, and J.~Toivanen {\em Phys. Lett. B},
  vol.~647, p.~128, 2007.

\bibitem{korte3}
M.~Kortelainen and J.~Suhonen {\em Phys. Rev. C}, vol.~76, p.~024315, 2007.

\bibitem{jun16}
J.~Terasaki {\em Phys. Rev. C}, vol.~93, p.~024317, 2016.

\bibitem{phfb13}
P.~Rath, R.~Chandra, K.~Chaturvedi, P.~Lohani, P.~Raina, and J.~Hirsch {\em
  Phys. Rev. C}, vol.~88, p.~064322, 2013.

\bibitem{NREDF10}
T.~Rodriguez and G.~Martinez-Pinedo {\em Phys. Rev. Lett}, vol.~105, p.~252503,
  2010.

\bibitem{schwenk16b}
S.~Stroberg, H.~Hergert, J.~Holt, S.~Bogner, and A.~Schwenk {\em Phys. Rev. C},
  vol.~93, p.~051301, 2016.

\bibitem{schwenk16a}
H.~Hergert, S.~K. Bogner, T.~D. Morris, A.~Schwenk, and K.~Tsukiyama {\em Phys.
  Rept.}, vol.~621, p.~165, 2016.

\bibitem{navratil14}
M.~Schuster, S.~Quaglioni, C.~Johnson, E.~Jurgenson, and P.~Navratil {\em Phys.
  Rev. C}, vol.~90, p.~011301, 2014.

\bibitem{navratil16}
P.~Navratil, S.~Quaglioni, G.~Hupin, C.~Romero-Redondo, and A.~Calci {\em
  Physica Scripta}, vol.~91, p.~053002, 2016.

\bibitem{qrpa1}
V.~Rodin, A.~Faessler, F.~{\v S}imkovic, and P.~Vogel {\em Phys. Rev. C},
  vol.~68, p.~044302, 2003.

\bibitem{qrpa2}
V.~Rodin, A.~Faessler, F.~{\v S}imkovic, and P.~Vogel {\em Nucl. Phys. A},
  vol.~766, p.~107, 2006.

\bibitem{Freeman12}
S.~Freeman and J.~Schiffer {\em J. Phys. G}, vol.~39, p.~124004, 2012.

\bibitem{SchifferGeb}
B.~Kay {\em et~al.} {\em Phys. Rev. C}, vol.~79, p.~121301, 2009.

\bibitem{SchifferMo}
J.~Thomas {\em et~al.} {\em Phys. Rev. C}, vol.~86, p.~047304, 2012.

\bibitem{SchifferTe}
B.~Kay {\em et~al.} {\em Phys. Rev. C}, vol.~87, p.~011302(R), 2013.

\bibitem{SchifferTeXe}
J.~Entwisle {\em et~al.} {\em Phys. Rev. C}, vol.~93, p.~064312, 2016.

\bibitem{Yako09}
K.~Yako {\em et~al.} {\em Phys. Rev. Lett.}, vol.~103, p.~012503, 2009.

\bibitem{VZ88}
P.~Vogel and M.~Zirnbauer {\em Phys. Rev. C}, vol.~37, p.~731, 1988.

\bibitem{Dusan15}
D.~{\v S}tef{\' a}nik, F.~{\v S}imkovic, and A.~Faessler {\em Phys. Rev. C},
  vol.~91, p.~064311, 2015.

\bibitem{poves12}
E.~Caurier and F.~Nowacki {\em Phys. Lett. B}, vol.~711, p.~62, 2012.

\bibitem{Barea13}
J.~Barea, J.~Kotila, and F.~Iachello {\em Phys. Rev. C}, vol.~87, p.~014315,
  2013.

\bibitem{lisi08}
A.~Faessler, G.~Fogli, E.~Lisi, V.~Rodin, A.~Rotunno, and F.~{\v S}imkovic {\em
  J. Phys. G}, vol.~35, p.~075104, 2008.

\bibitem{DeppischGA}
F.F. Deppisch and J. Suhonen, arXiv:1606.02908 [nucl-th].

\bibitem{gansm}
J.~Men{\'e}ndez, D.~Gazit, and A.~Schwenk {\em Phys. Rev. Lett.}, vol.~107,
  p.~062501, 2011.

\bibitem{gaqrpa}
J.~Engel, F.~{\v S}imkovic, and P.~Vogel {\em Phys. Rev. C}, vol.~89,
  p.~064308, 2014.

\bibitem{EngelReview}
J.~Engel {\em J. Phys. G}, vol.~42, p.~034017, 2015.

\bibitem{lisi09}
A.~Faessler, G.~Fogli, E.~Lisi, V.~Rodin, and F.~{\v S}imkovic {\em Phys. Rev.
  D}, vol.~79, p.~053001, 2009.

\bibitem{Lisi15}
E.~Lisi, A.~Rotunno, and F.~{\v S}imkovic {\em Phys. Rev. D}, vol.~92,
  p.~093004, 2015.

\end{thebibliography}

\end{document}